\documentclass[aps,prb,superscriptaddress,twocolumn,10pt]{revtex4-1}
\usepackage{graphicx}
\usepackage{wasysym}
\usepackage{color}
\usepackage{listings}
\lstdefinelanguage[10]{Mathematica}[5.2]{Mathematica}%
  {morekeywords={ImageAdjust,Image,ImageConvolve,GaussianMatrix,%
	 MaxDetect,CornerNeighbor,Binarize,ColorNegate,DiskMatrix,%
	DeleteSmallComponents,Dilation,MorphologicalComponents,%
	SelectComponents,GrowCutComponents,ImageSubtract,CornerNeighbors,%
	Colorize,ConstantArray,ComponentMeasurements,CrossMatrix,ImageCrop,%
	ImageData,Erosion,ImagePad,Parallelize,Normalize,NonlinearModelFit,%
	Quiet,FindPeaks}%
}

\begin{document}

\hyphenation{To-po-gra-phy}

\lstset{language=[10]Mathematica}

\newcommand{\comm}[1]{}
\newcommand{\didv}{\ensuremath{\mathrm{d}I/\mathrm{d}V}}
\newcommand{\iv}{\ensuremath{I(V)}}
\newcommand{\mV}{\mathrm{mV}}
\newcommand{\uV}{\ensuremath{\mu\mathrm{V}}}
\newcommand{\Vb}{\ensuremath{V_\mathrm{bias}}}
\newcommand{\CO}{(Color Online)\ }
\newcommand{\CB}[1]{\textsf{``CB{#1}''}}
\newcommand{\hl}[1]{\textcolor{red}{#1}}
\newcommand{\ot}[2]{\begin{tabular}{l} {#1}~\% \\ $#2^\circ$ \\ \end{tabular}}
\newcommand{\agir}{\ensuremath{(\sqrt{3}\times\sqrt{3})\,\mathrm{R}30^\circ}}
\newcommand{\agit}{\ensuremath{\mathit{triangular}}}
\newcommand{\agih}{\ensuremath{\mathit{hexagonal}}}
\newcommand{\ND}{NED} %Nearest neighbor network edge distribution

\makeatletter
\def\ignorecitefornumbering#1{%
     \begingroup
         \@fileswfalse
         #1%                     % do \cite comand
    \endgroup
}
\makeatother

\title{Imaging three phases of iodine on Ag (111) using low-temperature scanning tunneling microscopy}
\author{Michael Dreyer}
\email{dreyer@lps.umd.edu}
\author{Joseph Murray}
\author{Robert E. Butera}
\affiliation{Department of Physics, University of Maryland, College Park, MD 20742, USA.}
\affiliation{Laboratory for Physical Sciences, 8050 Greenmead Drive, College Park, MD 20740, USA.}

\date{\today}

\begin{abstract}
We investigated the adsorption of iodine on Ag (111) in ultra-high vacuum. Using low-temperature scanning tunneling microscopy (LT-STM) measurements we catalog the complex surface structures on the local scale. We identified three distinct phases with increasing iodine coverage which we tentatively associate with three phases previously reported in LEED experiments (\agir, \agit, \agih). We used Fourier space and real space analysis to fully characterize each phase. While Fourier analysis most easily connects our measurements to previous LEED studies, the real space inspection reveals local variations in the superstructures of the \agih\ and \agit\ phase. The latter, observed here for the first time by LT-STM,  appears to be stabilized by one or two adatoms sitting at the center of a rosette-like iodine reconstruction. The most stunning discovery is that variation in the adatom separation of the \agit\ phase reconstructs the Ag (111) surface lattice.
\end{abstract}

\maketitle

\section{Introduction}

%\cite{YAMADA1996321}
%Continuous variation of iodine adlattices on Ag(111) electrodes: in situ STM and ex situ LEED studies (1996)
%Single crystal
%in situ wet chemical STM, LEED, AES
%STM: only r3 like, LEED: r3, tri, hex depending on deposition potential

%\cite{DICENZO1982411}
%XPS STUDIES OF ADATOM-ADATOM INTERACTIONS: I/Ag( 111) AND I/Cu(lll) (1982)
%Ag (111), I vapor at 10-8 Torr
%LEED, XPS
%measure next-nearest neighbor interaction

%\cite{PhysRevLett.41.598.3}
%Extended X-Ray-Absorption Fine Structure of Surface Atoms on Single-Crystal Substrates: Iodine Adsorbed on Ag(111) (1978)
%I at 2*10-8 torr on single crystal, monitored by LEED
%surface-extended X-ray absorption fine structure (SEXAFS)
%measure bond length

Halogens on silver are of interest for catalytic purposes and for fundamental understanding of interaction of active gases with metal surfaces\cite{JONES198825,ANDRYUSHECHKIN201883}. Furthermore, on-surface synthesis
% has seeded an interest in using surfaces to
enables the formation of planar products and highly reactive substances that are inaccessible using traditional heterogeneous chemistry. To date, on-surface synthesis has focused on synthesizing carbon nanostructures from polycyclic hydrocarbons, with a large emphasis on using Ullmann-type reactions to couple organohalogens on metal surfaces such as Au, Ag and Cu. \cite{surfchem1} The structure of the metal surface \cite{surfchem2} and the pattern formed by self-assembly of the deposited precursors \cite{surfchem3} have both been shown to play a role in templating the nanostructure.

In the established on-surface Ullmann reactions, a high temperature anneal is required to activate the carbon--halogen bond and induce coupling of the pre-deposited precursors. This step also leaves a layer of halogen atoms on the substrate, which diffuses between the nanostructure and substrate resulting in both establishing a desirable electronic separation between the structure from the metal surface, and undesirably passivating the surface to further reaction. \cite{surfchem4} In fact, further work intentionally exposes the substrate post-reaction to additional iodine vapor to post-synthetically decouple the product from the substrate. \cite{surfchem5} 

A recent study overcomes the passivation challenge, by demonstrating that aryl coupling can be performed on a passivated surface, specifically an Ag(111) surface passivated with an inert chemisorbed iodine layer, by using an already activated aryl radical precursor in place of an aryl-halogen. \cite{surfchem6} The chemisorbed iodine layer was assumed to form the well-known hexagonal \agir\ I-Ag(111) lattice.\cite{ANDRYUSHECHKIN201883}. In this work, we demonstrate that the I-Ag(111) surface can be more complex than a single lattice structure. We show real-space STM images and analysis of three phases we associate with the three structures found by LEED\cite{BARDI1983145,YAMADA1996321} with increasing coverage. To our knowledge, we present the first real-space images of the \agit\ reconstruction. In light of observations that heating can transition the LEED pattern from \agit\ to \agih\ and back\cite{BARDI1983145} we discuss the possibility that the \agit\ phase might be responsible for both phases. Through real space analysis of nearest-neighbor networks of the top monolayer as well as the superstructures we can relate the \agih\ superstructure to the top iodine layer. In contrast, the \agit\ superstructure is tied to the Ag (111) surface atoms. We further show that for either phase iodine can easily be manipulated by the STM tip. As the surface is known to play a role in templating Ullmann reactions, understanding the morphology of the passivating halogen layer is important for pursuing radical reactions on halogen passivated metal substrates. 

\section{Experimental techniques}

STM characterization was performed in a home-built low-temperature STM system\cite{our_4K_system}. In this system the bias voltage is applied to the sample while the tip is held at virtual ground by the transimpedance amplifier. The Ag (111) single crystal was cleaned in ultra-high vacuum (UHV) by cycles of argon ion etching at an energy of 750~eV while heating the sample to 450~$^\circ$C as measured using an optical pyrometer. After the initial preparation, cleaning for $\sim30$ minutes proved sufficient to refresh the surface. The surface quality was verified by measuring scanning tunneling spectroscopy (STS) maps of the Ag surface state where any contamination will act as a scattering center for the two dimensional electron gas. After cleaning, the sample was left to cool on the manipulator for 10 minutes prior to iodine exposure. The temperature was below the detection limit of our optical pyrometer of 250~$^\circ$C, but was presumed to be above room temperature. Iodine was deposited from an AgI electrochemical cell\cite{iodinesource} at a rate of about $0.083$~ML/min. During the deposition the pressure inside the UHV chamber stayed in the low $10^{-10}$~mbar range.
% The sample was not deliberately heated during iodine exposure.
Afterward, the sample was transferred into the STM and imaged at a temperature of 77~K -- unless stated otherwise. For each iodine coverage the preparation process was repeated. During scanning iodine was randomly picked up by the tip which enhanced STM imaging.

\section{Results and discussion}

\begin{figure}%
\agir\hfill\agit\hfill\agih

\includegraphics[width=.32\columnwidth]{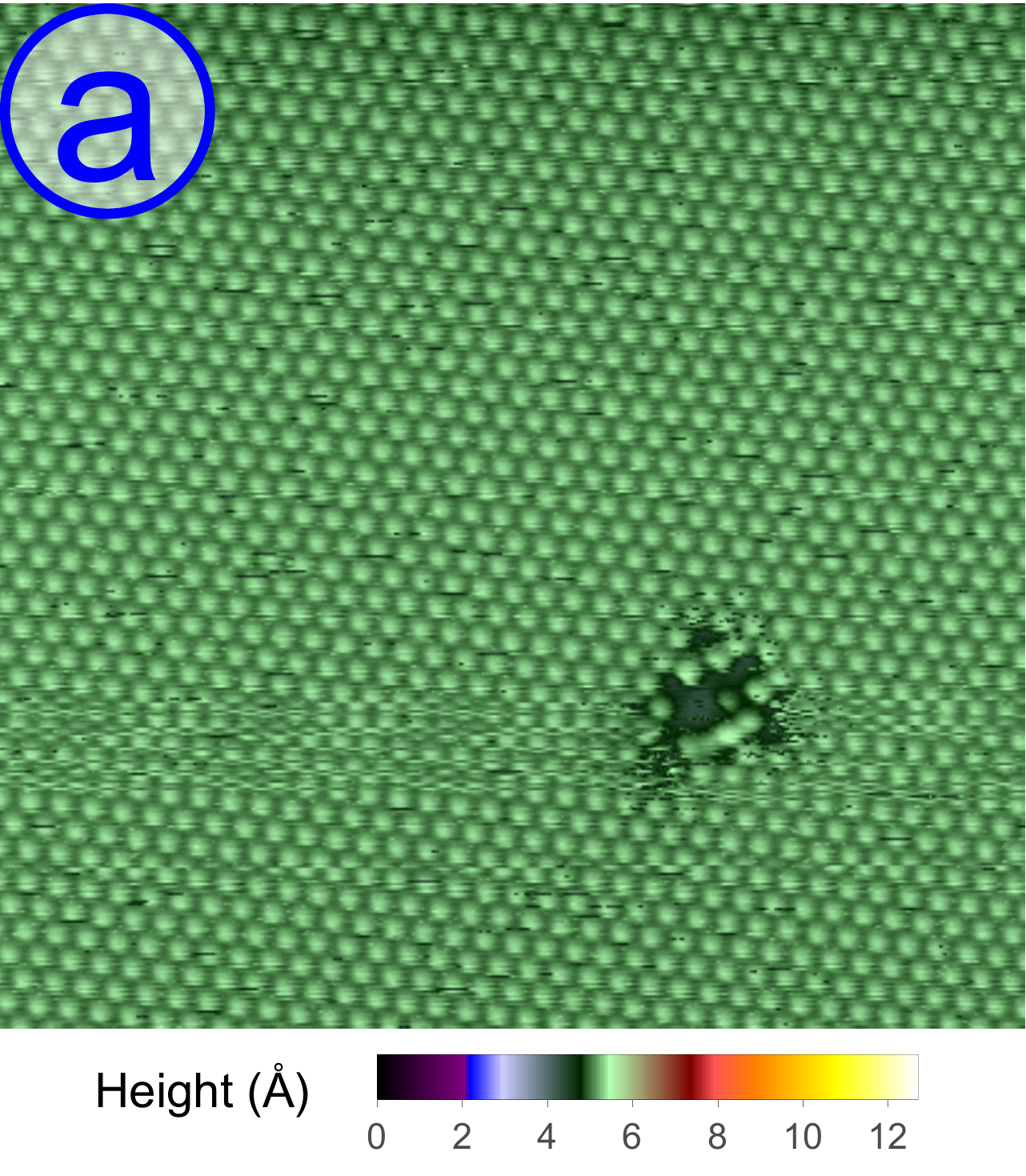}\hfill%
\includegraphics[width=.32\columnwidth]{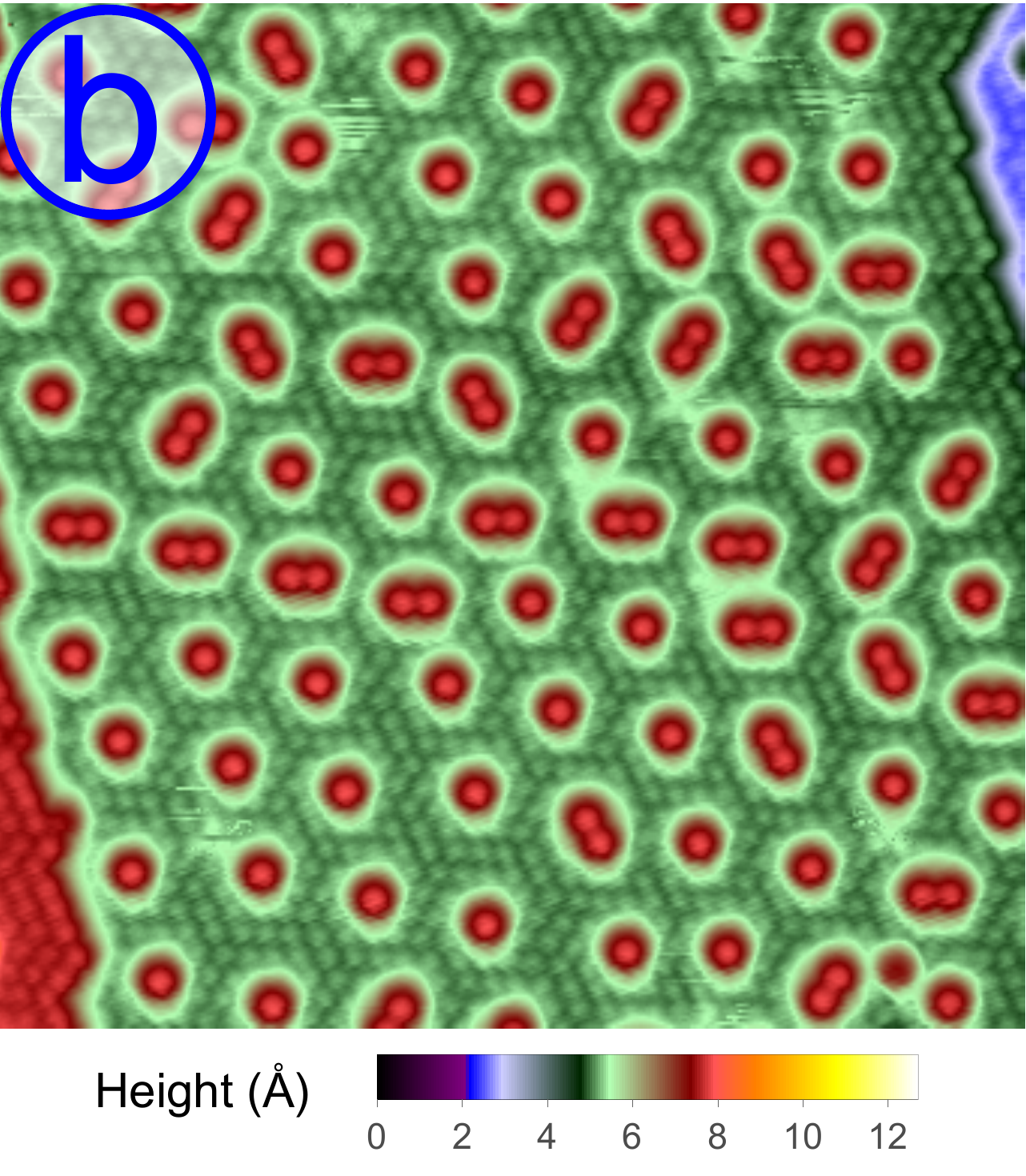}\hfill%
\includegraphics[width=.32\columnwidth]{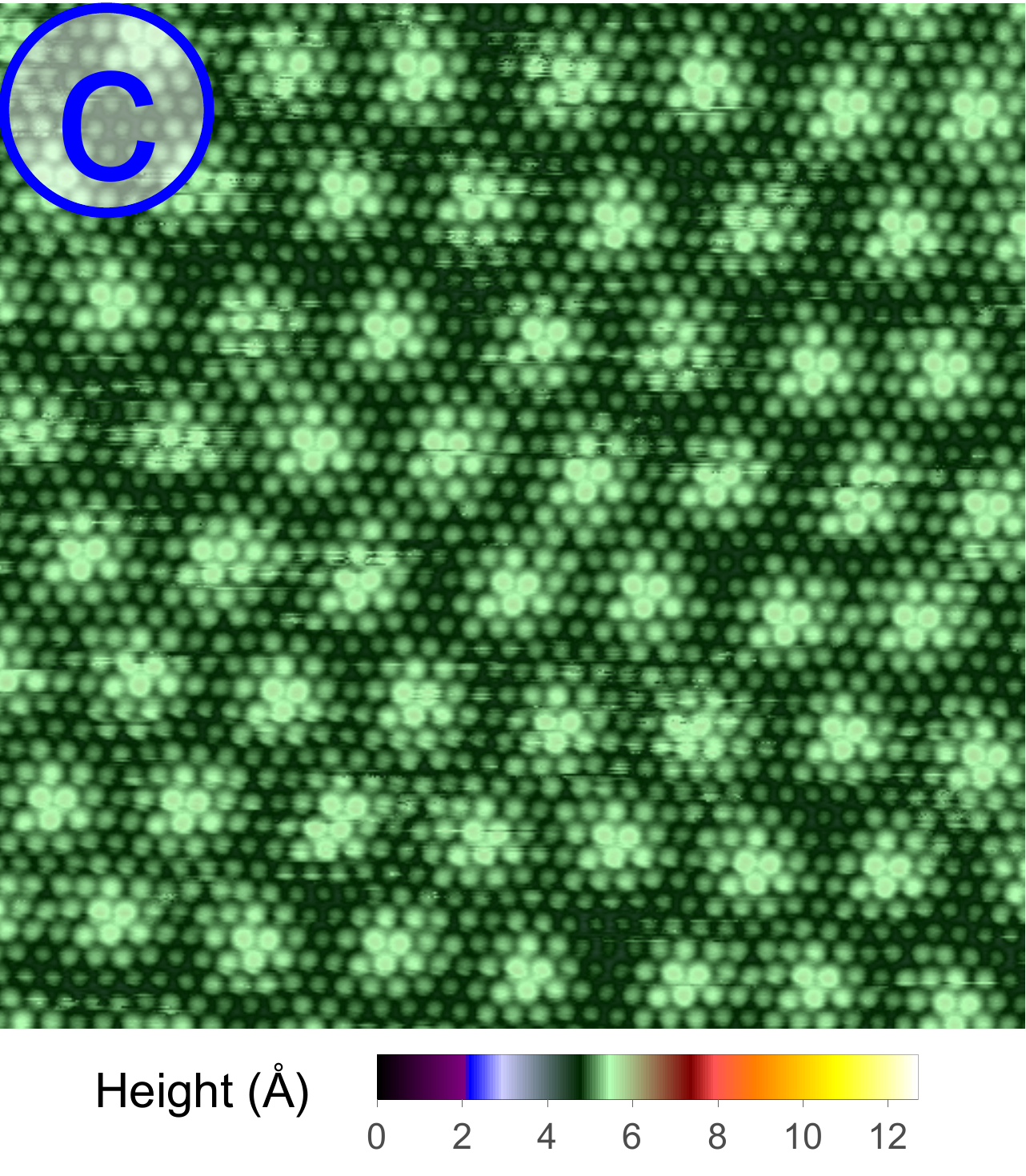}

\smallskip
\includegraphics[width=.32\columnwidth]{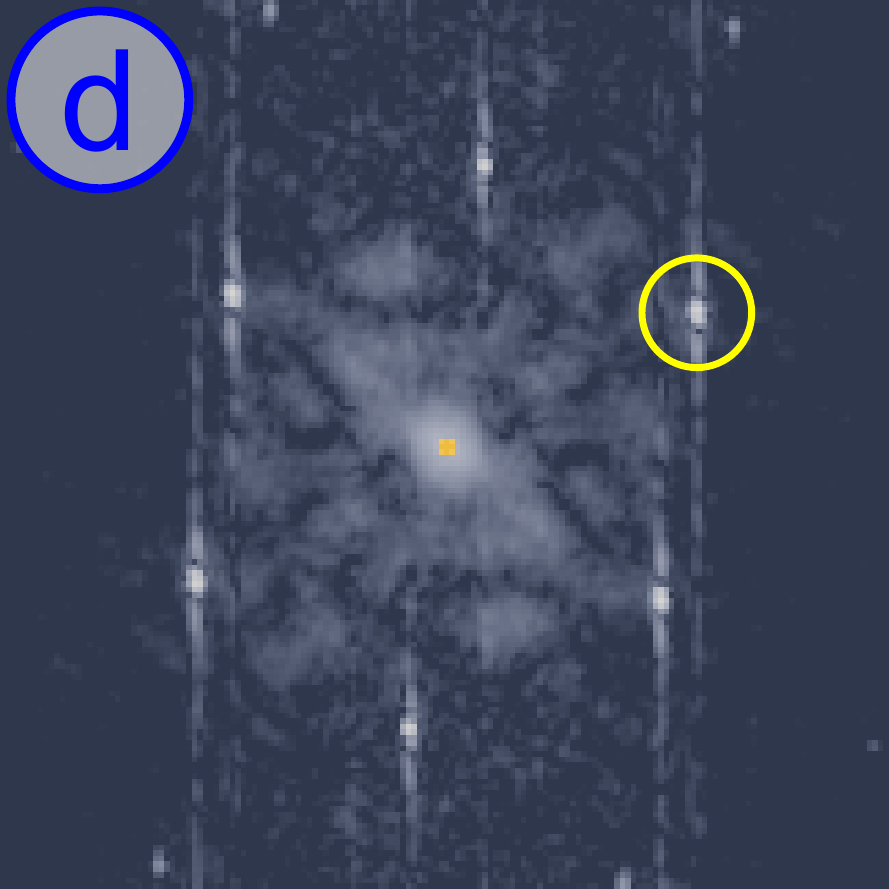}\hfill%
\includegraphics[width=.32\columnwidth]{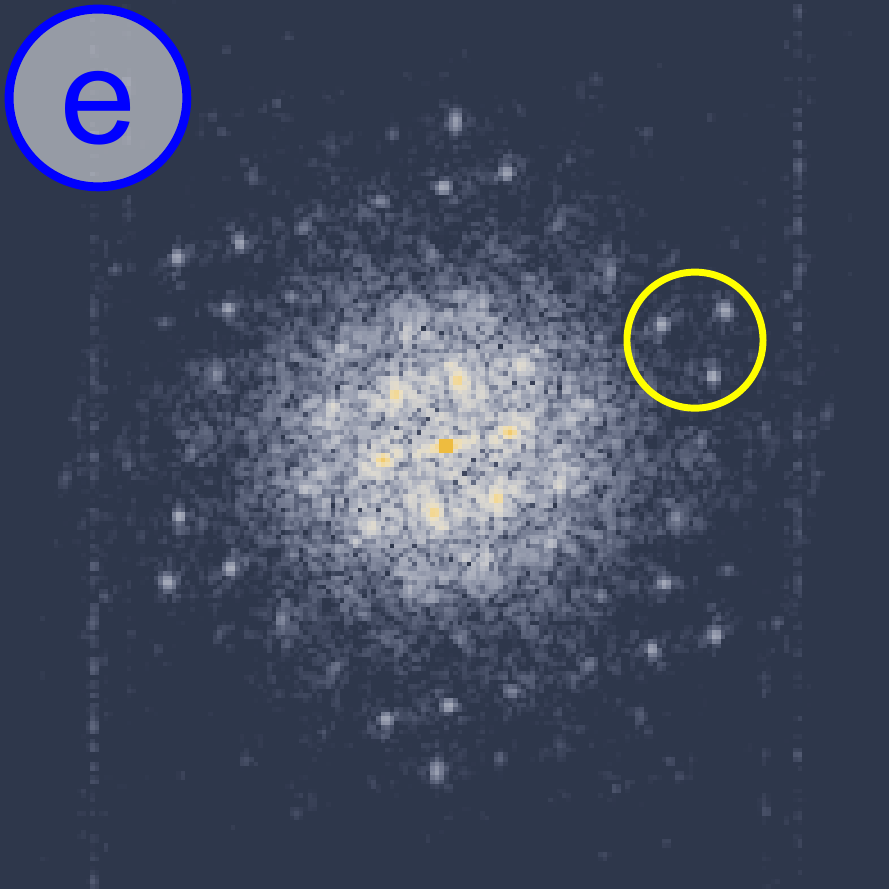}\hfill%
\includegraphics[width=.32\columnwidth]{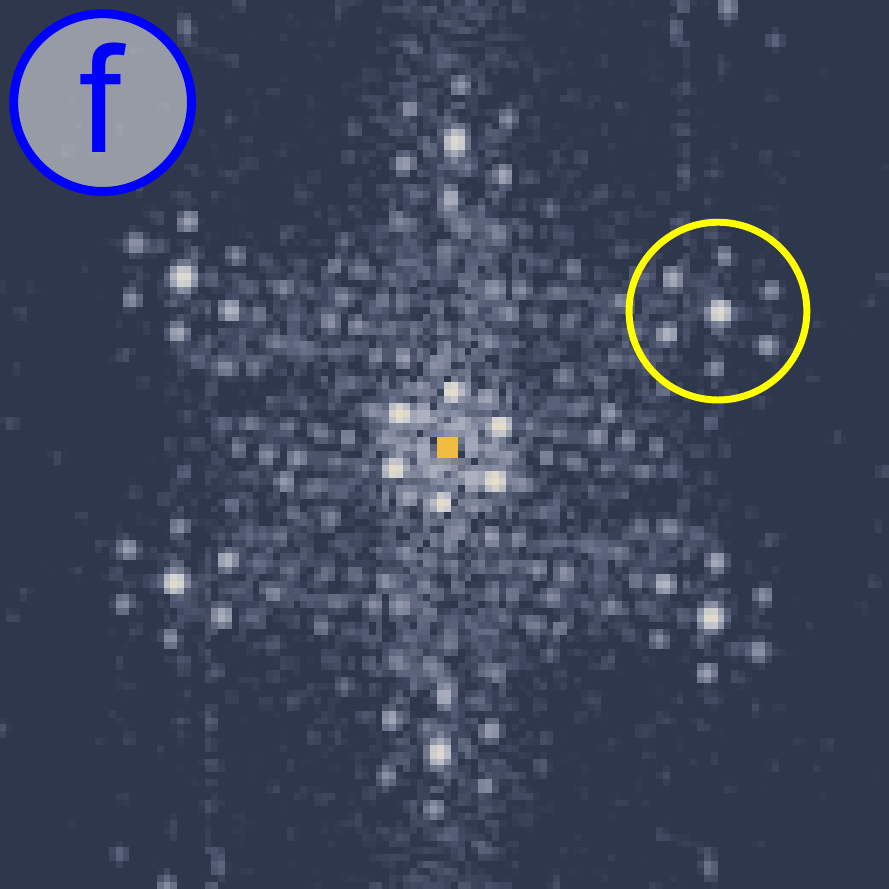}%
\caption{\CO Three phases of iodine on silver with increasing coverage. a)-c) show representative LT-STM images we associate with the \agir, \agit, and \agih\ phase, respectively. d)-f) show the center regions of the corresponding FFT images. (Imaging parameters: $20\times20$~nm$^2$ each, (a) $\Vb=-100$~mV, $I=1.5$~nA, (b) $\Vb=500$~mV, $I=200$~pA, (c) $\Vb=-50$~mV, $I=100$~pA)}\label{fig:AgIFFT}%
\end{figure}

\begin{figure}%
\includegraphics[width=.95\columnwidth]{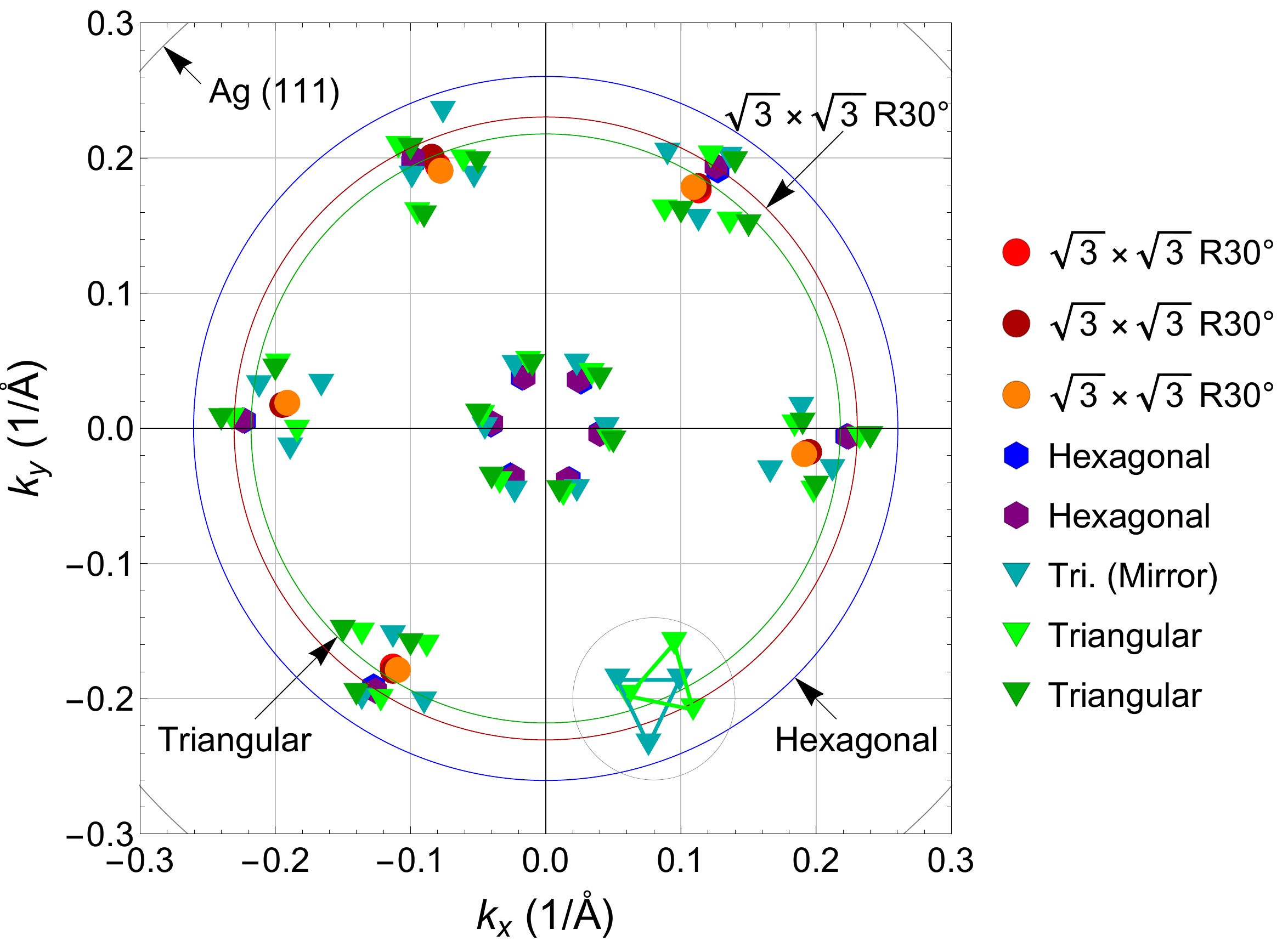}%
\caption{\CO Summary of major FFT spots for eight different STM images covering the three phases of iodine on Ag (111). The large colored circles mark the expected wave vectors according to the published LEED measurements. While the six spots of the \agih\ phase were averaged to reduce clutter, we included the three spots for the \agit\ phase at the respective outer position. The inner spots stem from the superstructure of the \agih\ and the \agit\ phase. In the bottom right position (gray circle) only the FFT spots of two \agit\ domains with opposite orientations are shown.}\label{fig:FFTsum}%
\end{figure}

%A fair amount of work, including STM, has been performed on the Cl/Ag(111) system.
The I/Ag(111) system has been extensively studied with electron diffraction techniques.  Multiple phases have been identified with LEED, namely \agir, \agit, and \agih\cite{BARDI1983145,YAMADA1996321} which were observed with increasing iodine coverage. However, relatively few experiments have sought to identify the corresponding atomic structure for each of these phases with real space imaging techniques, such as STM. Some STM studies were performed in ambient conditions\cite{doi:10.1021/j100052a049,doi:10.1021/la00014a024} as well as in ultra-high vacuum\cite{YAMADA1996321,doi:10.1021/jp0513465} and found the \agir\ structure. One study described a ``compressed \agir'' structure\cite{ANDRYUSHECHKIN201883} which, we believe, is consistent with our images of the \agih\ phase. For the \agir\ phase there is broad agreement of iodine sitting in the FCC (ABCA ...), not the HCP (ABA...), hollow sites of silver (111)\cite{FARRELL1981527,MAGLIETTA1982141,PhysRevLett.30.17}. The atomic structure of the \agit\ and \agih\ phase have not been reported using STM.

Figure~\ref{fig:AgIFFT} shows three representative STM images and the respective fast Fourier transformed (FFT) images which we associate with the three phases detected by LEED\cite{BARDI1983145}. The FFT images match the corresponding LEED patterns well. The measured length scales are also of the same order as the LEED observations. This can be seen in Fig.~\ref{fig:FFTsum} and Tab.~\ref{tab:LEED} which summarize the major FFT spots from eight different STM images. The FFT spots were rotated to compensate for the scan angle of the respective STM data set. The remaining differences in sample orientation are due to slight angle variations when locking the sample stud into the microscope. In particular the sample can be put into the microscope in two coarse orientations $180^\circ$ apart -- usually at random. This, however, should not make any difference in FFT for the (111) surface of a cubic close-packed crystal such as Ag, the \agir, \agih, or \agit\ structure. The \agit\ structure, however, appears in FFT with two different orientations of the triangle spots. We attribute this to a mirrored reconstruction. We will discuss these mirror domains and their consequences later. The inner FFT peaks of the \agit\ and \agih\ phases represent a superstructure. While the orientation is similar, the periodicity of the two superstructures with $\overline{l}_\mathrm{\agit}=19.9$~\AA\ and $\overline{l}_\mathrm{\agih}=24.0$~\AA\ is clearly different (see App.~\ref{app:FFTinner}, Fig~\ref{fig:FFTsum2}).

In addition to the FFT analysis we measured the individual atomic distances in real-space (See App.~\ref{app:realspace}). As shown in Tab.~\ref{tab:LEED}, the mean real-space distances reproduce the iodine-iodine distances ($d_\mathrm{I-I}$) from FFT for the \agir\ and \agih\ phase. However for the \agit\ phase we obtain the presumed FFT row-row distance ($d_\mathrm{R-R}$) which is in excellent agreement with the published LEED values. This indicates that, in this case, FFT does not measure a row to row distance, as is expected for a triangular lattice, but rather the inter-atomic distance directly. This is likely due to the adatoms concealing part of the atomic lattice leaving mostly long stripes of the monolayer visible. The real-space measurements of the inter-atomic distances for the three reconstructions did not reveal any additional structure (See App.~\ref{app:realspace}). In contrast, real-space measurements of the \agih\ and \agit\ superstructures show substructures of their own (Section \ref{sec:tandh}).

\begin{table}%
\begin{tabular}{l||c|c||c|c||c}
Reconstruction&\multicolumn{2}{c||}{LEED (\AA)}&\multicolumn{2}{c}{FFT (\AA)}&RS (\AA)\\
              &$d_\mathrm{I-I}$&$d_\mathrm{R-R}$&$d_\mathrm{R-R}$&$d_\mathrm{I-I}$&$d_\mathrm{I-I}$\\
\hline
\agir&5.00      &4.43      &4.88&5.64&5.79\\
\agih&4.43      &3.83      &4.44&5.13&5.19\\
\agit&4.59--5.00&3.98--4.43&4.82&5.57&4.87\\
\end{tabular}
\caption{Comparison of the periodicity according to LEED\cite{BARDI1983145} and obtained in this LT-STM study. $d_\mathrm{I-I}$ denotes the inter-atomic distance of iodine. $d_\mathrm{R-R}$ denotes the iodine row distance assuming an equilateral triangular lattice ($d_\mathrm{R-R}=\sqrt{3}/2\cdot d_\mathrm{I-I}$). The last column (RS) gives the mean value of real-space distance measurements.}
\label{tab:LEED}
\end{table}

\subsection{Up to $1/3$~ML (\protect\agir)}

\begin{figure}%
%\includegraphics[width=.32\columnwidth]{IAg_sml.pdf}\hfill%
%\includegraphics[width=.32\columnwidth]{IAg_1Ml1.pdf}\hfill%
%\includegraphics[width=.32\columnwidth]{IAg_1Ml2.pdf}%
%\caption{\CO Iodine on silver at increasing coverage $\theta$ up to $1/3$~ML. a) Initial edge decoration including adsorption at an Ag screw dislocation. b) $\theta <1/3$~Ml, edges stabilize the iodine structure on the terraces. c) $\theta \sim 1/3$~Ml, mostly stable iodine with localized iodine motion ({\em e.g.} arrow). (Imaging parameters: (a) $44\times44$~nm$^2$, $\Vb=-500$~mV, $I=100$~pA, (b) $25\times25$~nm$^2$, $\Vb=-100$~mV, $I=500$~pA, (c) $50\times50$~nm$^2$, $\Vb=-80$~mV, $I=500$~pA)}\label{fig:AgIthin}%
\includegraphics[width=.495\columnwidth]{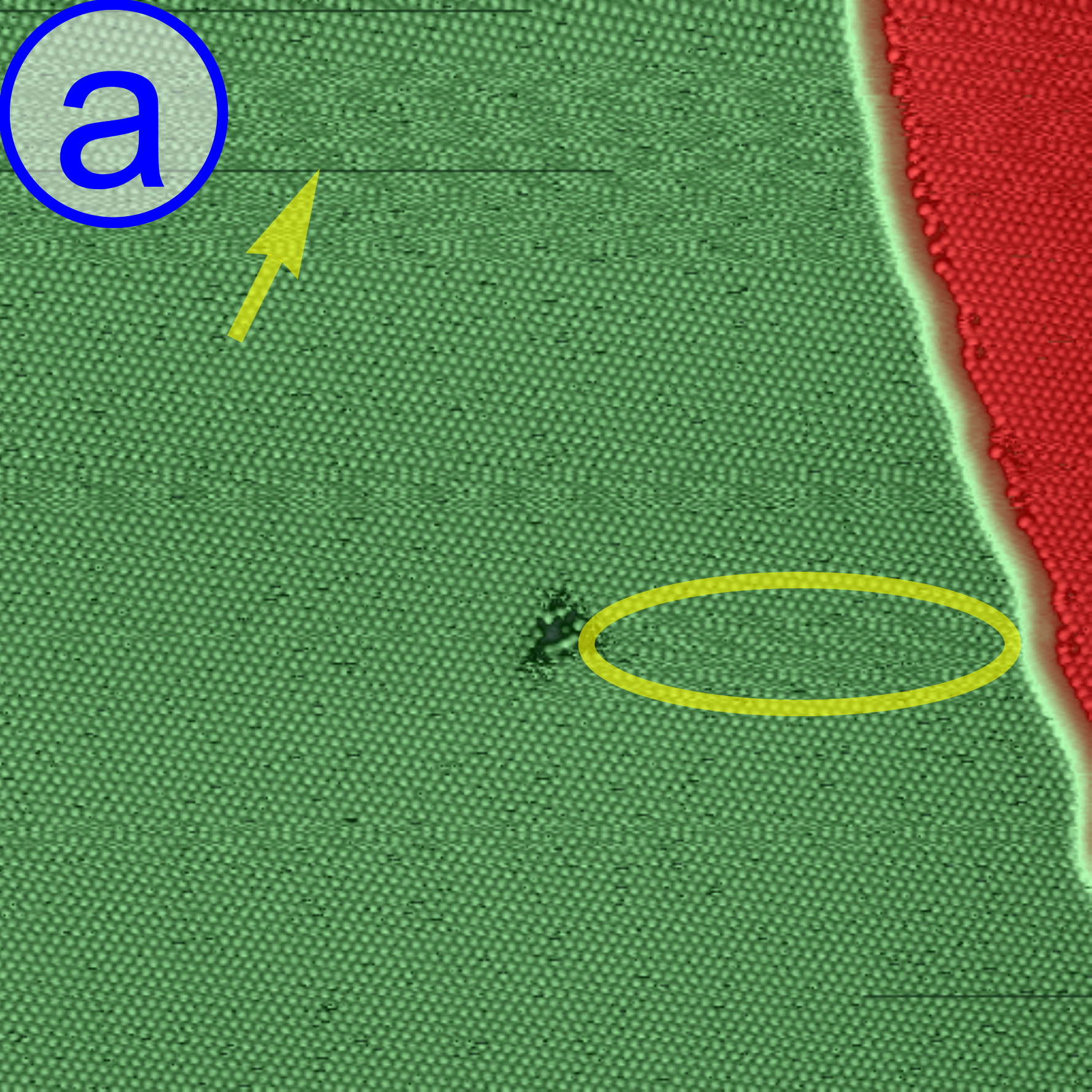}\hfill%
\includegraphics[width=.495\columnwidth]{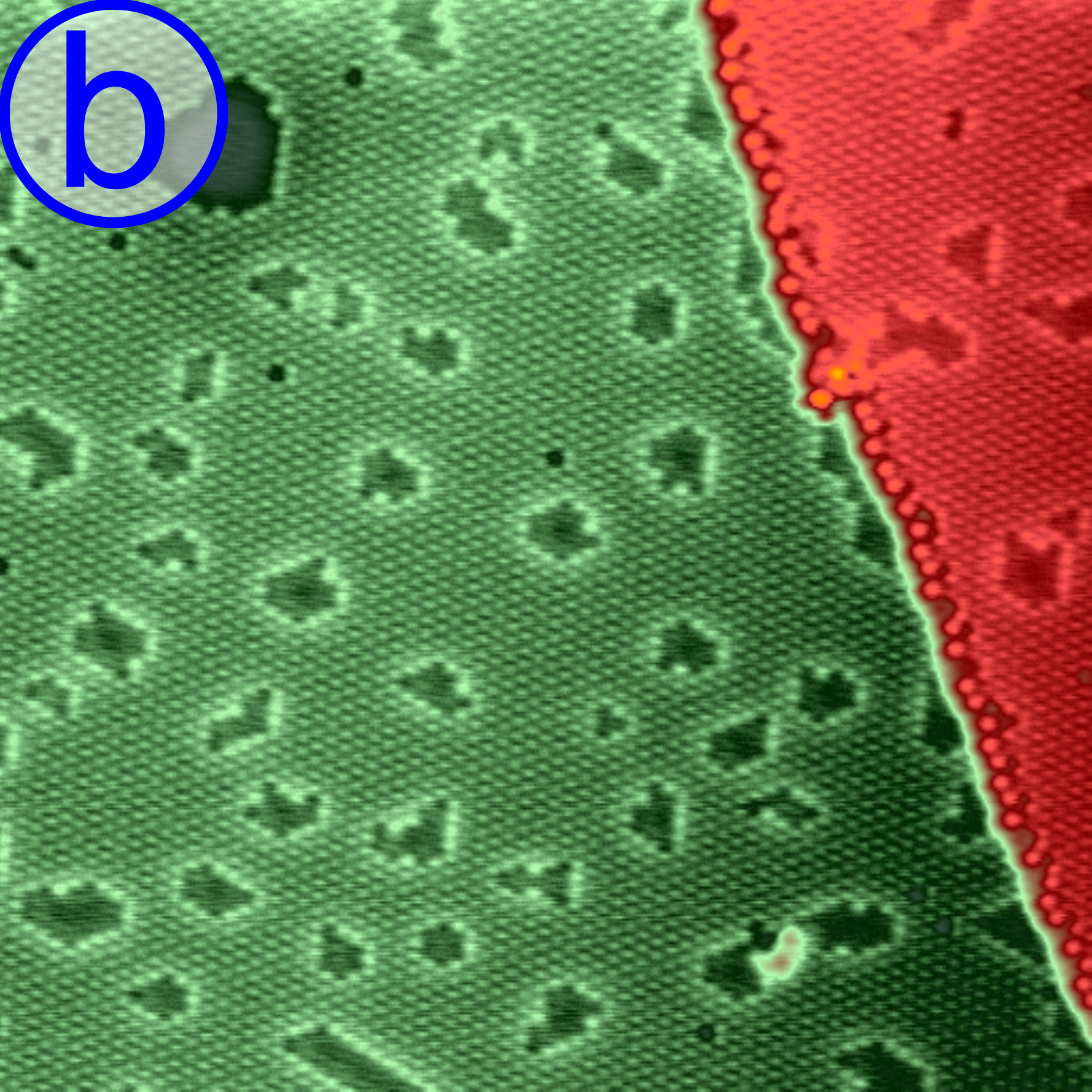}%
\caption{\CO Two Ag (111) samples with slightly different iodine coverage measured at temperatures of a) 77~K and b) 4~K, respectively. a) Mostly stable iodine with localized iodine motion ({\em e.g.} ellipse) at almost $1/3$~ML coverage. The arrow indicates a temporary tip change. b) in contrast, at 4~K iodine is stable even below $1/3$~ML with holes exposing the Ag surface. (Imaging parameters: (a) $50\times50$~nm$^2$, $\Vb=-80$~mV, $I=500$~pA, (b) $41\times41$~nm$^2$, $\Vb=1000$~mV, $I=100$~pA)}\label{fig:AgIthin}%
\end{figure}

Figure~\ref{fig:AgIthin} shows representative images of the \agir\ reconstruction found at coverages of iodine up to $1/3$~ML. The images were taken at temperatures of (a) 77~K and (b) 4~K. Initially iodine is highly mobile at 77~K although it is difficult to distinguish between tip induced and thermal motion. Consequently, iodine is at first only visible as edge decoration.
% and stuck to dislocation lines.
As the coverage  approaches $1/3$~ML stable regions can be found on small Ag terraces. Larger terraces appear noisy due to the iodine motion. Even close to $1/3$~ML, where most areas show atomic order, some regions exhibit signs of motion as the patterns shift while scanning. One example is indicated by the ellipse in Fig.~\ref{fig:AgIthin}a). To the left of the defect the iodine lattice is intact while on the right side it appears distorted. The image also shows dark lines (arrow) which indicate tip changes where a presumed iodine atom at the tip apex gets pushed out of the way leading to a height change of $0.7$~\AA. This is distinctly different from the iodine motion on the sample surface. Additionally, the tunneling current became increasingly noisy for bias voltages above $1.0$~V (applied to the sample) indicating tip induced iodine motion. No such effect exists for negative bias voltages.
  
In contrast, at 4~K (Fig.~\ref{fig:AgIthin}b), iodine forms stable areas even at coverages well below $1/3$~ML. Since the sample was prepared at room temperature, the ordering must occur while the sample is cooling down after being placed into the STM. Since such stable regions were not observed at 77~K, the iodine motion appears to freeze out between 77~K (6.6~meV) and 4~K (0.34~meV).

The \agir\ phase has been observed multiple times by STM in different environments\cite{doi:10.1021/j100052a049,doi:10.1021/la00014a024,YAMADA1996321,doi:10.1021/jp0513465}. It shows no superstructure and has established iodine positions on Ag (111). However, the island structures found at a temperature of 4~K for the first time might deserve further study in terms of tip-induced iodine motion and the presumably size-dependent electronic structure. 
 
\subsection{\protect\agit\ and \protect\agih\ phases}
\label{sec:tandh}

\begin{figure}%
\includegraphics[width=.62\columnwidth]{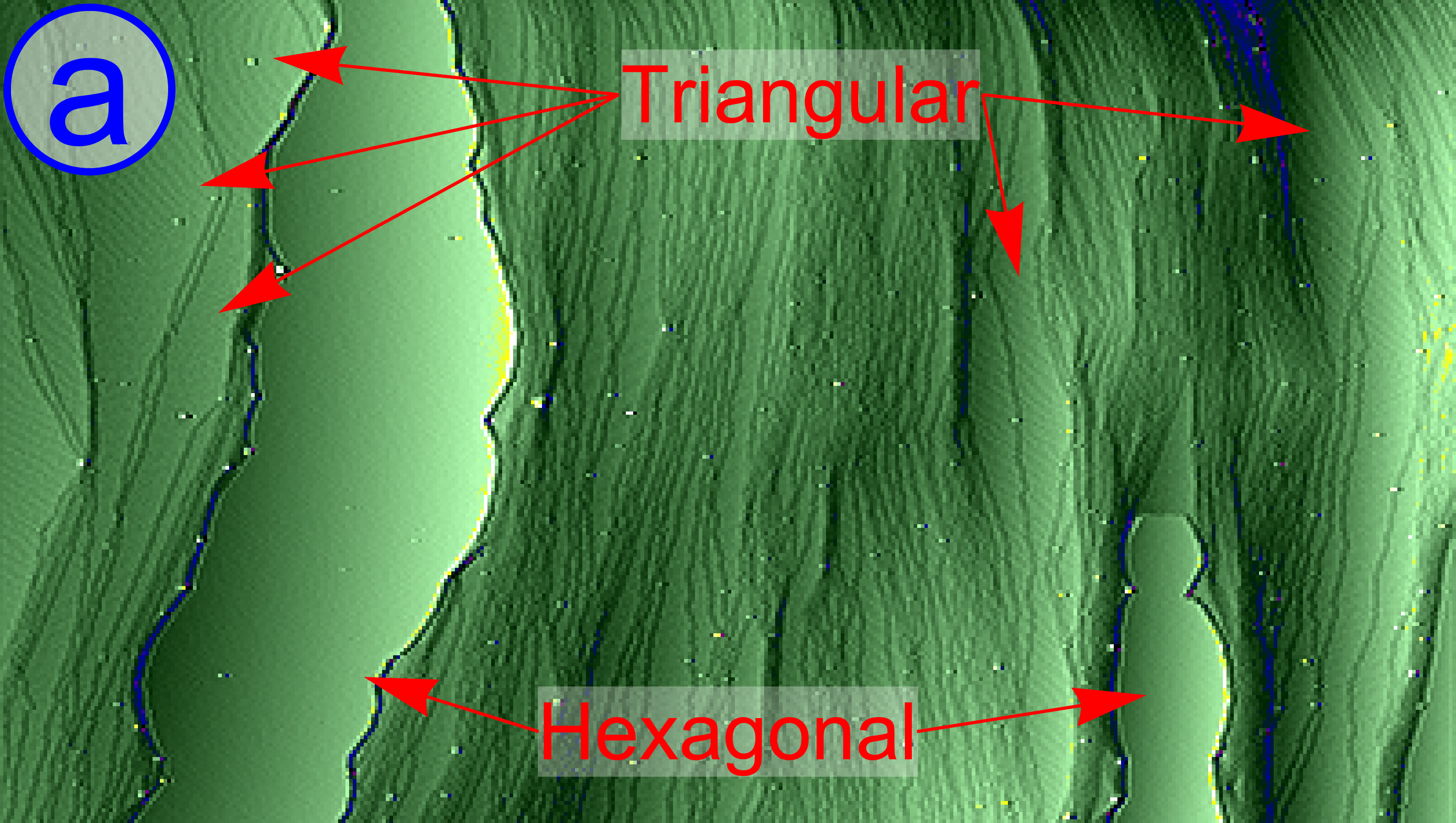}\hfill%
\includegraphics[width=.35\columnwidth]{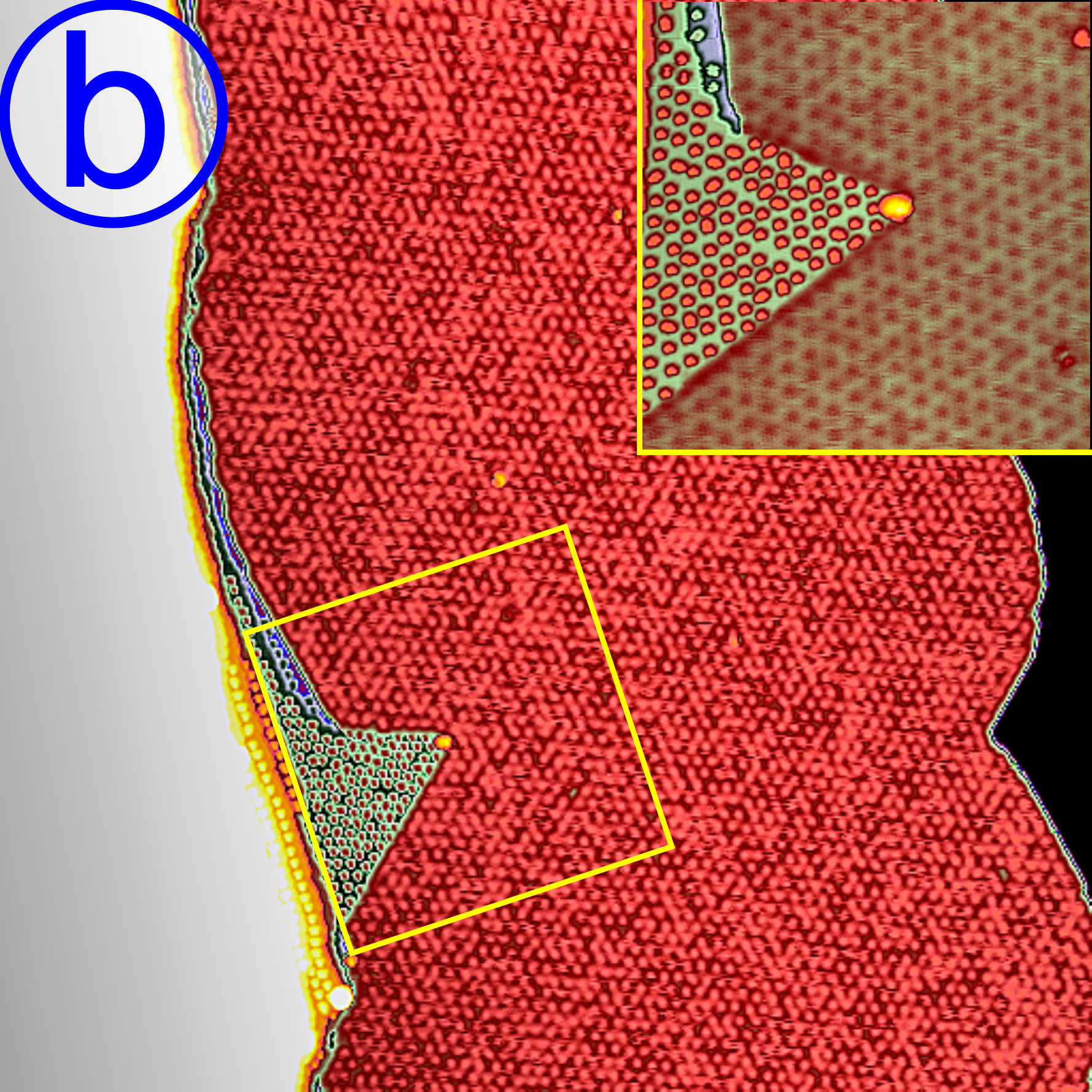}%
\caption{\CO Coexistence of the \agit\ and \agih\ phase at coverages above $1/3$~ML. (Imaging parameters: (a) $955\times625$~nm$^2$, $\Vb=2.5$~V, $I=100$~pA, (b) $200\times200$~nm$^2$, $\Vb=2.0$~V, $I=100$~pA, inset: $50\times50$~nm$^2$, $\Vb=1.25$~V, $I=200$~pA, )}\label{fig:AgIMixed}%
\end{figure}

At coverages above $1/3$~ML we usually observed the coexistence of two reconstructions (phases) as shown in Fig.~\ref{fig:AgIMixed}. No attempt was made to create a uniform coverage of either phase since it was not necessary for this STM study. The \agit\ phase is preferentially found on the smaller terraces. The step edges remain straight with $120^\circ$ angles as is typical for a Ag (111) single crystal surface. Conversely, the \agih\ phase is found on larger terraces with curved edges. This distinction allows one to identify either phase in larger STM scans that lack atomic resolution. Fig.~\ref{fig:AgIMixed}b) shows one of the few cases where the \agih\ phase grew up to the edge of a terrace covered by the \agit\ phase. The expansion of the \agih\ phase was likely hampered by a particle on the surface. The apparent step height between the two phases in Fig.~\ref{fig:AgIMixed}b) varies nonlinear with bias voltage from $0.8$~\AA\ to $2.6$~\AA\ (See App.~\ref{app:hexsuper}, Fig.~\ref{fig:HvsV}). This indicates, that the \agih\ phase is more insulating than the \agit\ phase. 

\subsubsection{\protect\agih\ phase}

\begin{figure}%
\includegraphics[width=.48\columnwidth]{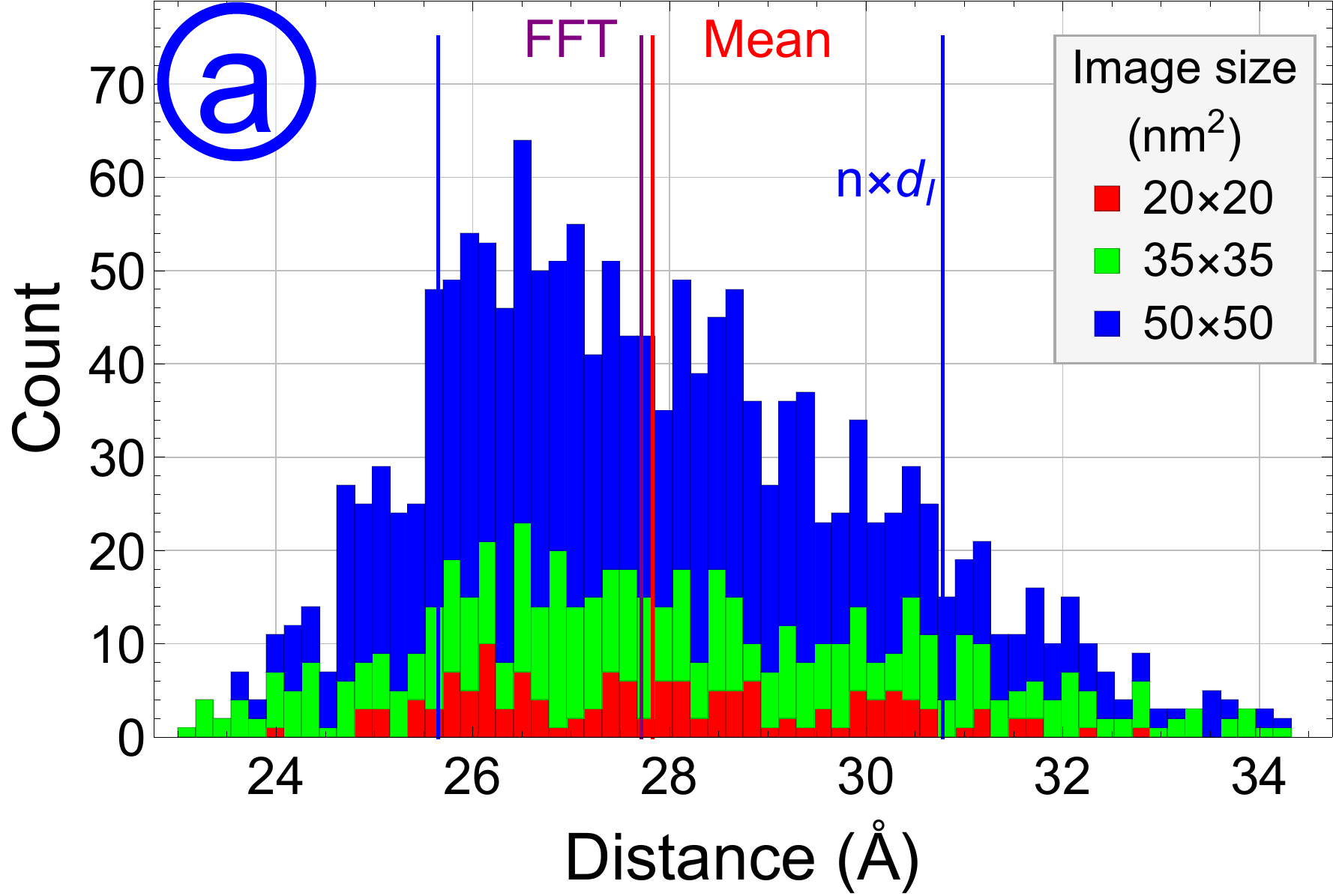}\hfill%
\includegraphics[width=.495\columnwidth]{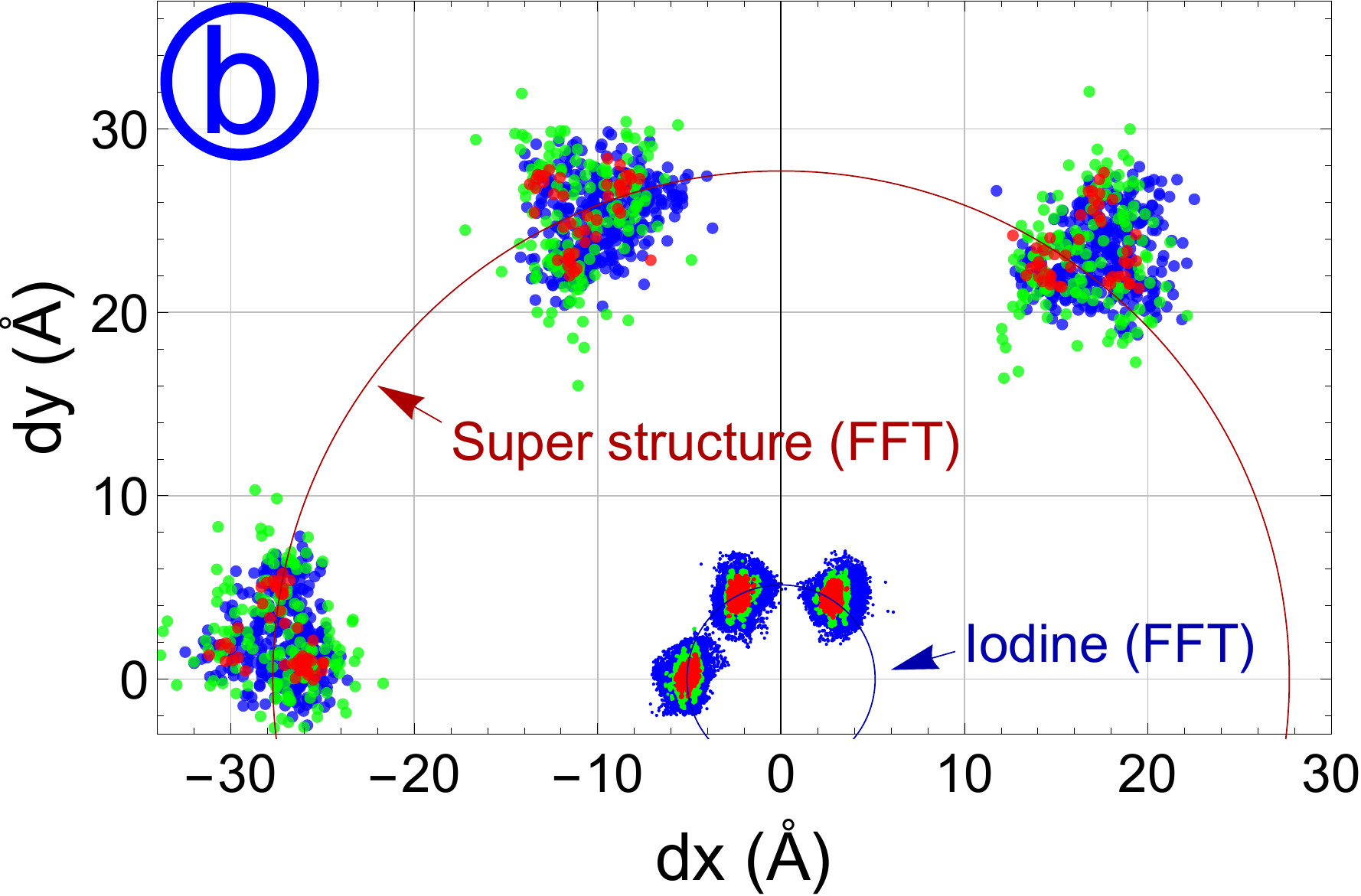}%
\caption{\CO a) Histogram of the super structure lattice constant measured from three STM data sets with lateral scales of $20 \times 20$~nm$^2$, $35 \times 35$~nm$^2$, and $50 \times 50$~nm$^2$ (See App.~\ref{app:hexsuper}, Fig.~\ref{fig:AgIHexSource}). b) Scatter plot of the real space distances (Nearest neighbor network vector distribution) between superstructure maxima alongside the iodine inter-atomic distances. The circles represent the average length scale as measured by FFT.}\label{fig:AgIhd1}%
\end{figure}

\begin{figure}%
\includegraphics[width=.6\columnwidth]{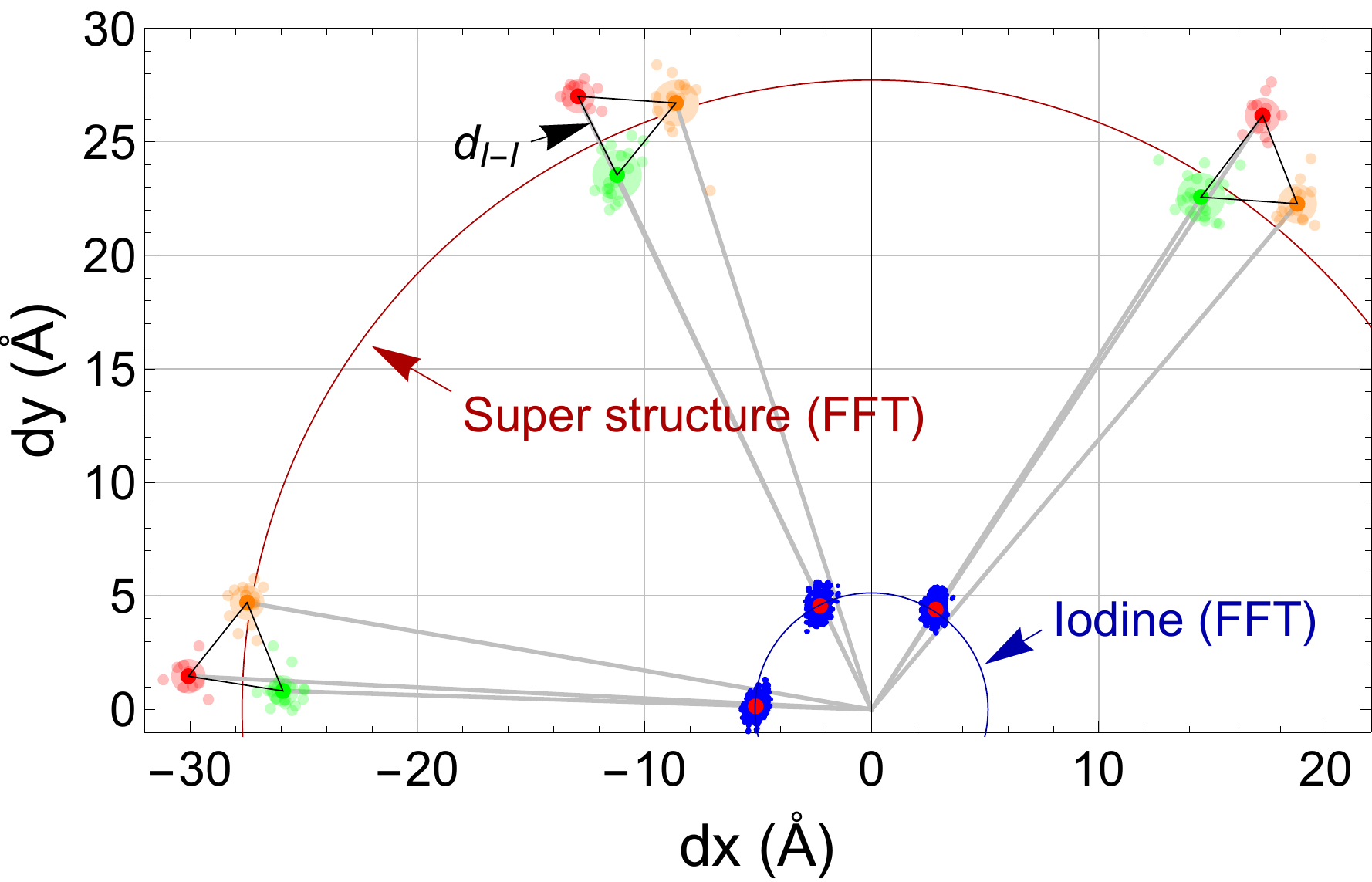}
\includegraphics[width=.35\columnwidth]{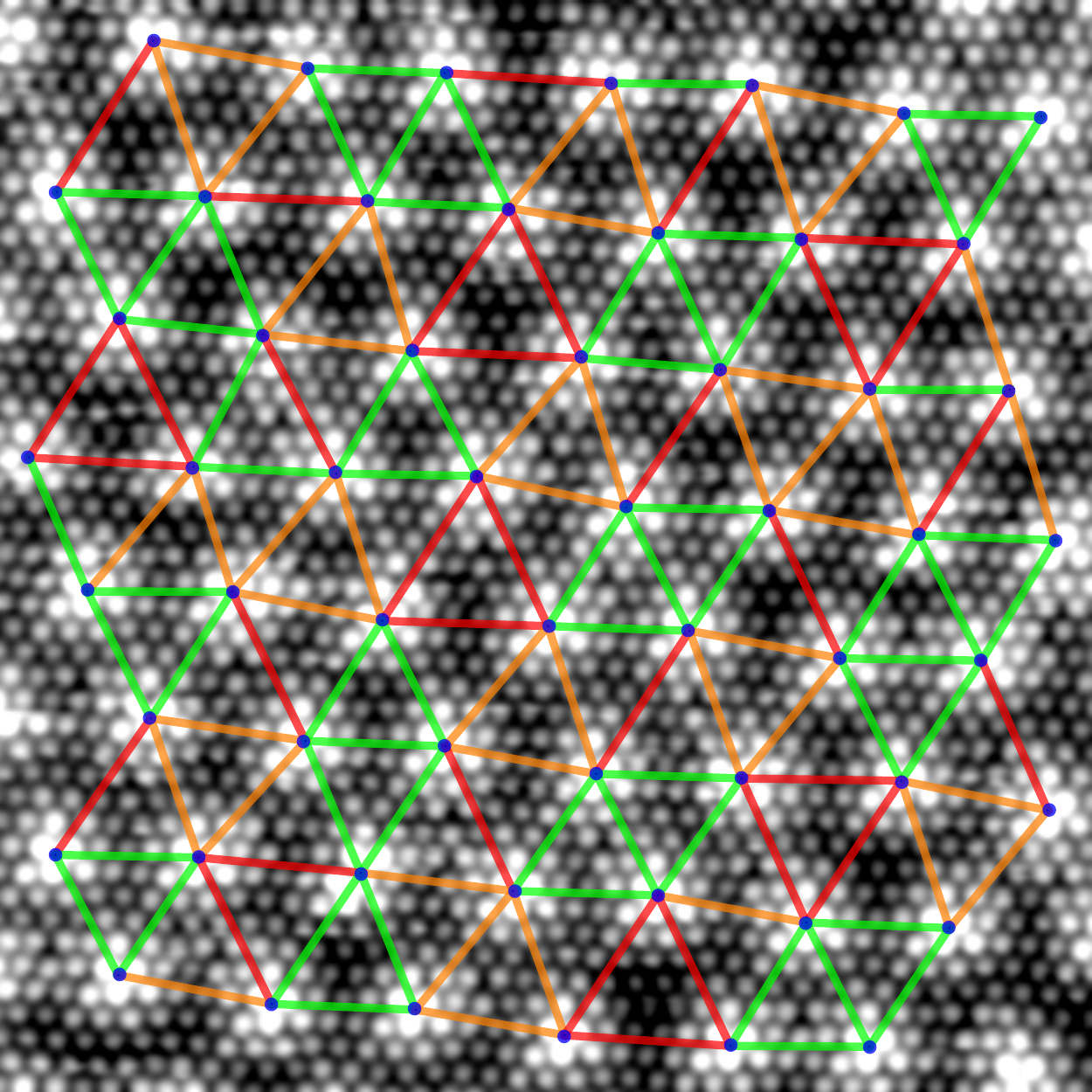}
\caption{\CO Nearest neighbor network vector distribution of the \agih\ superstructure  (Red, Green, Orange) and the iodine lattice (Blue). The circles represent the respective average length scale measured by FFT. At this resolution the superstructure shows three well separated spots for each direction. This allows one to colorize the nearest neighbor network of the superstructure showing three distinct types of triangles (Imaging parameters: $20\times20$~nm$^2$, $\Vb=50$~mV, $I=0.1$~nA).}\label{fig:AgIhl1}%
\end{figure}

The \agih\ structure has likely been observed before in STM\cite{ANDRYUSHECHKIN201883}. As shown in Fig.~\ref{fig:AgIFFT}c), the structure consists of atoms arranged in a triangular lattice with a Moir\'e like superstructure not unlike a charge density wave\cite{cdwreview}. The interatomic distance between iodine atoms as measured by FFT was $5.13$~\AA\ The superstructure had an average period of $24.0$~\AA, while showing some irregularities in real space images. FFT also showed an average angle offset of $2.6^\circ$ between the atomic structure and the superstructure.

We separately determined the real-space lattice of the superstructure and the atomic structure (See App.~\ref{app:NNNhowto}) from appropriately low-pass and high-pass filtered versions of three STM images obtained at different resolutions ($20 \times 20$~nm$^2$, $35 \times 35$~nm$^2$, and $50 \times 50$~nm$^2$, each at $512 \times 512$ pixels, App.~\ref{app:hexsuper}, Fig.~\ref{fig:AgIHexSource}). Higher resolution enhances the precision but limits the statistics, and vice versa. Fig.~\ref{fig:AgIhd1}a) shows the combined histogram of the superstructure lattice constant. As expected, the mean value agrees well with the one determined by FFT. However, the histogram spreads over more than one iodine lattice constant which is wider than expected from noise. Another way to examine the lattice is to plot the distances from one maximum to its neighbors in a $(dx,dy)$ scatter plot (Fig.~\ref{fig:AgIhd1}b). The plot  shows mainly three blobs stemming from the three lattice directions. The blobs appear triangular in shape rather then ellipsoidal. The latter would be expected for a random shift caused by imaging noise as seen in Fig.~\ref{fig:AgIhd1}b) for the atomic lattice of the \agih\ phase. Examining the highest resolution data set separately (Fig.~\ref{fig:AgIhl1}) sheds light on this structure. Here we find three distinct point groups at each of the three superlattice positions. The distance between the groups corresponds to the observed atomic distance within the surface and matches vectors of $(5,0)$, $(6,0)$ and $(5,1)$ in the atomic lattice basis. We colorized the edges of the superstructure lattice according to the location in the scatter plot and overlaid it onto the STM image (Fig.~\ref{fig:AgIhl1}). Here we find smaller ``green'', larger ``red'' and rotated ``orange'' triangles as the main building blocks of the superlattice. The superlattice vertices are preferentially found between three surface atoms within a radius of 1.4~\AA. This is more concisely shown in the appendix (App.~\ref{app:hexsuper}).

The \agih\ structure becomes unstable at bias voltages $\Vb\geq 3.0$~V leading to pit formation when the STM tip repeatedly passes over an area. This allows one to measure a layer thickness of 6~\AA. Cross sections also show a height change of $0.65$~\AA\ from the \agit\ to the \agih\ phase (green). However, the measured heights vary with voltage (App.~\ref{app:hexsuper}, Fig.~\ref{fig:HvsV}). This might indicate a reduced density of states of the \agih\ phase and makes it difficult to ascertain the true layer thickness. It remains an open question whether Ag is mixing into the \agih\ phase. Intermixing has been postulated for I/Ag(100) to explain a reconstruction observed on the (100) surface\cite{PhysRevB.80.125409}. AgI can grow in a Wurtzite structure with a lattice constant of $a=4.58$~\AA\ and $c=7.494$~\AA. This form also has a band gap of $2.63$~eV in the bulk crystal. Since the dimensions are close to the observed values and we suspect a reduced density of states on this surface, AgI remains a possible candidate for the \agih\ phase.

\subsubsection{\protect\agit\ phase}

\begin{figure}%
%\includegraphics[width=.3\columnwidth]{IAg_tristeps.pdf}\hfill%
%(a) $100\times100$~nm$^2$, $\Vb=250$~mV, $I=200$~pA, (b)
\includegraphics[width=.85\columnwidth]{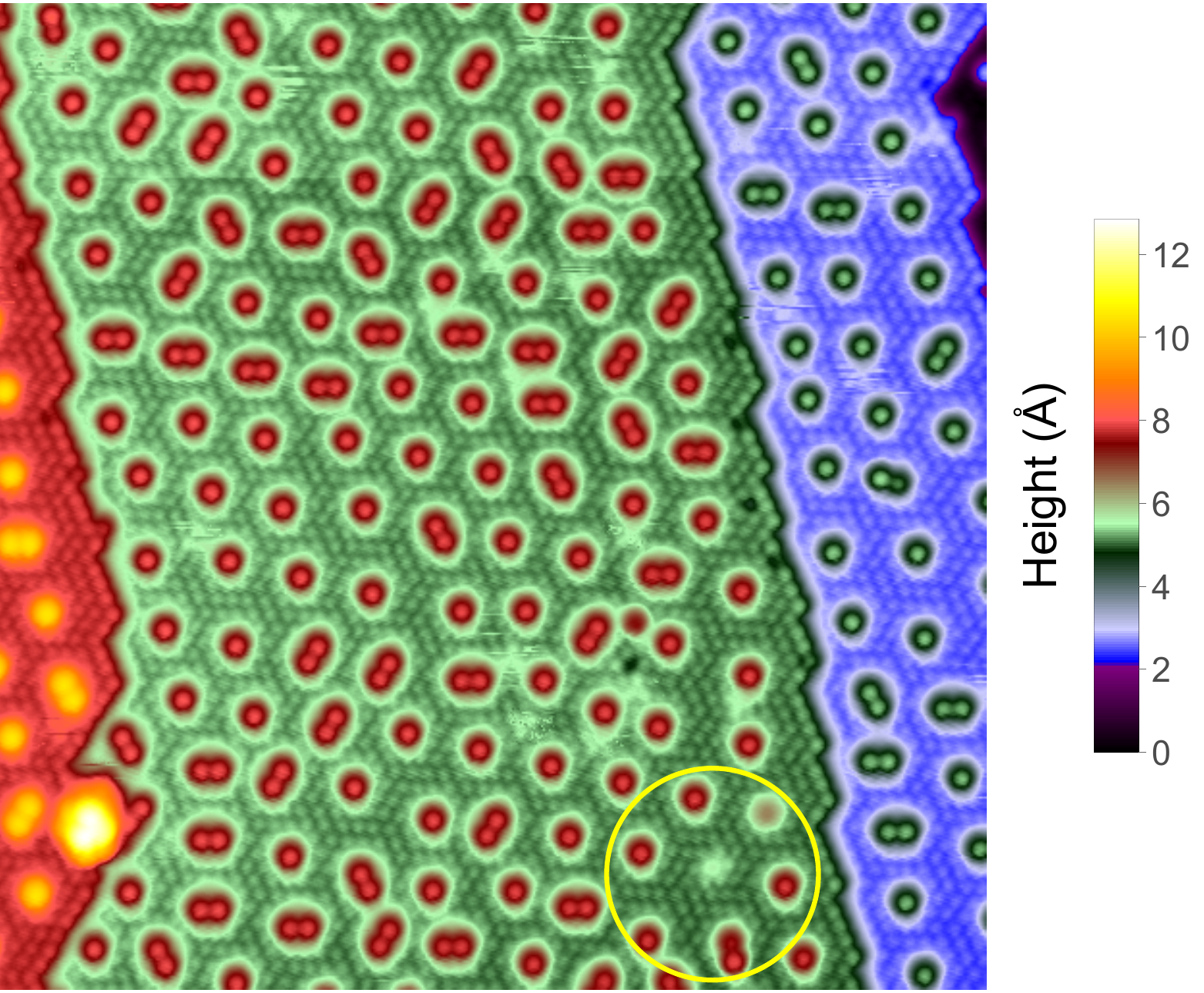}\\[2mm]
\includegraphics[width=.65\columnwidth]{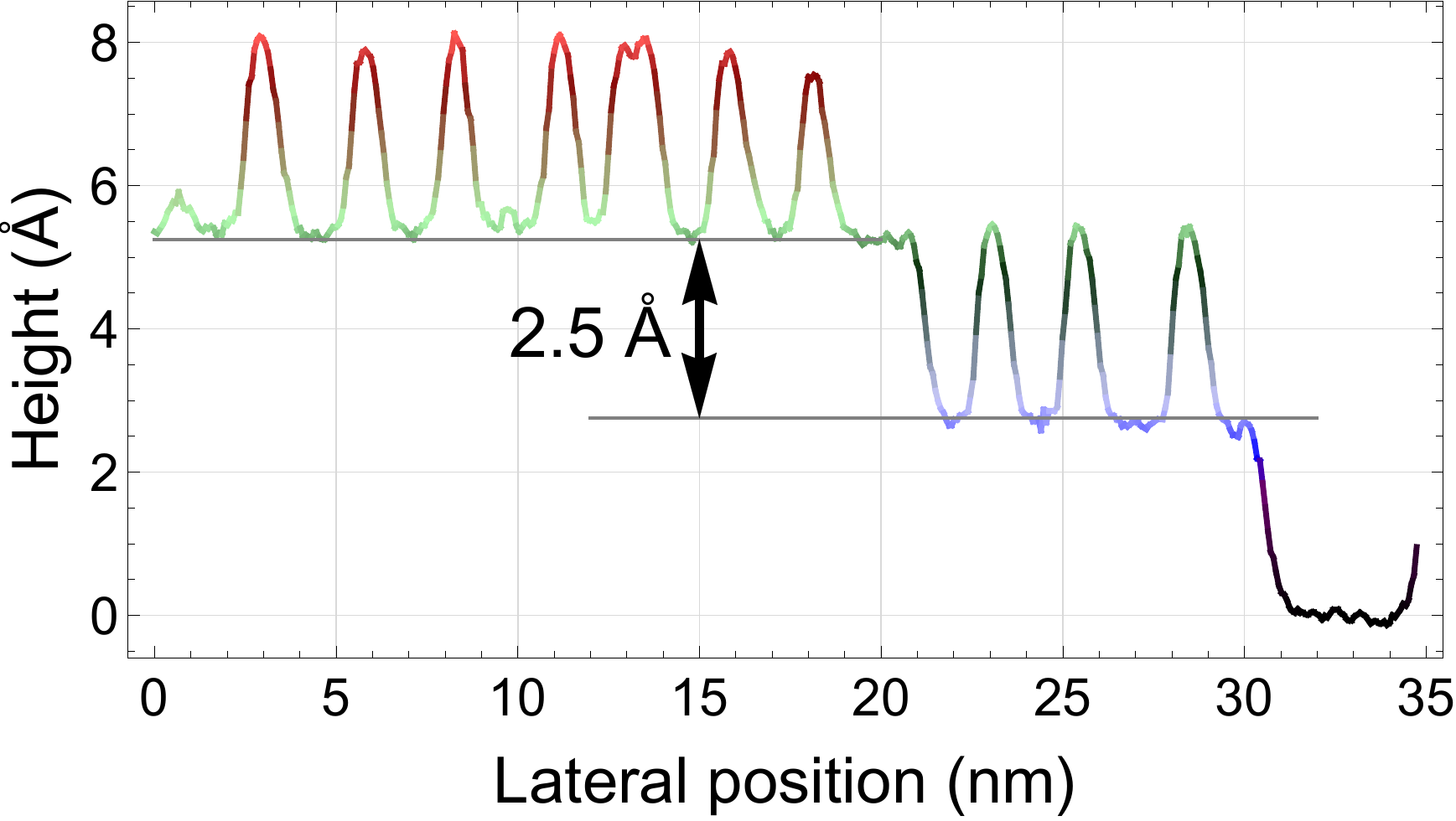}\hfill%
\caption{\CO Atomically resolved STM images of the \agit\ phase showing the monolayer structure and adatoms across three terraces. The yellow circle marks a missing adatom. The section shows a height of 2.5~\AA\ for the adatoms. (Imaging parameters: $30\times30$~nm$^2$, $\Vb=500$~mV, $I=200$~pA)}\label{fig:Agtri1}%
\end{figure}

The \agit\ phase is most peculiar. To our knowledge, no prior STM measurements of this phase have been reported. It consists of an iodine monolayer showing a rosette-like reconstruction with one or two adatoms sitting at the center of the each rosette. Fig.~\ref{fig:Agtri1} shows an example across three terraces. The cross-section shows the normal step height of Ag (111) and an adatom height of about $2.5$~\AA. On each terrace, the pattern of the iodine monolayer is atomically resolved all the way to the step edge. We note that the step edge is aligned with the reconstruction. The image also shows the rare occasion of an uncovered rosette structure in the lower right corner (yellow circle). The center appears brighter, likely due to an electronic effect that attracts the adatoms which we believe to be iodine. Selective FFT filtering identifies the rosette reconstruction as the source of the triangular spots in FFT while the six spots near the center represent the main components of the adatom structure (See App.~\ref{app:triselfilt}).

We performed a statistical analysis of the adatom structure over five images from two different sample preparations (details in App.~\ref{app:adatoms}). We measured the ratio of singlets to doublets and analyzed the direction of the doublets. The results are summarized in Tab.~\ref{tab:adatoms}. We found about $2/3$ of sites as singlets and $1/3$ as doublets. The ratio might depend on the coverage unless excess iodine is easily absorbed in the coexisting \agih\ phase. The doublets are aligned in three distinct directions commensurate with the iodine monolayer. The distribution of the three possible angles was found to be approximately uniform.

\begin{figure}%
\includegraphics[width=.95\columnwidth]{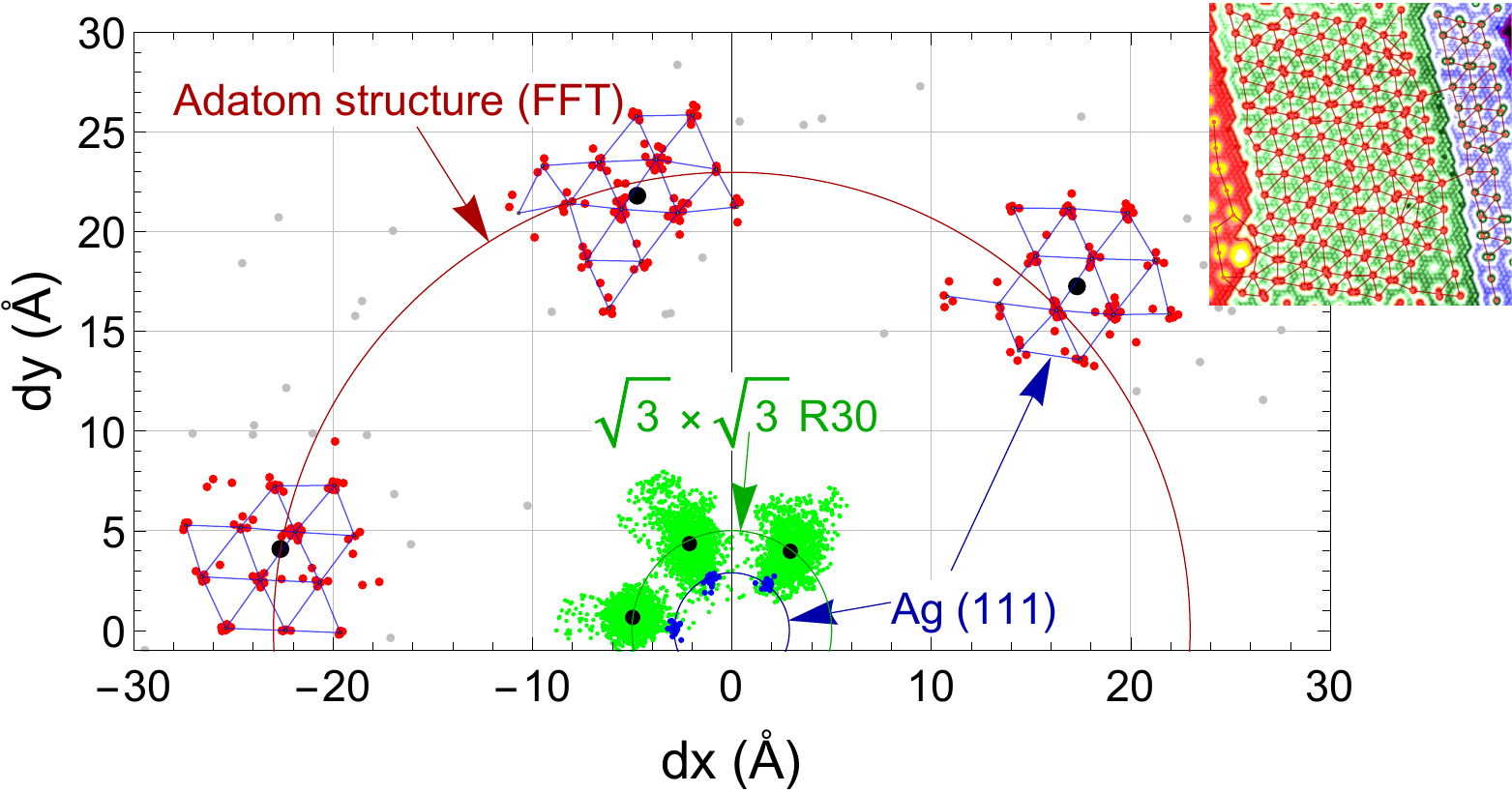}%
\caption{\CO Summary of the nearest neighbor network vector distributions for the adatom centroids (red) and the atomic lattice (green). The later split into three regions with an average length scale (black dot) of the iodine monolayer as expected. The three principle regions of the adatom lattice themselves form triangular lattices with a lattice constant of $2.9$~\AA. These lattice vectors are represented by blue points.}\label{fig:AgItl1}%
\end{figure}

A closer examination reveals that the adatom lattice structure, and --- by extension --- the underlying rosette structure, is not very regular (Fig.~\ref{fig:Agtri1}a). We tested this by constructing a nearest neighbor network from the adatom centroid locations and separately from the positions of the iodine monolayer. Fig.~\ref{fig:AgItl1} shows the edge distribution in a $(dx,dy)$ scatter plot. While the monolayer (green) shows a broad distribution without internal structure, the adatom distribution (red) was quite surprising. Instead of three somewhat broad spots for the three lattice directions we found three regions which resolve into a distinct triangular lattice of its own. Its average lattice constant is $\sim 2.9$~\AA\ which is close to the atomic spacing of the Ag (111) surface ($d_\mathrm{Ag-Ag}\approx 2.89$~\AA) and distinct from the measured inter-iodine spacing of $d_\mathrm{I-I}\approx 5.57$~\AA. Therefore the centers of the rosettes are tied to the Ag (111) lattice and not, as in the case of the \agih\ phase, to the inter-iodine distance.

The mean lattice vector of the adatoms is close to $(8,1)$ in the Ag (111) basis (App.~\ref{app:trisuper}, Fig.~\ref{fig:Trimodel}). This corresponds to a theoretical angle of $6.2^\circ$ between the Ag (111) surface vector and adatom lattice vectors which is close to the measured angle of $5.6^\circ$. The structure and thereby possible correlations between superlattice vectors is obviously complicated since there are more than ten sub-spots for each direction. One correlation, however, concerns the double occupancy of adatom positions. It appears that the double occupancy preferentially occurs on distorted lattice sites, shifted by one Ag (111) lattice vector away from the most prominent adatom lattice position (App.~\ref{app:trisuper}, Fig.~\ref{fig:Tridouble}). The shift also correlates with the orientation of the adatom dimer.

\begin{figure}%
\includegraphics[width=.325\columnwidth]{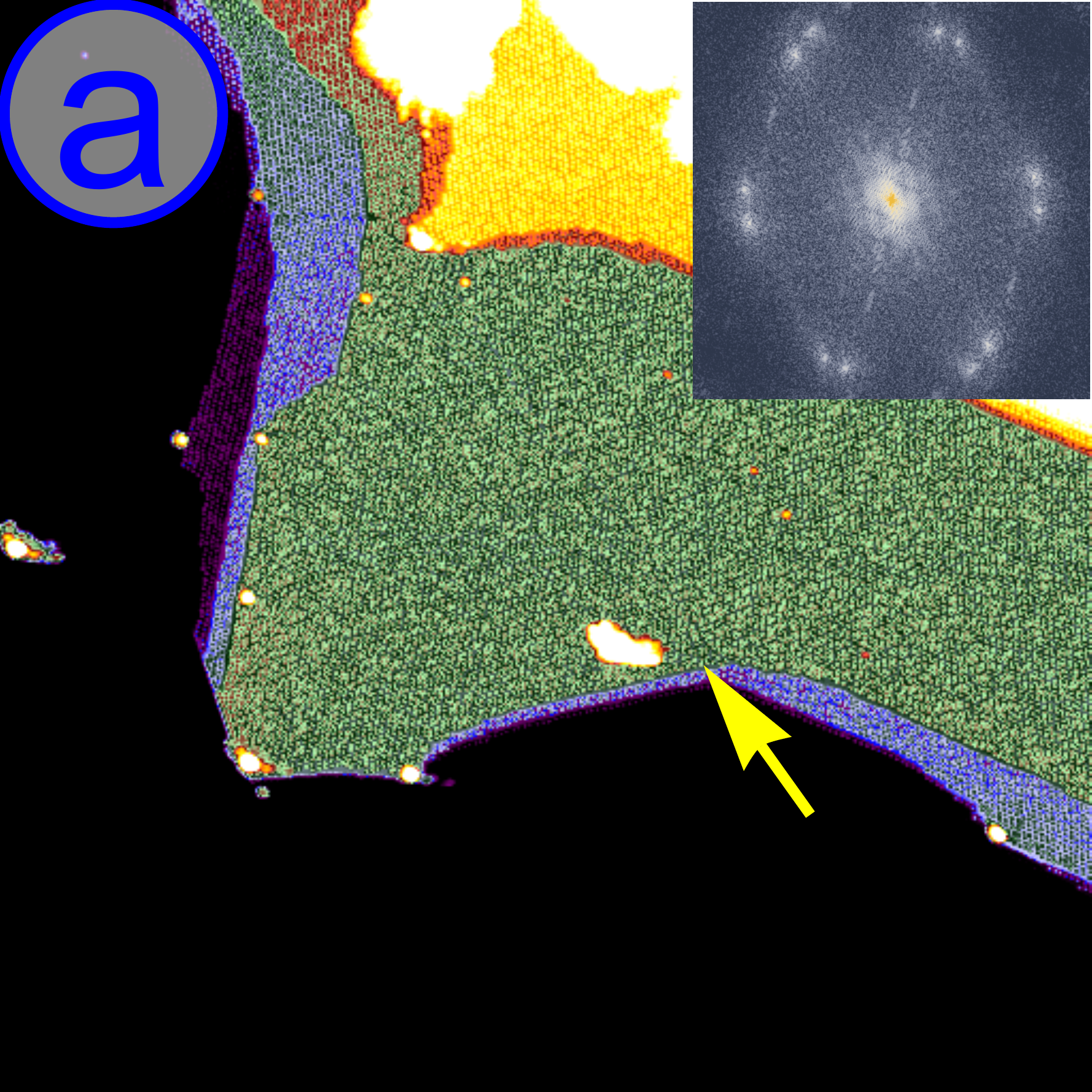}\hfill%
\includegraphics[width=.325\columnwidth]{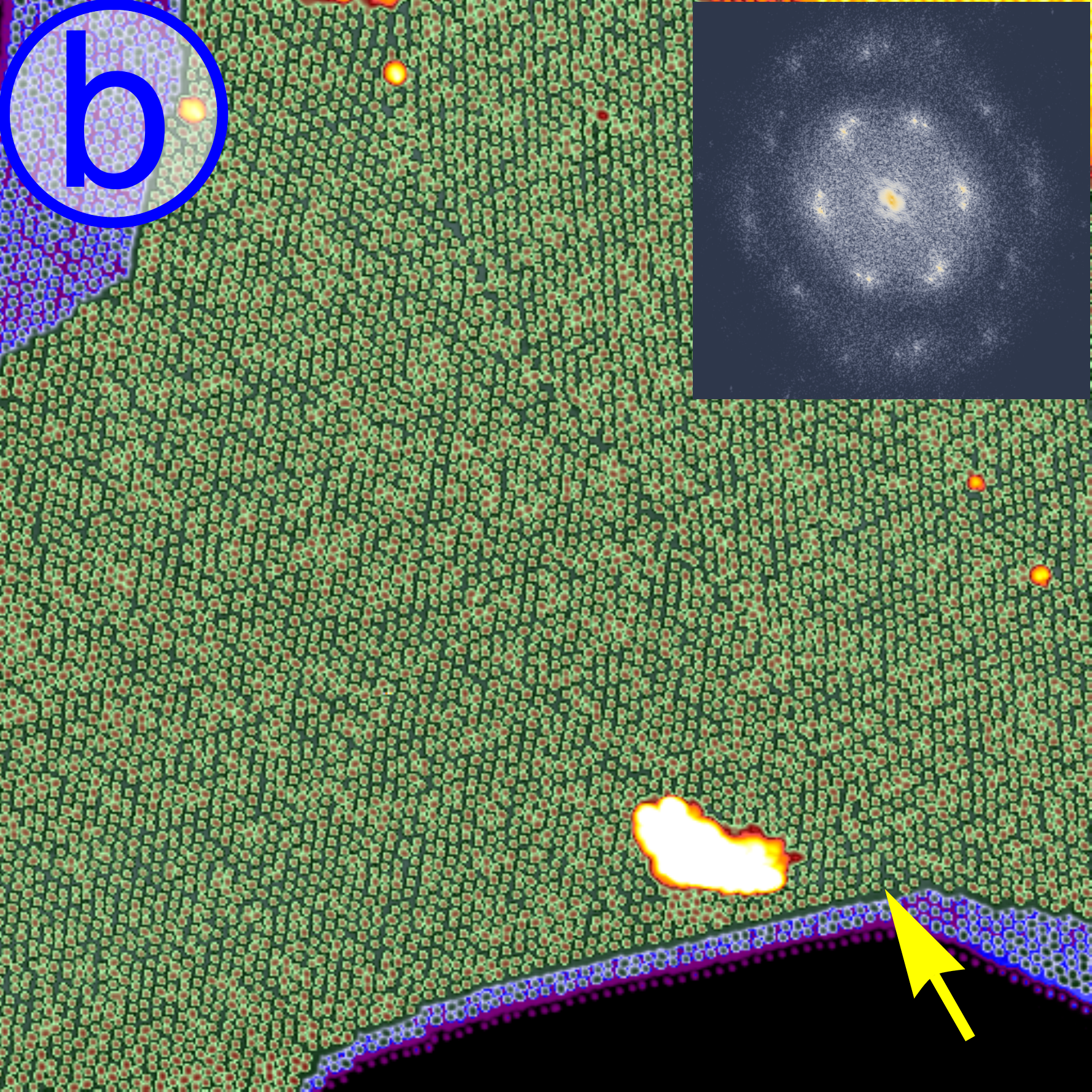}\hfill%
\includegraphics[width=.325\columnwidth]{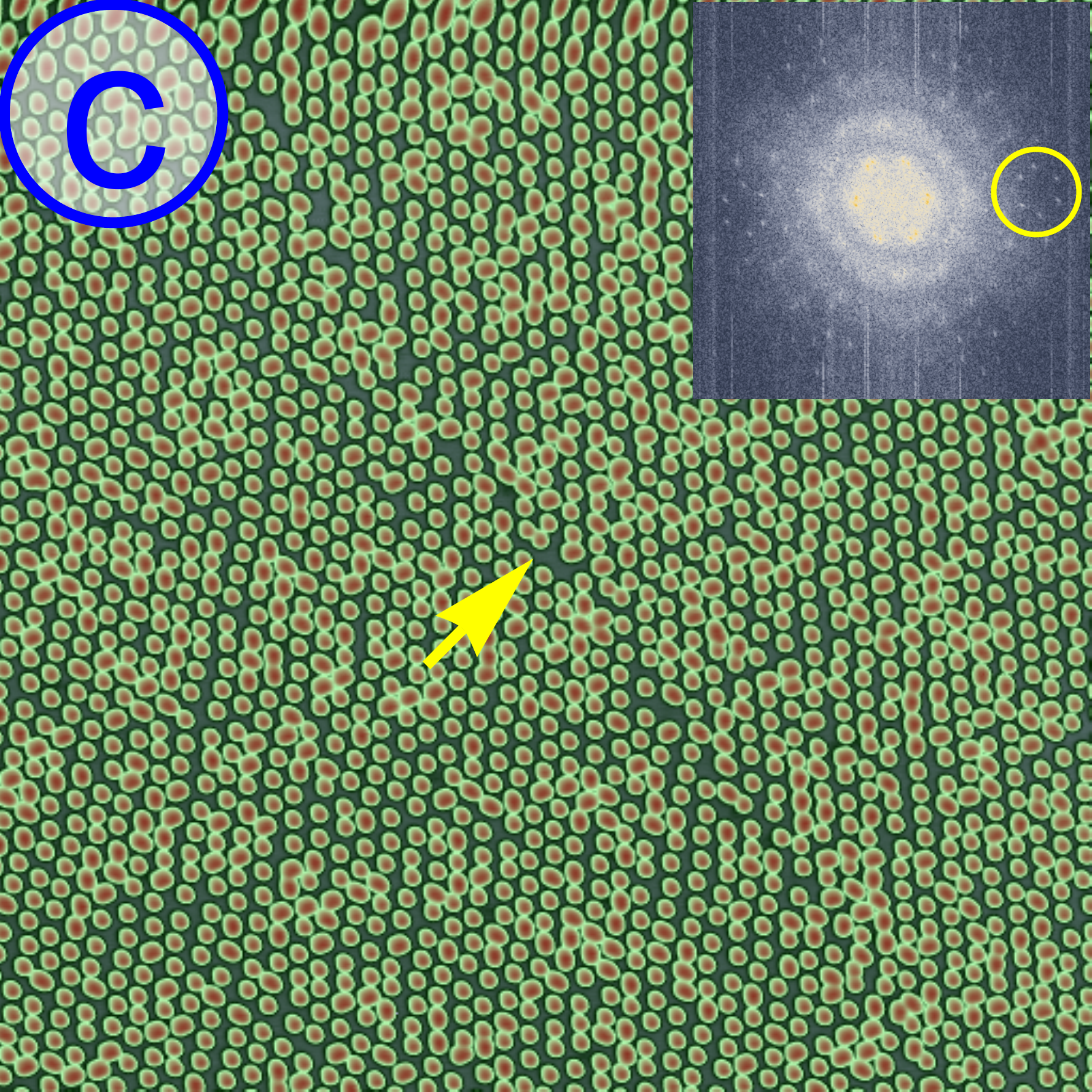}%
\caption{\CO STM images of a domain boundary between two triangular domains marked by yellow arrows. The insets show images of the respective FFT. (Imaging parameters: (a) $400\times400$~nm$^2$, $\Vb=500$~mV, $I=100$~pA, (b) $200\times200$~nm$^2$, $\Vb=500$~mV, $I=100$~pA), (c) $100\times100$~nm$^2$, $\Vb=500$~mV, $I=100$~pA)}\label{fig:Agtridom}%
\end{figure}

\begin{figure}%
\includegraphics[width=.33\columnwidth]{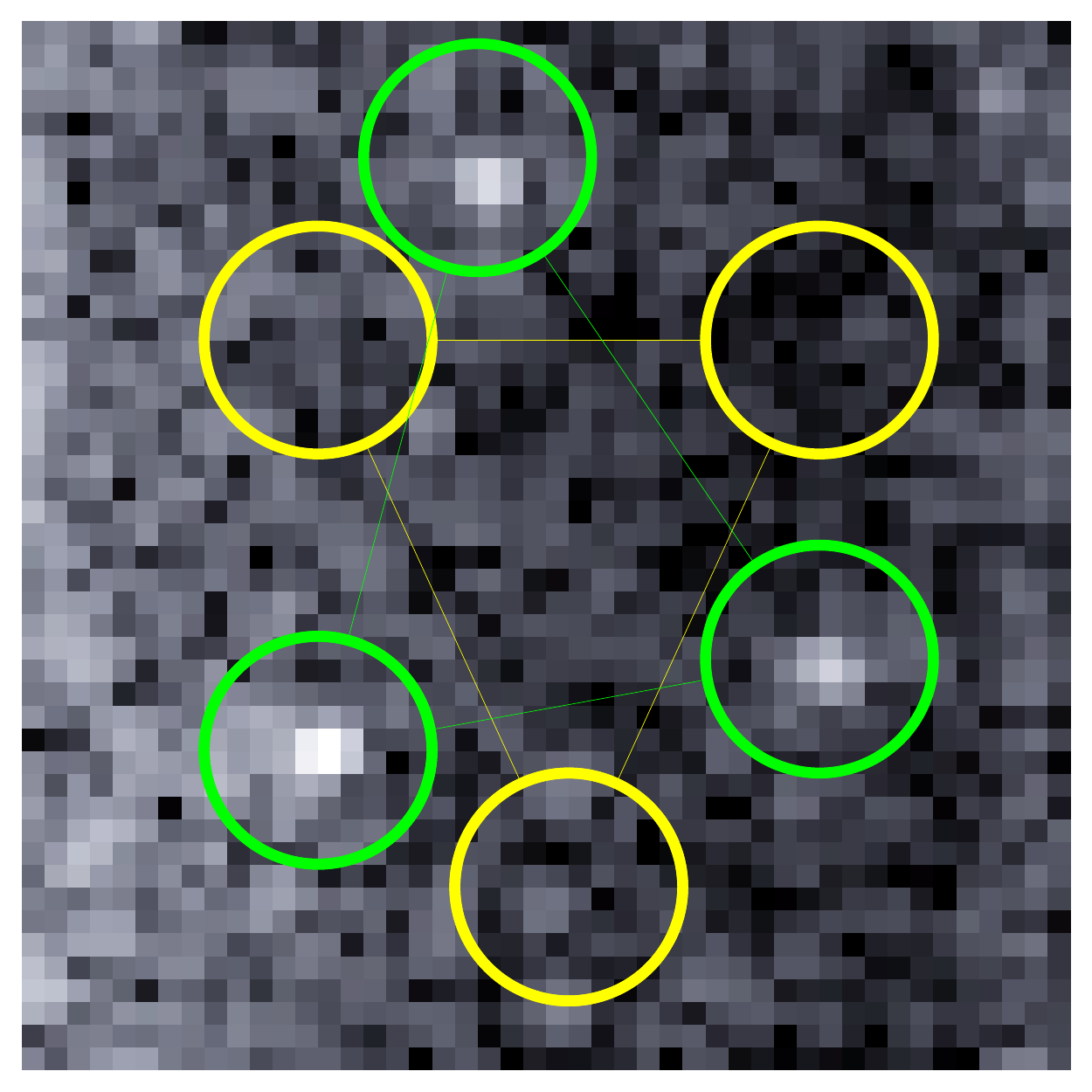}\hfill%
\includegraphics[width=.33\columnwidth]{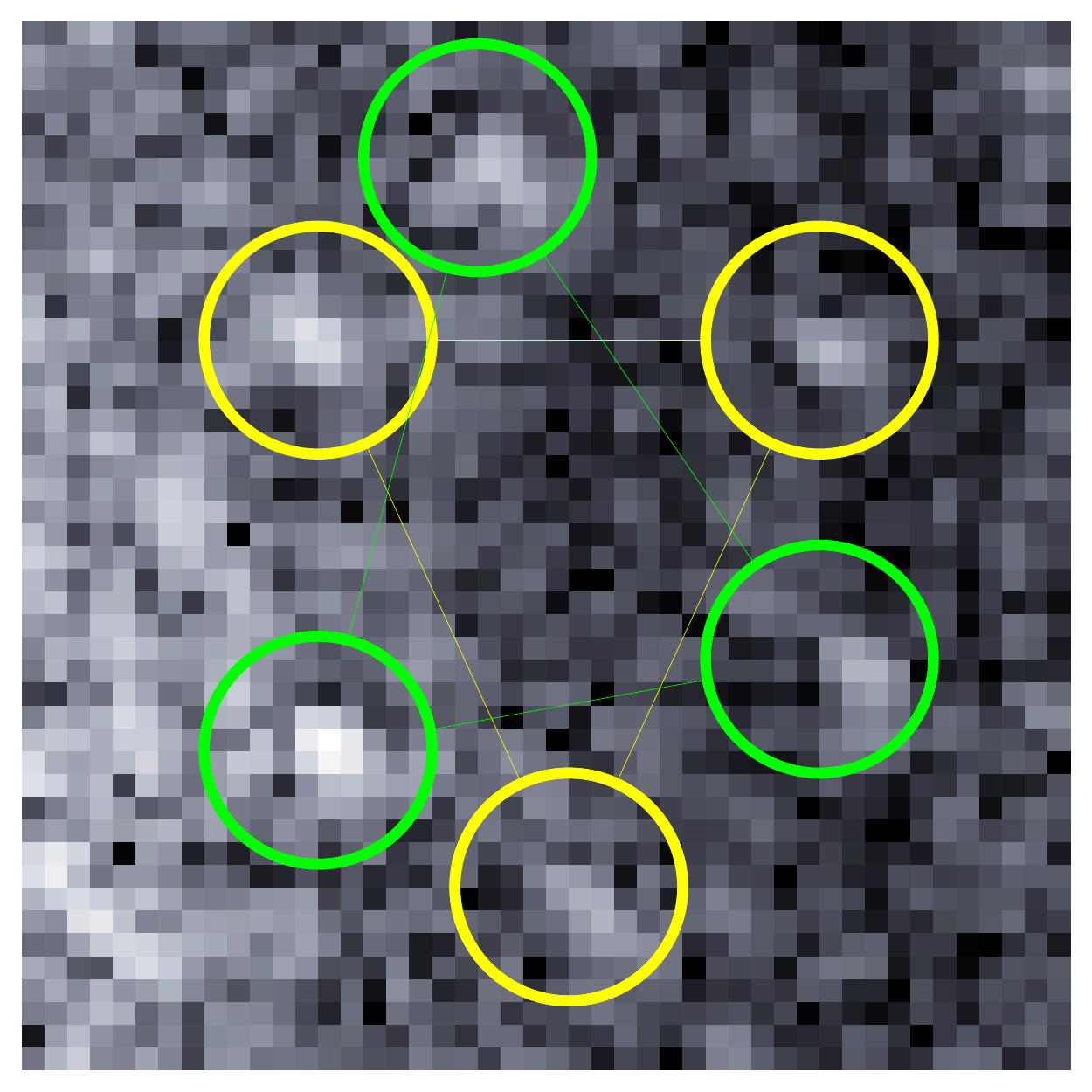}\hfill%
\includegraphics[width=.33\columnwidth]{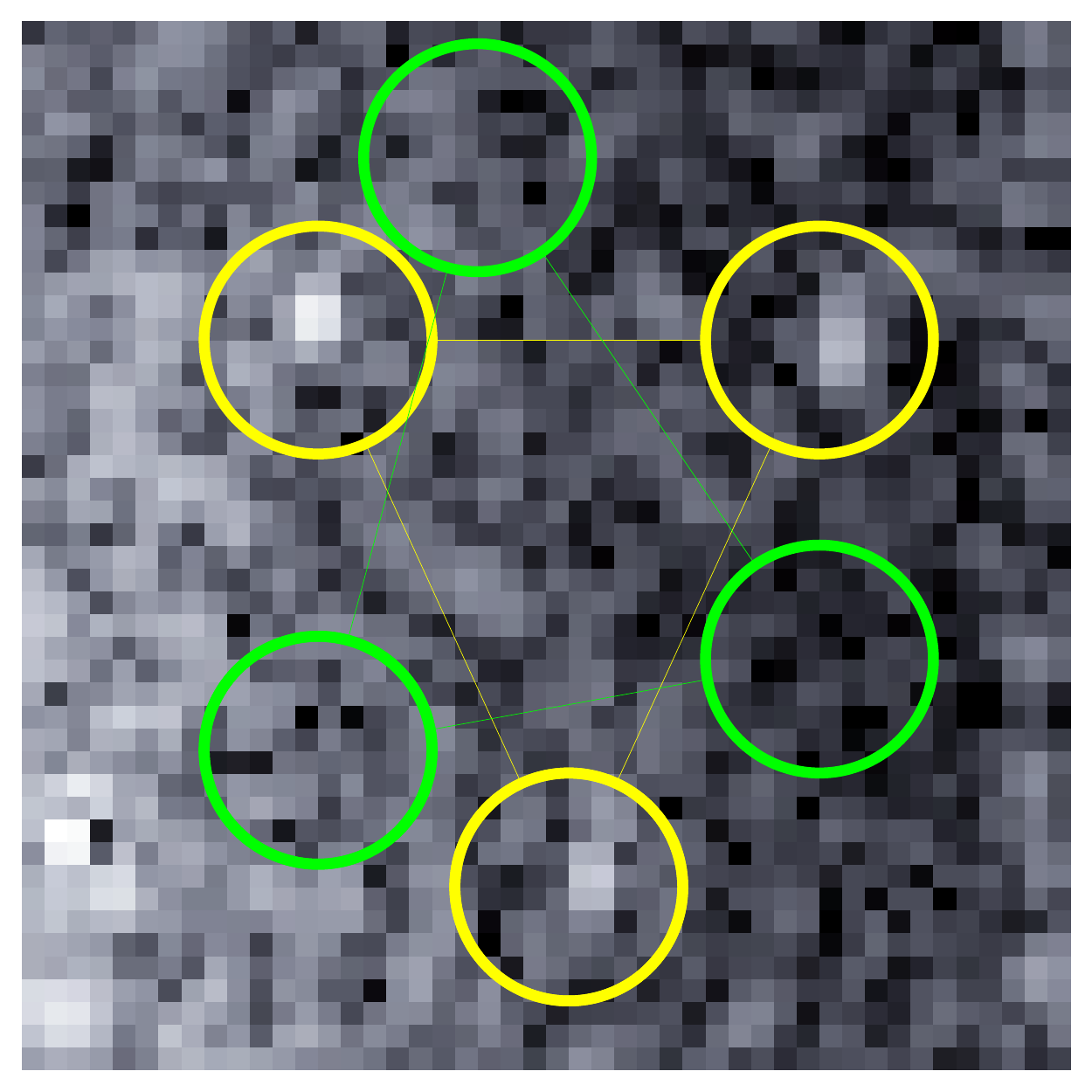}%
\caption{\CO FFT images showing only the \agit\ structure (yellow circle in Fig.~\ref{fig:Agtridom}c) of $1/4$ sized regions taken from image Fig.~\ref{fig:Agtridom}c). Left to right: lower left, center, and upper right corner, respectively. }\label{fig:Agtridom2}%
\end{figure}

On a rare larger Ag terrace we found a mirror domain boundary of the \agit\ phase. Mirror domains represent a change in chirality of the rosettes where the mean lattice vector should change from $(8,1)$ to $(8,-1)$ in the Ag (111) basis. The expected angle between the domains is $12.4^\circ$. Fig.~\ref{fig:Agtridom} shows a sequence of zoomed in images of the domain boundary. The location of the boundary is marked in each image by a yellow arrow. The respective FFT (inset) show an average angle of $\sim 9.7^\circ$ between the respective adatom superstructures, close to the expected angle for two mirror domains. At the highest resolution (Fig.~\ref{fig:Agtridom}c) we find six spots in FFT (inset, yellow circle) for the underlying rosette reconstruction of the iodine monolayer. As shown in Fig.~\ref{fig:Agtridom2}, the spots are the composition of two opposing triangles from the left and right side of the domain boundary (cf. Fig.~\ref{fig:FFTsum}). This immediately raises the question whether the LEED patterns of the \agih\ phase is merely the average over multiple \agit\ domains. We found, however, that mixed domains are only allowed on larger terraces and that the chirality is tied to the mean direction of the underlying Ag steps. The linkage was confirmed in an area where the step direction changed by $\sim 30^\circ$ and the FFT image showed split adatom peaks (App.~\ref{app:tridom}, Fig.~\ref{fig:apptds}). Selective filtering ties the chirality of the domains to the respective step direction. A forcing effect then might result from the average orientation of the Ag step edges with respect to the Ag crystal lattice which is also given by the misorientation of the surface with respect to the (111) normal with the azimuth component defining the average step orientation. Given that for an angle of $0.5^\circ$ the average width of the Ag terraces is only $\overline{w_\mathrm{Ag}}=\cot(0.5^\circ)\cdot 2.36$~\AA$\approx 27.0$~nm we assume that the LEED experiments would show a single domain and hence a triangular structure. This proposed coupling of the prevalent step direction of the single crystal to the chirality of the iodine reconstruction might serve as an explanation of the LEED results. 

\begin{figure}%
\includegraphics[width=.95\columnwidth]{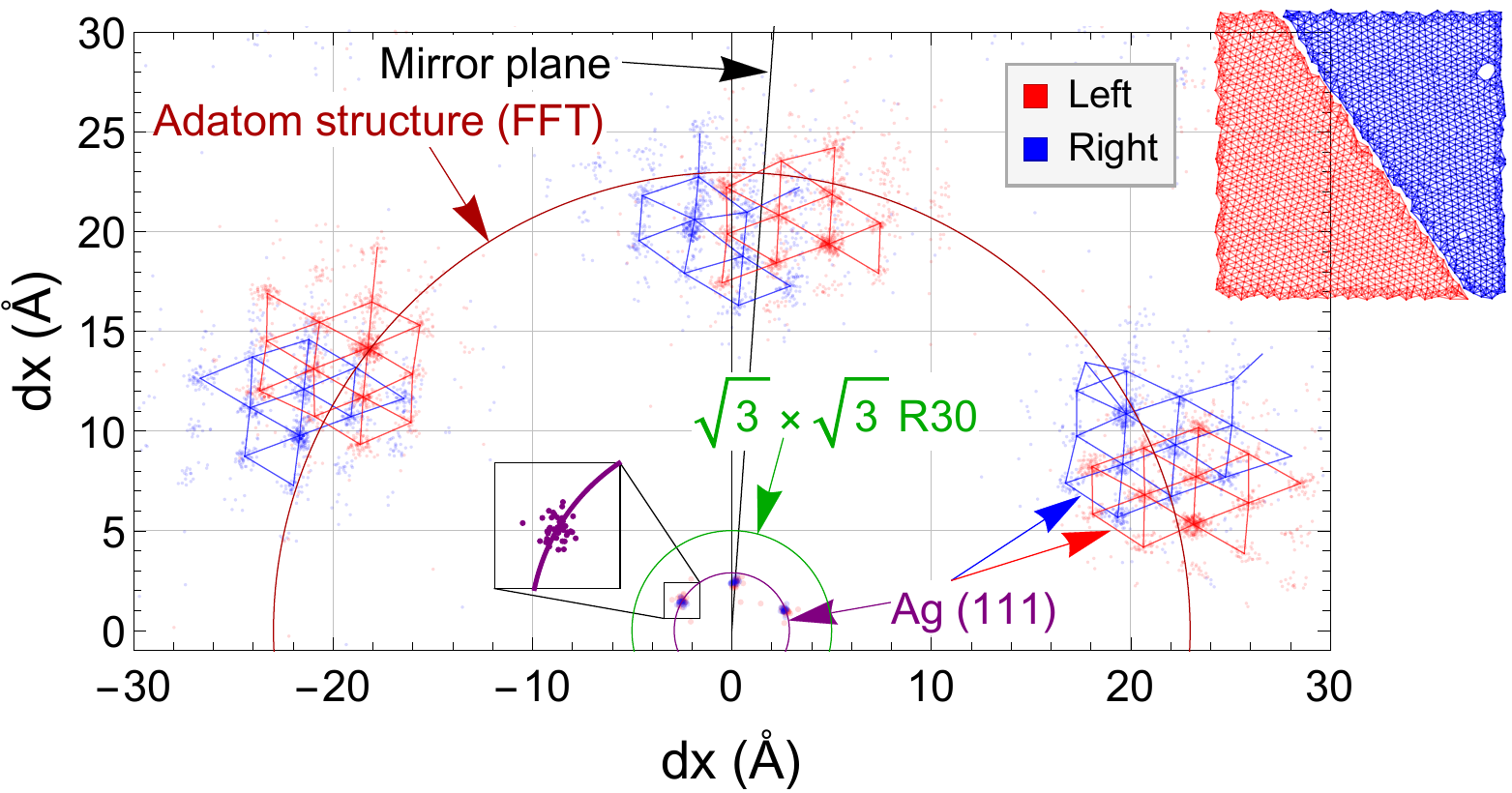}\hfill\\
\includegraphics[width=.85\columnwidth]{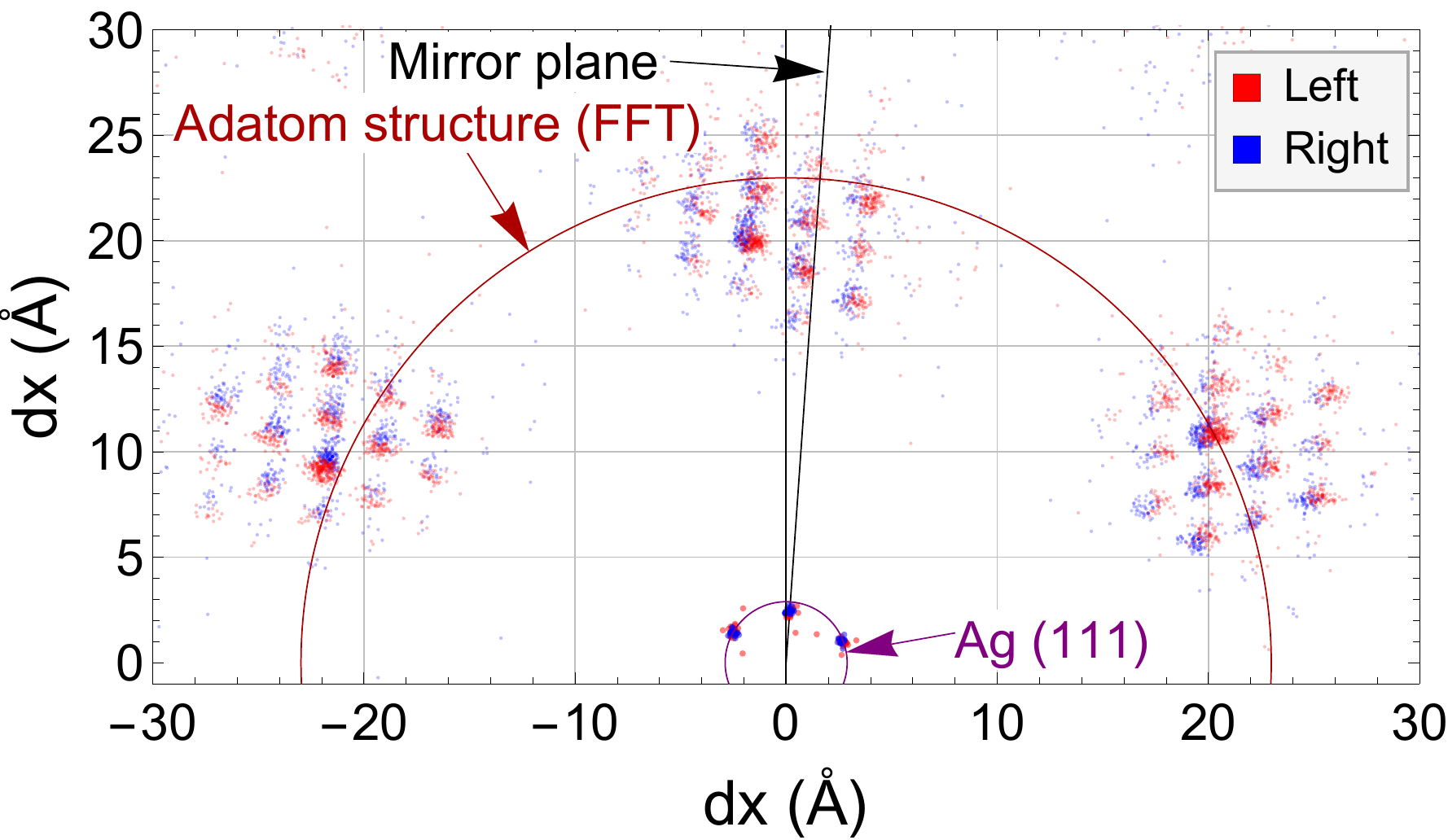}\hfill
\caption{\CO Nearest neighbor network vector distribution of the adatoms shown in Fig.~\ref{fig:Agtridom}c) separately for the two chiral domains (Blue, red). The variations in the adatom-adatom distances reproduces a single Ag lattice as expected. Mirroring the left domain on the mirror plane produces coinciding spots with similar distribution patterns.}\label{fig:AgItlm1}%
\end{figure}

The mirror domains allow deeper insight into the alignment of the rosette reconstruction with the Ag (111) surface atoms. We separately constructed the adatom lattice for either domain and plotted the edge distribution in Fig.~\ref{fig:AgItlm1}. The edge distribution shows two distinct sets of interwoven triangular lattices, each similar to the one shown in Fig.~\ref{fig:AgItl1}. Extracting the edge distribution from the interwoven lattice leads to three spots in a scatter plot at the Ag lattice constant, confirming a single underlying Ag lattice as expected. The interwoven nature of the edge distributions requires the rosette centers to sit on Ag (111) hollow sites. This is shown in more detail in appendix \ref{app:trisuper} (Fig.~\ref{fig:Trimirat}). Mirroring the edge distribution from one domain at an angle of $9.7^\circ$ (the angle offset found for the adatom structure in FFT) leads to overlapping point groups --- even with a similar structure. This confirms that the two domains are indeed of the same structure. Fig.~\ref{fig:AgItlm1} also shows that the mirror plane (black line) runs along one of the Ag (111) lattice vectors. 

\begin{figure}%
\includegraphics[width=.49\columnwidth]{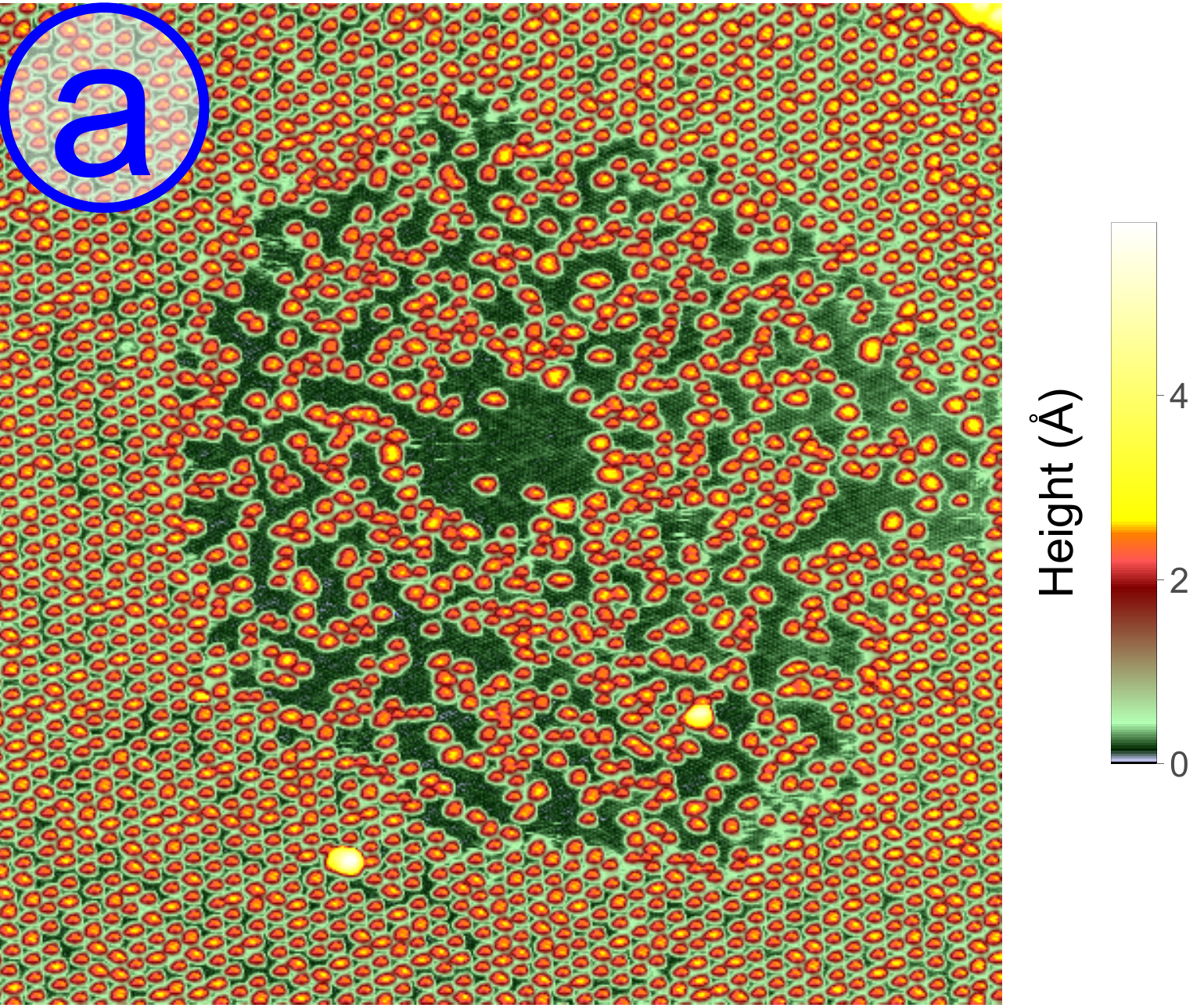}\hfill%
\includegraphics[width=.49\columnwidth]{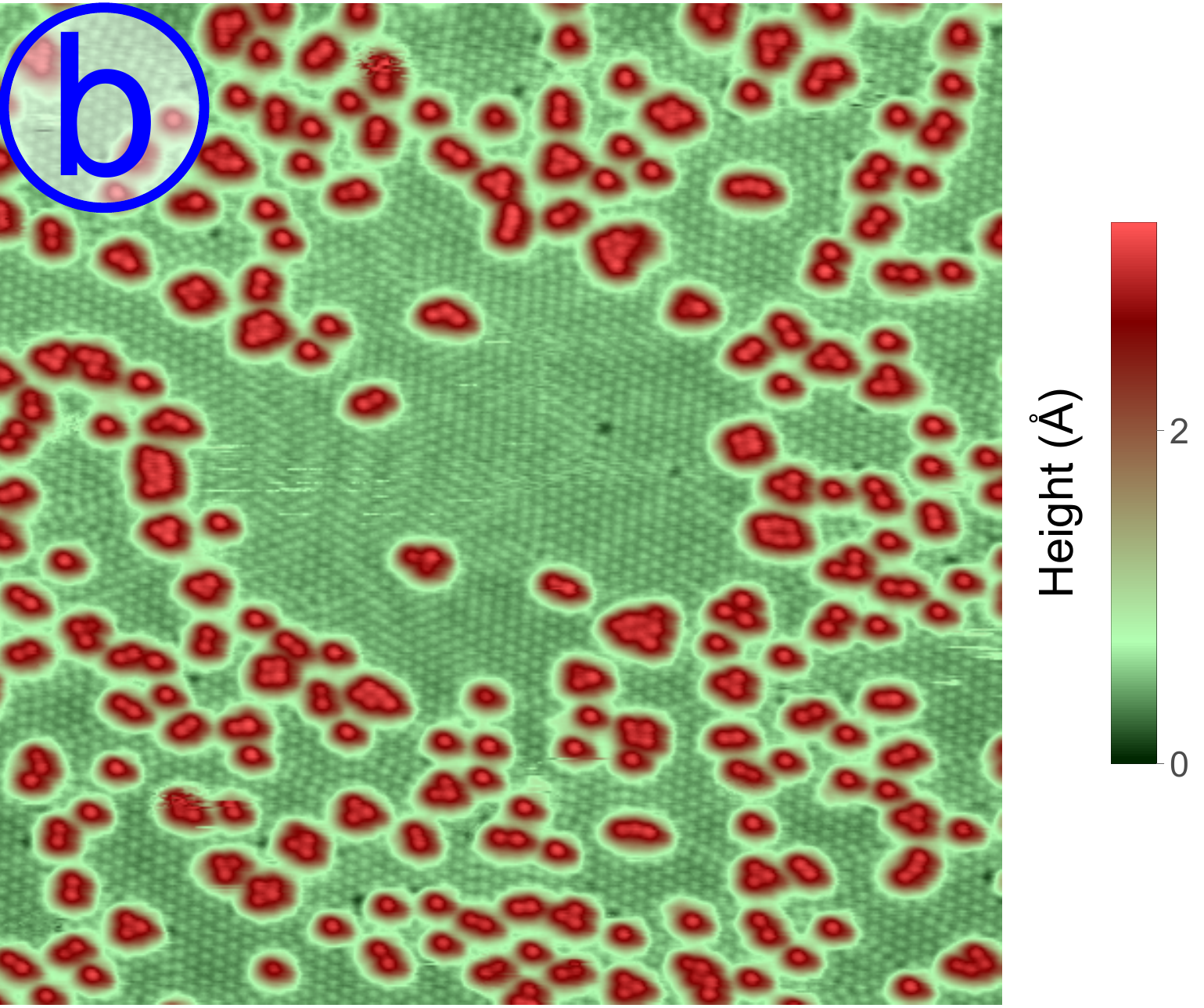}%
\caption{\CO Voltage pulse induced modification of the \agit\ phase. a) disturbed adatom structure. b) The iodine monolayer was also altered. (Imaging parameters: (a) $112\times112$~nm$^2$, $\Vb=750$~mV, $I=200$~pA, (b) $40\times40$~nm$^2$, $\Vb=750$~mV, $I=200$~pA)}\label{fig:AgImod1}%
\end{figure}

Finally, the adatom structure can be disturbed by the STM tip. Fig.~\ref{fig:AgImod1}a) shows the result of a single $6$~V
% (?)
pulse applied to the STM junction. The adatom structure was altered within a radius of $45$~nm of the tip position. Outside the radius, the regular adatom structure remained. A closer inspection of the exposed iodine monolayer (Fig.~\ref{fig:AgImod1}b) does not show the expected rosette pattern. Instead, a slightly wavy superstructure of the monolayer, which was stable over multiple scans, is now visible. This raises the question whether the adatom structure is necessary to stabilize the \agit\ structure, or whether the pulse was disruptive enough to also disturb the rosette reconstruction of the monolayer. In a different region we inadvertently removed part of the adatoms while taking an STS map (Fig.~\ref{fig:AgImod2}a). This also led to a local change of the rosette pattern. We therefore assume that the adatoms are integral to the \agit\ structure.

\section{Conclusion}

We observed three phases of iodine on silver (111) using low temperature STM which tentatively match the \agir, \agit, and \agih\ phases found in early LEED experiments. The \agih\ phase shows the worst match in terms of lattice constant compared to LEED. 

The main findings for the three phases are:
\begin{description}
\item[\agir] At 77 K, iodine is highly mobile until the full coverage is reached. Step edges stabilize the motion. We found no superstructure. At 4 K iodine mobility is reduced and we find stable islands or holes in the iodine layer depending on coverage.
\item[\agih] This structure, found on larger terraces, could be a double layer of iodine or the onset of AgI growth. It shows a charge-density-wave-like superstructure with maxima located at hollow sites of the top layer atoms. The distance of the superstructure maxima corresponds to 4 or 5 atomic distances, respectively, leading to a somewhat disordered appearance. Atoms can easily be removed by voltage pulses and larger vacancies formed this way distort the superstructure.
\item[\agit] This is the most intriguing structure. Iodine forms a mosaic of rosettes which apparently are stabilized by one or two adatoms at the center of each rosette. Removing the adatoms over larger areas with the STM tip also destroys the rosette structure. The adatom positions and, by extension, the centers of the rosettes are aligned with hollow sites of the underlying Ag lattice. The vectors of the adatom lattice vary by lattice vectors of the silver surface. The chirality of the rosettes is bound to the predominant Ag step direction. The latter is given by the miss-cut of the Ag crystal and thus varies little on well prepared surfaces. This explains why a triangular pattern is observed in LEED studies where small domains of random chirality would lead to a hexagonal appearance.
\end{description}

% The \agit\ phase was the most interesting. The lattice constant is a good match compared to LEED and the rosette structure leads to a triangular appearance in FFT. The chirality of the rosettes seems to be tied to the Ag (111) step direction. This explains how LEED is able to shows a triangular arrangement, where averaging over mirror domains would otherwise lead to spots arranged as hexagons. In addition we observed an adatom structure with lattice vectors tied to the underlying Ag (111).  

\bibliographystyle{apsrev4-1}
\bibliography{AgI}

\begin{thebibliography}{20}%
\makeatletter
\providecommand \@ifxundefined [1]{%
 \@ifx{#1\undefined}
}%
\providecommand \@ifnum [1]{%
 \ifnum #1\expandafter \@firstoftwo
 \else \expandafter \@secondoftwo
 \fi
}%
\providecommand \@ifx [1]{%
 \ifx #1\expandafter \@firstoftwo
 \else \expandafter \@secondoftwo
 \fi
}%
\providecommand \natexlab [1]{#1}%
\providecommand \enquote  [1]{``#1''}%
\providecommand \bibnamefont  [1]{#1}%
\providecommand \bibfnamefont [1]{#1}%
\providecommand \citenamefont [1]{#1}%
\providecommand \href@noop [0]{\@secondoftwo}%
\providecommand \href [0]{\begingroup \@sanitize@url \@href}%
\providecommand \@href[1]{\@@startlink{#1}\@@href}%
\providecommand \@@href[1]{\endgroup#1\@@endlink}%
\providecommand \@sanitize@url [0]{\catcode `\\12\catcode `\$12\catcode
  `\&12\catcode `\#12\catcode `\^12\catcode `\_12\catcode `\%12\relax}%
\providecommand \@@startlink[1]{}%
\providecommand \@@endlink[0]{}%
\providecommand \url  [0]{\begingroup\@sanitize@url \@url }%
\providecommand \@url [1]{\endgroup\@href {#1}{\urlprefix }}%
\providecommand \urlprefix  [0]{URL }%
\providecommand \Eprint [0]{\href }%
\providecommand \doibase [0]{http://dx.doi.org/}%
\providecommand \selectlanguage [0]{\@gobble}%
\providecommand \bibinfo  [0]{\@secondoftwo}%
\providecommand \bibfield  [0]{\@secondoftwo}%
\providecommand \translation [1]{[#1]}%
\providecommand \BibitemOpen [0]{}%
\providecommand \bibitemStop [0]{}%
\providecommand \bibitemNoStop [0]{.\EOS\space}%
\providecommand \EOS [0]{\spacefactor3000\relax}%
\providecommand \BibitemShut  [1]{\csname bibitem#1\endcsname}%
\let\auto@bib@innerbib\@empty
%</preamble>
\bibitem [{\citenamefont {Jones}(1988)}]{JONES198825}%
  \BibitemOpen
  \bibfield  {author} {\bibinfo {author} {\bibfnamefont {R.~G.}\ \bibnamefont
  {Jones}},\ }\href {\doibase https://doi.org/10.1016/0079-6816(88)90013-5}
  {\bibfield  {journal} {\bibinfo  {journal} {Progress in Surface Science}\
  }\textbf {\bibinfo {volume} {27}},\ \bibinfo {pages} {25 } (\bibinfo {year}
  {1988})}\BibitemShut {NoStop}%
\bibitem [{\citenamefont {Andryushechkin}\ \emph {et~al.}(2018)\citenamefont
  {Andryushechkin}, \citenamefont {Pavlova},\ and\ \citenamefont
  {Eltsov}}]{ANDRYUSHECHKIN201883}%
  \BibitemOpen
  \bibfield  {author} {\bibinfo {author} {\bibfnamefont {B.}~\bibnamefont
  {Andryushechkin}}, \bibinfo {author} {\bibfnamefont {T.}~\bibnamefont
  {Pavlova}}, \ and\ \bibinfo {author} {\bibfnamefont {K.}~\bibnamefont
  {Eltsov}},\ }\href {\doibase https://doi.org/10.1016/j.surfrep.2018.03.001}
  {\bibfield  {journal} {\bibinfo  {journal} {Surface Science Reports}\
  }\textbf {\bibinfo {volume} {73}},\ \bibinfo {pages} {83 } (\bibinfo {year}
  {2018})}\BibitemShut {NoStop}%
\bibitem [{\citenamefont {Held}\ \emph {et~al.}(2017)\citenamefont {Held},
  \citenamefont {Fuchs},\ and\ \citenamefont {Studer}}]{surfchem1}%
  \BibitemOpen
  \bibfield  {author} {\bibinfo {author} {\bibfnamefont {P.~A.}\ \bibnamefont
  {Held}}, \bibinfo {author} {\bibfnamefont {H.}~\bibnamefont {Fuchs}}, \ and\
  \bibinfo {author} {\bibfnamefont {A.}~\bibnamefont {Studer}},\ }\href
  {\doibase https://doi.org/10.1002/chem.201604047} {\bibfield  {journal}
  {\bibinfo  {journal} {Chemistry – A European Journal}\ }\textbf {\bibinfo
  {volume} {23}},\ \bibinfo {pages} {5874} (\bibinfo {year}
  {2017})}\BibitemShut {NoStop}%
\bibitem [{\citenamefont {Ammon}\ \emph {et~al.}(2019)\citenamefont {Ammon},
  \citenamefont {Haller}, \citenamefont {Sorayya},\ and\ \citenamefont
  {Maier}}]{surfchem2}%
  \BibitemOpen
  \bibfield  {author} {\bibinfo {author} {\bibfnamefont {M.}~\bibnamefont
  {Ammon}}, \bibinfo {author} {\bibfnamefont {M.}~\bibnamefont {Haller}},
  \bibinfo {author} {\bibfnamefont {S.}~\bibnamefont {Sorayya}}, \ and\
  \bibinfo {author} {\bibfnamefont {S.}~\bibnamefont {Maier}},\ }\href
  {\doibase https://doi.org/10.1002/cphc.201900347} {\bibfield  {journal}
  {\bibinfo  {journal} {ChemPhysChem}\ }\textbf {\bibinfo {volume} {20}},\
  \bibinfo {pages} {2333} (\bibinfo {year} {2019})}\BibitemShut {NoStop}%
\bibitem [{\citenamefont {Grill}\ \emph {et~al.}(2007)\citenamefont {Grill},
  \citenamefont {Dyer}, \citenamefont {Lafferentz}, \citenamefont {Persson},
  \citenamefont {Peters},\ and\ \citenamefont {Hecht}}]{surfchem3}%
  \BibitemOpen
  \bibfield  {author} {\bibinfo {author} {\bibfnamefont {L.}~\bibnamefont
  {Grill}}, \bibinfo {author} {\bibfnamefont {M.}~\bibnamefont {Dyer}},
  \bibinfo {author} {\bibfnamefont {L.}~\bibnamefont {Lafferentz}}, \bibinfo
  {author} {\bibfnamefont {M.}~\bibnamefont {Persson}}, \bibinfo {author}
  {\bibfnamefont {M.~V.}\ \bibnamefont {Peters}}, \ and\ \bibinfo {author}
  {\bibfnamefont {S.}~\bibnamefont {Hecht}},\ }\href {\doibase
  10.1038/nnano.2007.346} {\bibfield  {journal} {\bibinfo  {journal} {Nature
  Nanotechnology}\ }\textbf {\bibinfo {volume} {2}},\ \bibinfo {pages} {687}
  (\bibinfo {year} {2007})}\BibitemShut {NoStop}%
\bibitem [{\citenamefont {Eder}\ \emph {et~al.}(2013)\citenamefont {Eder},
  \citenamefont {Smith}, \citenamefont {Cebula}, \citenamefont {Heckl},
  \citenamefont {Beton},\ and\ \citenamefont {Lackinger}}]{surfchem4}%
  \BibitemOpen
  \bibfield  {author} {\bibinfo {author} {\bibfnamefont {G.}~\bibnamefont
  {Eder}}, \bibinfo {author} {\bibfnamefont {E.~F.}\ \bibnamefont {Smith}},
  \bibinfo {author} {\bibfnamefont {I.}~\bibnamefont {Cebula}}, \bibinfo
  {author} {\bibfnamefont {W.~M.}\ \bibnamefont {Heckl}}, \bibinfo {author}
  {\bibfnamefont {P.~H.}\ \bibnamefont {Beton}}, \ and\ \bibinfo {author}
  {\bibfnamefont {M.}~\bibnamefont {Lackinger}},\ }\href {\doibase
  10.1021/nn400337v} {\bibfield  {journal} {\bibinfo  {journal} {ACS Nano}\
  }\textbf {\bibinfo {volume} {7}},\ \bibinfo {pages} {3014} (\bibinfo {year}
  {2013})}\BibitemShut {NoStop}%
\bibitem [{\citenamefont {Rastgoo-Lahrood}\ \emph {et~al.}(2016)\citenamefont
  {Rastgoo-Lahrood}, \citenamefont {Bj{\"o}rk}, \citenamefont {Lischka},
  \citenamefont {Eichhorn}, \citenamefont {Kloft}, \citenamefont {Fritton},
  \citenamefont {Strunskus}, \citenamefont {Samanta}, \citenamefont
  {Schmittel}, \citenamefont {Heckl},\ and\ \citenamefont
  {Lackinger}}]{surfchem5}%
  \BibitemOpen
  \bibfield  {author} {\bibinfo {author} {\bibfnamefont {A.}~\bibnamefont
  {Rastgoo-Lahrood}}, \bibinfo {author} {\bibfnamefont {J.}~\bibnamefont
  {Bj{\"o}rk}}, \bibinfo {author} {\bibfnamefont {M.}~\bibnamefont {Lischka}},
  \bibinfo {author} {\bibfnamefont {J.}~\bibnamefont {Eichhorn}}, \bibinfo
  {author} {\bibfnamefont {S.}~\bibnamefont {Kloft}}, \bibinfo {author}
  {\bibfnamefont {M.}~\bibnamefont {Fritton}}, \bibinfo {author} {\bibfnamefont
  {T.}~\bibnamefont {Strunskus}}, \bibinfo {author} {\bibfnamefont
  {D.}~\bibnamefont {Samanta}}, \bibinfo {author} {\bibfnamefont
  {M.}~\bibnamefont {Schmittel}}, \bibinfo {author} {\bibfnamefont {W.~M.}\
  \bibnamefont {Heckl}}, \ and\ \bibinfo {author} {\bibfnamefont
  {M.}~\bibnamefont {Lackinger}},\ }\href {\doibase
  https://doi.org/10.1002/anie.201600684} {\bibfield  {journal} {\bibinfo
  {journal} {Angewandte Chemie International Edition}\ }\textbf {\bibinfo
  {volume} {55}},\ \bibinfo {pages} {7650} (\bibinfo {year}
  {2016})}\BibitemShut {NoStop}%
\bibitem [{\citenamefont {Galeotti}\ \emph {et~al.}(2020)\citenamefont
  {Galeotti}, \citenamefont {Fritton},\ and\ \citenamefont
  {Lackinger}}]{surfchem6}%
  \BibitemOpen
  \bibfield  {author} {\bibinfo {author} {\bibfnamefont {G.}~\bibnamefont
  {Galeotti}}, \bibinfo {author} {\bibfnamefont {M.}~\bibnamefont {Fritton}}, \
  and\ \bibinfo {author} {\bibfnamefont {M.}~\bibnamefont {Lackinger}},\ }\href
  {\doibase https://doi.org/10.1002/anie.202010833} {\bibfield  {journal}
  {\bibinfo  {journal} {Angewandte Chemie International Edition}\ }\textbf
  {\bibinfo {volume} {59}},\ \bibinfo {pages} {22785} (\bibinfo {year}
  {2020})}\BibitemShut {NoStop}%
\bibitem [{\citenamefont {Bardi}\ and\ \citenamefont
  {Rovida}(1983)}]{BARDI1983145}%
  \BibitemOpen
  \bibfield  {author} {\bibinfo {author} {\bibfnamefont {U.}~\bibnamefont
  {Bardi}}\ and\ \bibinfo {author} {\bibfnamefont {G.}~\bibnamefont {Rovida}},\
  }\href {\doibase https://doi.org/10.1016/S0039-6028(83)80023-5} {\bibfield
  {journal} {\bibinfo  {journal} {Surface Science}\ }\textbf {\bibinfo {volume}
  {128}},\ \bibinfo {pages} {145 } (\bibinfo {year} {1983})}\BibitemShut
  {NoStop}%
\bibitem [{\citenamefont {Yamada}\ \emph {et~al.}(1996)\citenamefont {Yamada},
  \citenamefont {Ogaki}, \citenamefont {Okubo},\ and\ \citenamefont
  {Itaya}}]{YAMADA1996321}%
  \BibitemOpen
  \bibfield  {author} {\bibinfo {author} {\bibfnamefont {T.}~\bibnamefont
  {Yamada}}, \bibinfo {author} {\bibfnamefont {K.}~\bibnamefont {Ogaki}},
  \bibinfo {author} {\bibfnamefont {S.}~\bibnamefont {Okubo}}, \ and\ \bibinfo
  {author} {\bibfnamefont {K.}~\bibnamefont {Itaya}},\ }\href {\doibase
  https://doi.org/10.1016/S0039-6028(96)00880-1} {\bibfield  {journal}
  {\bibinfo  {journal} {Surface Science}\ }\textbf {\bibinfo {volume} {369}},\
  \bibinfo {pages} {321 } (\bibinfo {year} {1996})}\BibitemShut {NoStop}%
\bibitem [{\citenamefont {Dreyer}\ \emph {et~al.}(2010)\citenamefont {Dreyer},
  \citenamefont {Lee}, \citenamefont {Wang},\ and\ \citenamefont
  {Barker}}]{our_4K_system}%
  \BibitemOpen
  \bibfield  {author} {\bibinfo {author} {\bibfnamefont {M.}~\bibnamefont
  {Dreyer}}, \bibinfo {author} {\bibfnamefont {J.}~\bibnamefont {Lee}},
  \bibinfo {author} {\bibfnamefont {H.}~\bibnamefont {Wang}}, \ and\ \bibinfo
  {author} {\bibfnamefont {B.}~\bibnamefont {Barker}},\ }\href {\doibase
  10.1063/1.3427217} {\bibfield  {journal} {\bibinfo  {journal} {Review of
  Scientific Instruments}\ }\textbf {\bibinfo {volume} {81}},\ \bibinfo {eid}
  {053703} (\bibinfo {year} {2010})}\BibitemShut {NoStop}%
\bibitem [{\citenamefont {Furman}\ and\ \citenamefont
  {Harrington}(1996)}]{iodinesource}%
  \BibitemOpen
  \bibfield  {author} {\bibinfo {author} {\bibfnamefont {S.~A.}\ \bibnamefont
  {Furman}}\ and\ \bibinfo {author} {\bibfnamefont {D.~A.}\ \bibnamefont
  {Harrington}},\ }\href {\doibase 10.1116/1.579929} {\bibfield  {journal}
  {\bibinfo  {journal} {Journal of Vacuum Science \& Technology A}\ }\textbf
  {\bibinfo {volume} {14}},\ \bibinfo {pages} {256} (\bibinfo {year}
  {1996})}\BibitemShut {NoStop}%
\bibitem [{\citenamefont {Schott}\ and\ \citenamefont
  {White}(1994)}]{doi:10.1021/j100052a049}%
  \BibitemOpen
  \bibfield  {author} {\bibinfo {author} {\bibfnamefont {J.~H.}\ \bibnamefont
  {Schott}}\ and\ \bibinfo {author} {\bibfnamefont {H.~S.}\ \bibnamefont
  {White}},\ }\href {\doibase 10.1021/j100052a049} {\bibfield  {journal}
  {\bibinfo  {journal} {The Journal of Physical Chemistry}\ }\textbf {\bibinfo
  {volume} {98}},\ \bibinfo {pages} {291} (\bibinfo {year} {1994})},\ \Eprint
  {http://arxiv.org/abs/https://doi.org/10.1021/j100052a049}
  {https://doi.org/10.1021/j100052a049} \BibitemShut {NoStop}%
\bibitem [{\citenamefont {Hossick~Schott}\ and\ \citenamefont
  {White}(1994)}]{doi:10.1021/la00014a024}%
  \BibitemOpen
  \bibfield  {author} {\bibinfo {author} {\bibfnamefont {J.}~\bibnamefont
  {Hossick~Schott}}\ and\ \bibinfo {author} {\bibfnamefont {H.~S.}\
  \bibnamefont {White}},\ }\href {\doibase 10.1021/la00014a024} {\bibfield
  {journal} {\bibinfo  {journal} {Langmuir}\ }\textbf {\bibinfo {volume}
  {10}},\ \bibinfo {pages} {486} (\bibinfo {year} {1994})},\ \Eprint
  {http://arxiv.org/abs/https://doi.org/10.1021/la00014a024}
  {https://doi.org/10.1021/la00014a024} \BibitemShut {NoStop}%
\bibitem [{\citenamefont {Bushell}\ \emph {et~al.}(2005)\citenamefont
  {Bushell}, \citenamefont {Carley}, \citenamefont {Coughlin}, \citenamefont
  {Davies}, \citenamefont {Edwards}, \citenamefont {Morgan},\ and\
  \citenamefont {Parsons}}]{doi:10.1021/jp0513465}%
  \BibitemOpen
  \bibfield  {author} {\bibinfo {author} {\bibfnamefont {J.}~\bibnamefont
  {Bushell}}, \bibinfo {author} {\bibfnamefont {A.~F.}\ \bibnamefont {Carley}},
  \bibinfo {author} {\bibfnamefont {M.}~\bibnamefont {Coughlin}}, \bibinfo
  {author} {\bibfnamefont {P.~R.}\ \bibnamefont {Davies}}, \bibinfo {author}
  {\bibfnamefont {D.}~\bibnamefont {Edwards}}, \bibinfo {author} {\bibfnamefont
  {D.~J.}\ \bibnamefont {Morgan}}, \ and\ \bibinfo {author} {\bibfnamefont
  {M.}~\bibnamefont {Parsons}},\ }\href {\doibase 10.1021/jp0513465} {\bibfield
   {journal} {\bibinfo  {journal} {The Journal of Physical Chemistry B}\
  }\textbf {\bibinfo {volume} {109}},\ \bibinfo {pages} {9556} (\bibinfo {year}
  {2005})},\ \bibinfo {note} {pMID: 16852150},\ \Eprint
  {http://arxiv.org/abs/https://doi.org/10.1021/jp0513465}
  {https://doi.org/10.1021/jp0513465} \BibitemShut {NoStop}%
\bibitem [{\citenamefont {Farrell}\ \emph {et~al.}(1981)\citenamefont
  {Farrell}, \citenamefont {Traum}, \citenamefont {Smith}, \citenamefont
  {Royer}, \citenamefont {Woodruff},\ and\ \citenamefont
  {Johnson}}]{FARRELL1981527}%
  \BibitemOpen
  \bibfield  {author} {\bibinfo {author} {\bibfnamefont {H.}~\bibnamefont
  {Farrell}}, \bibinfo {author} {\bibfnamefont {M.}~\bibnamefont {Traum}},
  \bibinfo {author} {\bibfnamefont {N.}~\bibnamefont {Smith}}, \bibinfo
  {author} {\bibfnamefont {W.}~\bibnamefont {Royer}}, \bibinfo {author}
  {\bibfnamefont {D.}~\bibnamefont {Woodruff}}, \ and\ \bibinfo {author}
  {\bibfnamefont {P.}~\bibnamefont {Johnson}},\ }\href {\doibase
  https://doi.org/10.1016/0039-6028(81)90044-3} {\bibfield  {journal} {\bibinfo
   {journal} {Surface Science}\ }\textbf {\bibinfo {volume} {102}},\ \bibinfo
  {pages} {527 } (\bibinfo {year} {1981})}\BibitemShut {NoStop}%
\bibitem [{\citenamefont {Maglietta}\ \emph {et~al.}(1982)\citenamefont
  {Maglietta}, \citenamefont {Zanazzi}, \citenamefont {Bardi}, \citenamefont
  {Sondericker}, \citenamefont {Jona},\ and\ \citenamefont
  {Marcus}}]{MAGLIETTA1982141}%
  \BibitemOpen
  \bibfield  {author} {\bibinfo {author} {\bibfnamefont {M.}~\bibnamefont
  {Maglietta}}, \bibinfo {author} {\bibfnamefont {E.}~\bibnamefont {Zanazzi}},
  \bibinfo {author} {\bibfnamefont {U.}~\bibnamefont {Bardi}}, \bibinfo
  {author} {\bibfnamefont {D.}~\bibnamefont {Sondericker}}, \bibinfo {author}
  {\bibfnamefont {F.}~\bibnamefont {Jona}}, \ and\ \bibinfo {author}
  {\bibfnamefont {P.}~\bibnamefont {Marcus}},\ }\href {\doibase
  https://doi.org/10.1016/0039-6028(82)90136-4} {\bibfield  {journal} {\bibinfo
   {journal} {Surface Science}\ }\textbf {\bibinfo {volume} {123}},\ \bibinfo
  {pages} {141 } (\bibinfo {year} {1982})}\BibitemShut {NoStop}%
\bibitem [{\citenamefont {Forstmann}\ \emph {et~al.}(1973)\citenamefont
  {Forstmann}, \citenamefont {Berndt},\ and\ \citenamefont
  {B\"uttner}}]{PhysRevLett.30.17}%
  \BibitemOpen
  \bibfield  {author} {\bibinfo {author} {\bibfnamefont {F.}~\bibnamefont
  {Forstmann}}, \bibinfo {author} {\bibfnamefont {W.}~\bibnamefont {Berndt}}, \
  and\ \bibinfo {author} {\bibfnamefont {P.}~\bibnamefont {B\"uttner}},\ }\href
  {\doibase 10.1103/PhysRevLett.30.17} {\bibfield  {journal} {\bibinfo
  {journal} {Phys. Rev. Lett.}\ }\textbf {\bibinfo {volume} {30}},\ \bibinfo
  {pages} {17} (\bibinfo {year} {1973})}\BibitemShut {NoStop}%
\bibitem [{\citenamefont {Wilson}\ \emph {et~al.}(1975)\citenamefont {Wilson},
  \citenamefont {Salvo},\ and\ \citenamefont {Mahajan}}]{cdwreview}%
  \BibitemOpen
  \bibfield  {author} {\bibinfo {author} {\bibfnamefont {J.}~\bibnamefont
  {Wilson}}, \bibinfo {author} {\bibfnamefont {F.~D.}\ \bibnamefont {Salvo}}, \
  and\ \bibinfo {author} {\bibfnamefont {S.}~\bibnamefont {Mahajan}},\ }\href
  {\doibase 10.1080/00018737500101391} {\bibfield  {journal} {\bibinfo
  {journal} {Advances in Physics}\ }\textbf {\bibinfo {volume} {24}},\ \bibinfo
  {pages} {117} (\bibinfo {year} {1975})}\BibitemShut {NoStop}%
\bibitem [{\citenamefont {Andryushechkin}\ \emph {et~al.}(2009)\citenamefont
  {Andryushechkin}, \citenamefont {Zhidomirov}, \citenamefont {Eltsov},
  \citenamefont {Hladchanka},\ and\ \citenamefont
  {Korlyukov}}]{PhysRevB.80.125409}%
  \BibitemOpen
  \bibfield  {author} {\bibinfo {author} {\bibfnamefont {B.~V.}\ \bibnamefont
  {Andryushechkin}}, \bibinfo {author} {\bibfnamefont {G.~M.}\ \bibnamefont
  {Zhidomirov}}, \bibinfo {author} {\bibfnamefont {K.~N.}\ \bibnamefont
  {Eltsov}}, \bibinfo {author} {\bibfnamefont {Y.~V.}\ \bibnamefont
  {Hladchanka}}, \ and\ \bibinfo {author} {\bibfnamefont {A.~A.}\ \bibnamefont
  {Korlyukov}},\ }\href {\doibase 10.1103/PhysRevB.80.125409} {\bibfield
  {journal} {\bibinfo  {journal} {Phys. Rev. B}\ }\textbf {\bibinfo {volume}
  {80}},\ \bibinfo {pages} {125409} (\bibinfo {year} {2009})}\BibitemShut
  {NoStop}%
\end{thebibliography}%
\pagebreak
\appendix

\section{Real-space analysis}
\label{app:NNNhowto}
\begin{figure}%
\includegraphics[width=.90\columnwidth]{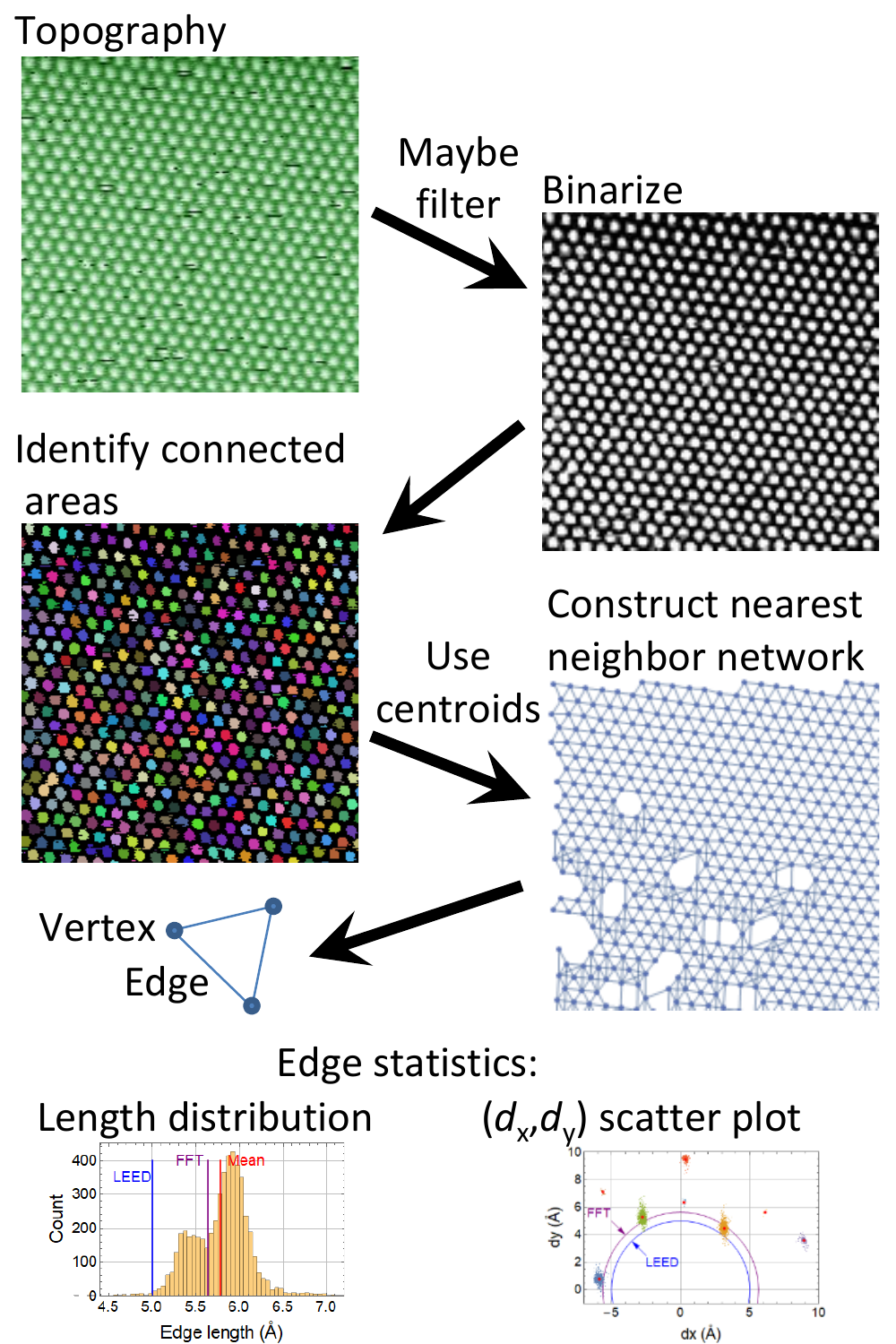}
\caption{\CO Flow diagram of extracting real space statistics from nearest neighbor networks.}\label{fig:NNNhowto}%
\end{figure}
Analyzing lattices in real space allows one to detect and classify local changes in lattice parameters and to identify systematic variations. Furthermore, relations between the atomic lattice and a superstructure can be evaluated at the local level. The process is shown in Fig.~\ref{fig:NNNhowto}. We started with an STM data set. The data was sometimes filtered to eliminate steps or to separate the atomic structure from a superstructure. In general, preprocessing was applied as sparingly as possible to avoid the possible introduction of artifacts. The data was then binarized with a chosen threshold level resulting in well separated cohesive regions. For each of the regions the centroid position, area, and sometimes elongation were extracted. Using the centroids positions was sufficiently accurate. More sophisticated approaches, such as using the topographic data as weights when calculating the centroids, gave little or no improvement. The results were post-selected based on area to eliminate spurious features due to noise or residual step edges. The centroid positions were then used as vertices to generate a nearest neighbor network graph, typically with six neighbors assuming a triangular lattice. Again, post-selection was used to exclude graph edges that were clearly too long or too short. Other defects such as crossing graph edges were left in place. These defects were rare enough to remain statistically insignificant. Throughout the process all back references were maintained to identify the source of any anomaly within the original STM data. The distribution of the edge length typically had a mean value close to the lattice constant determined by FFT, while a histogram width and shape give hints as to the existence of substructures. A second instructive visualization was to mark each edge vector $(d_\mathrm{x},d_\mathrm{y})$ directly in a scatter plot. For a regular triangular lattice we expect three ellipsoidal clusters. The size and asymmetry of a cluster are due to imaging noise and scanner distortions, respectively. Clusters at combinations of lattice vectors stem from missing or missed nearest neighbors so that next-nearest neighbors were included during the graph construction. Similarly, single points were errors most likely from binarizing the image creating spurious connected regions. In contrast, clusters at positions incommensurate with the expected lattice vector hint at a physical cause. This is the case for the \agih\ and \agit\ superstructure.

\section{Supplemental: real-space measurements of the interatomic distances}
\label{app:realspace}

\begin{figure}%
\includegraphics[width=.495\columnwidth]{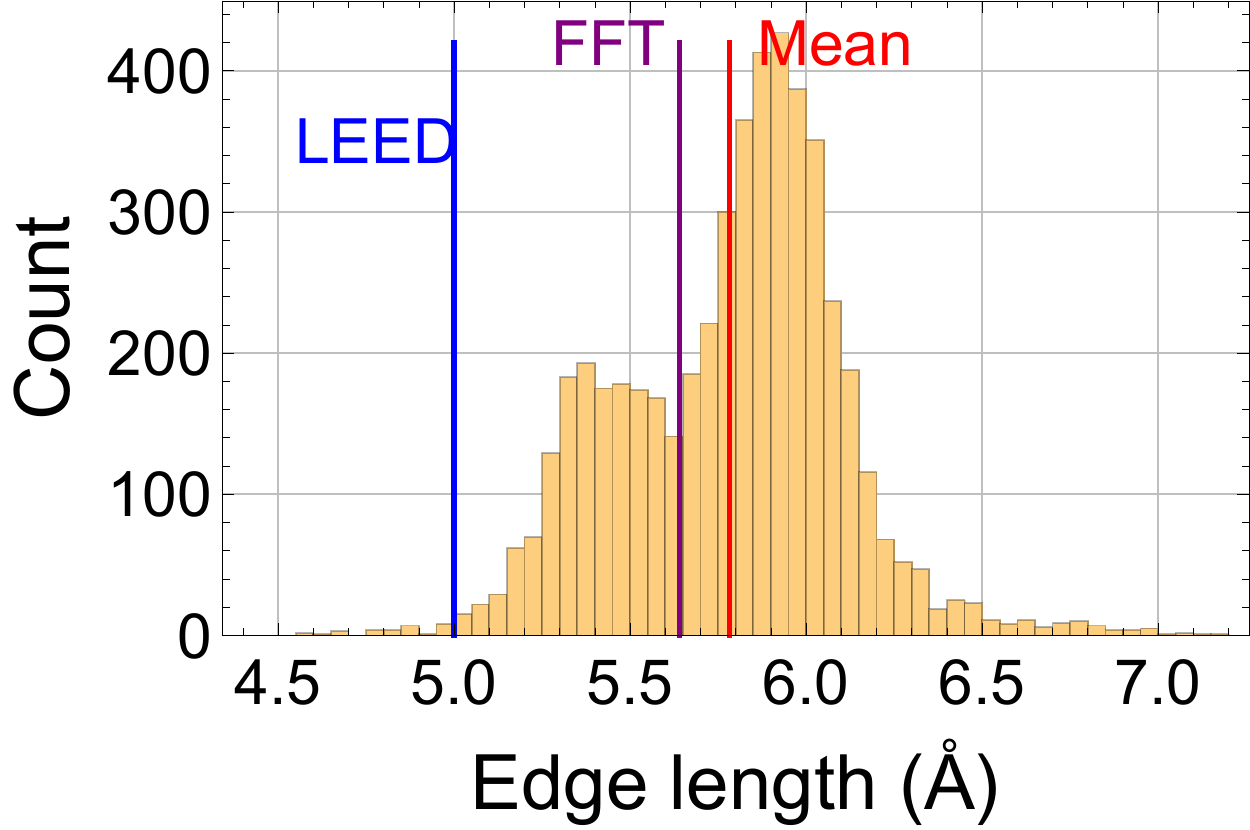}\hfill%
\includegraphics[width=.495\columnwidth]{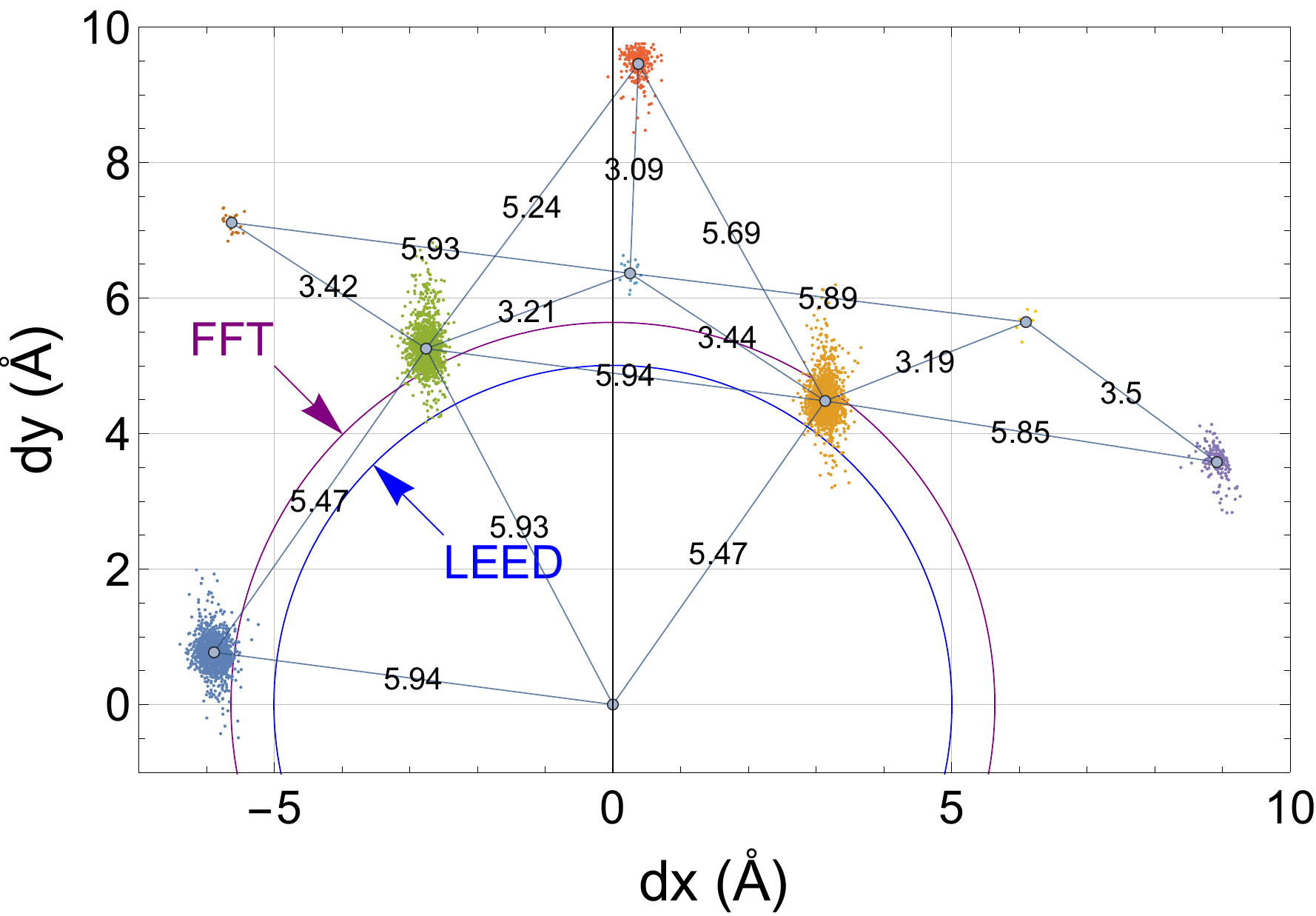}
\caption{\CO Histogram and distance vector plot of real-space measurements of the inter-atomic distances for the atomic lattice of the \agir\ phase. We marked the distances according to LEED (blue) as well as FFT (purple). The distance plot also contains the distances within the major clusters as an example.}\label{fig:monodist}%
\end{figure}

\begin{figure}%
\includegraphics[width=.45\columnwidth]{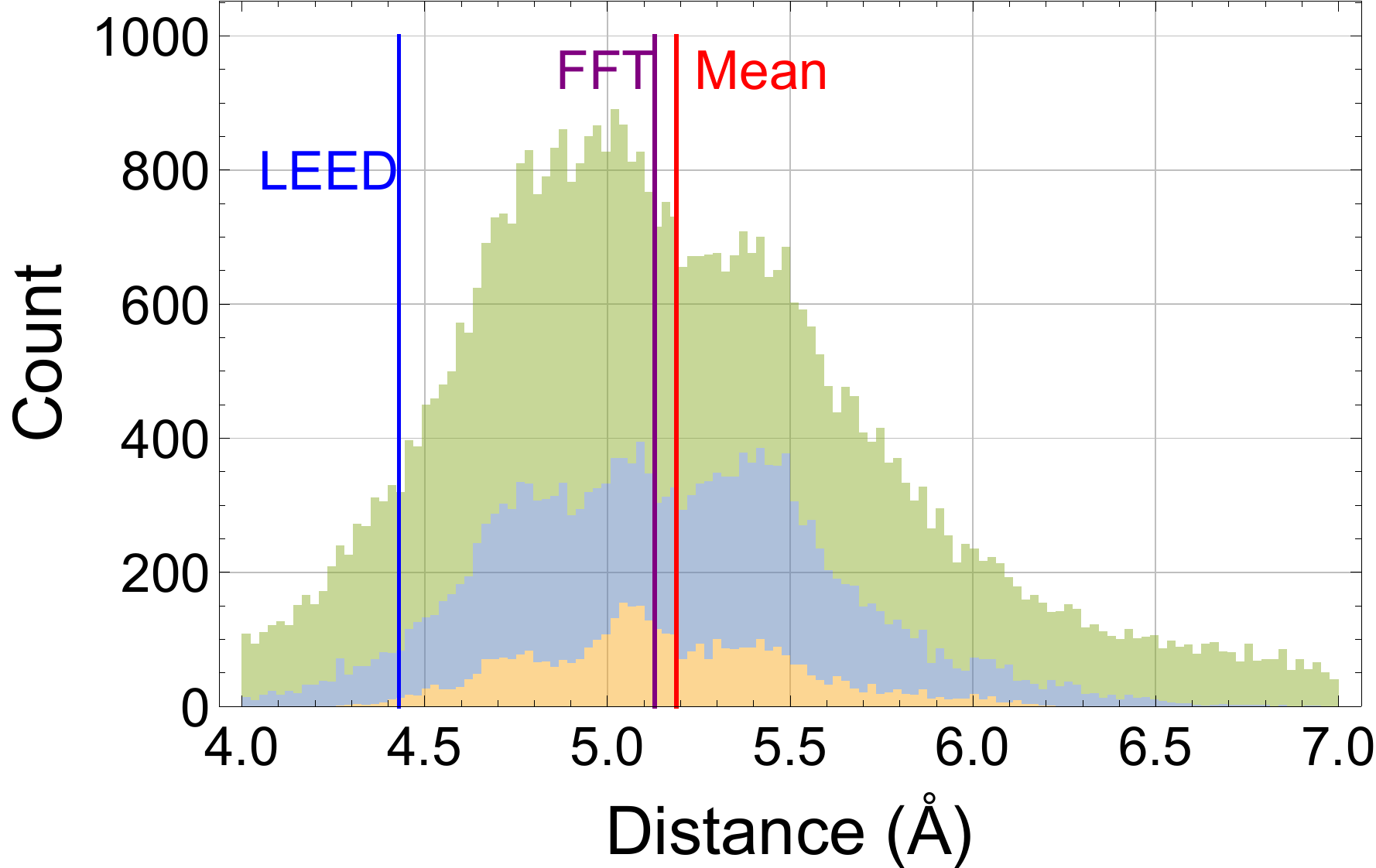}\hfill%
\includegraphics[width=.45\columnwidth]{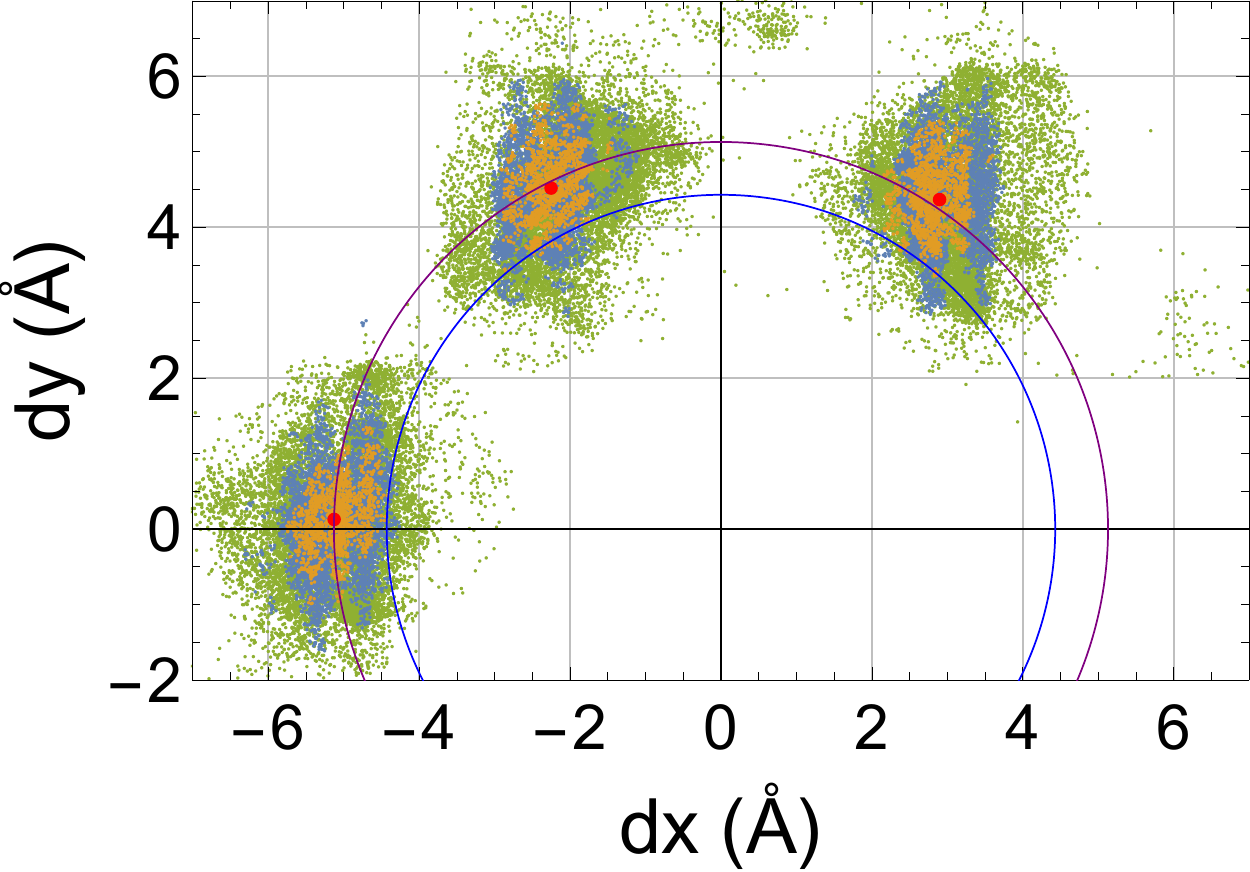}%
\caption{\CO Cumulative histogram and distance vector plot of real-space measurements of the inter-atomic distances of the \agih\ phase. We examined three STM images measured at different resolutions of $20\times20$~nm$^2$ (yellow), $35\times35$~nm$^2$ (blue), and $50\times50$~nm$^2$ (green), respectively. While lower resolution increases the accuracy, it reduces the statistics. We marked the distances according to LEED (blue), FFT (purple) as well as the mean value of this data set (red).}\label{fig:apphat}%
\end{figure}

\begin{figure}%
\includegraphics[width=.495\columnwidth]{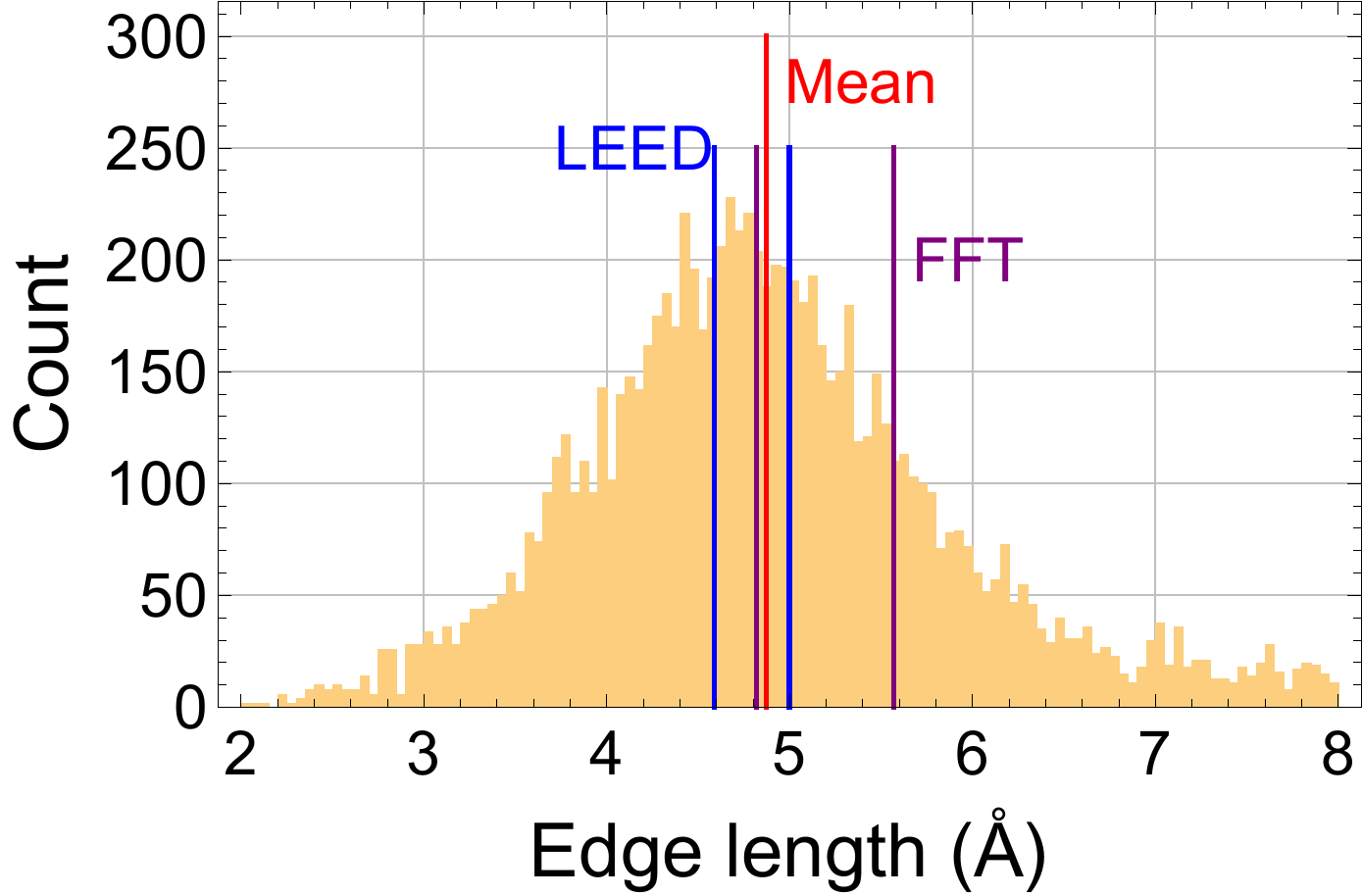}\hfill%
\includegraphics[width=.495\columnwidth]{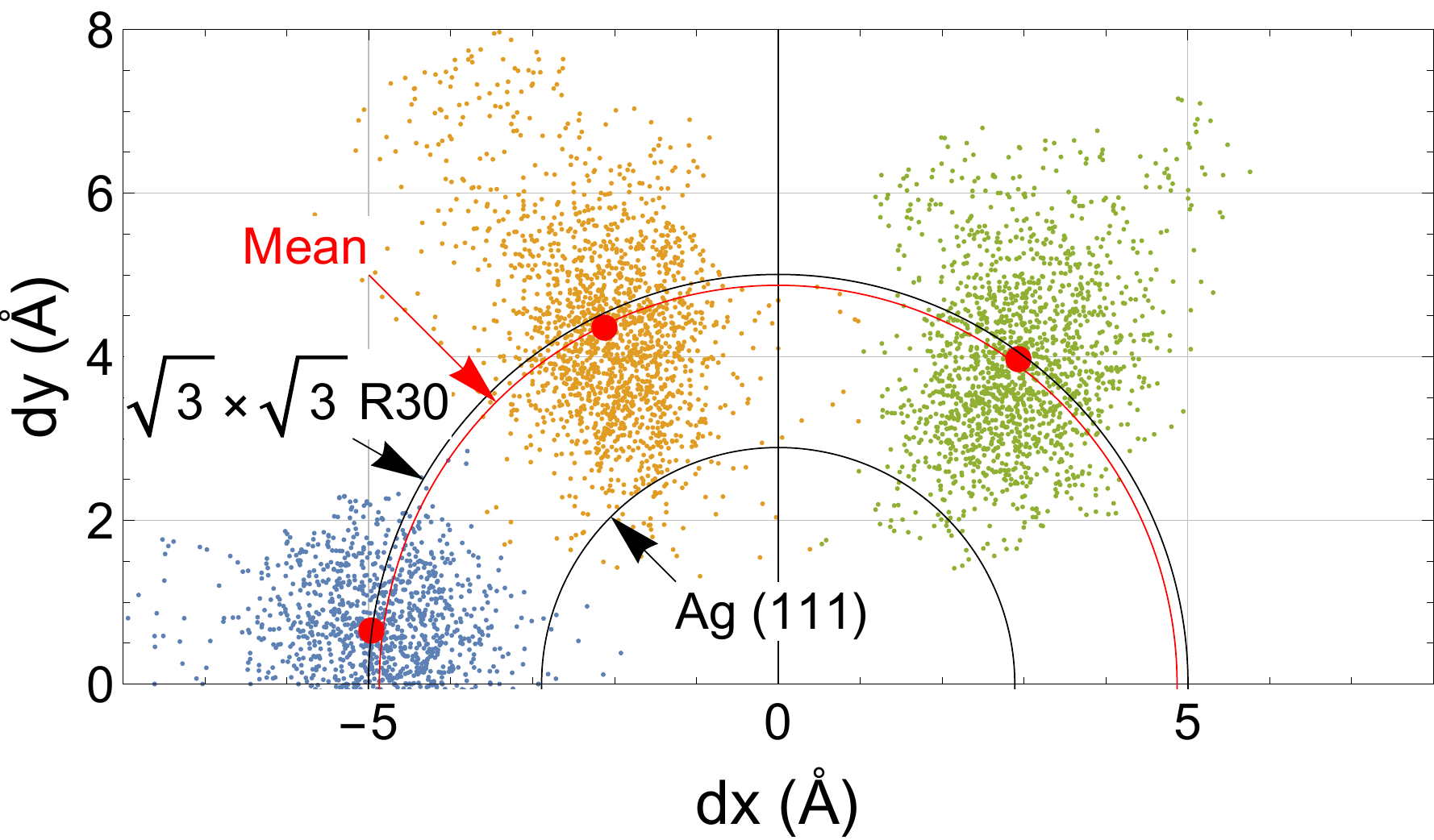}
\caption{\CO  Histogram and distance vector plot of real-space measurements of the inter-atomic distances for the atomic lattice of the \agit\ phase. We marked the \agir\ and Ag (111) inter-atomic distances (Black) as well as the mean value of this data set (red).}\label{fig:TriMonoLay}%
\end{figure}

We measured the inter-atomic distances while disregarding any existing superstructure for the three phases of iodine on Ag (111) by appropriately filtering the images. After finding the atom locations we constructed a nearest neighbor graph with a maximum of six neighbors within a cut-off radius of $\approx 9$~\AA, but no further restraints. The histogram in Fig.~\ref{fig:monodist}--\ref{fig:TriMonoLay} shows the length distribution of the respective graph edges. As expected, the mean value of the distribution agrees with the value measured by FFT. The multiple peaks are likely an artifact of the STM measurement due to the x-y asymmetry of the piezo scanner. Using the x and y distances of all graph edges we generated the respective scatter plot. The three main spots represent the primary axis of the triangular lattice. They appear just as elliptical structureless blobs. The respective centroids give us the atomic lattice vectors.

\section{FFT peaks of the \protect\agih\ and \protect\agit\ superlattice structure}
\label{app:FFTinner}

\begin{figure}%
\includegraphics[width=.75\columnwidth]{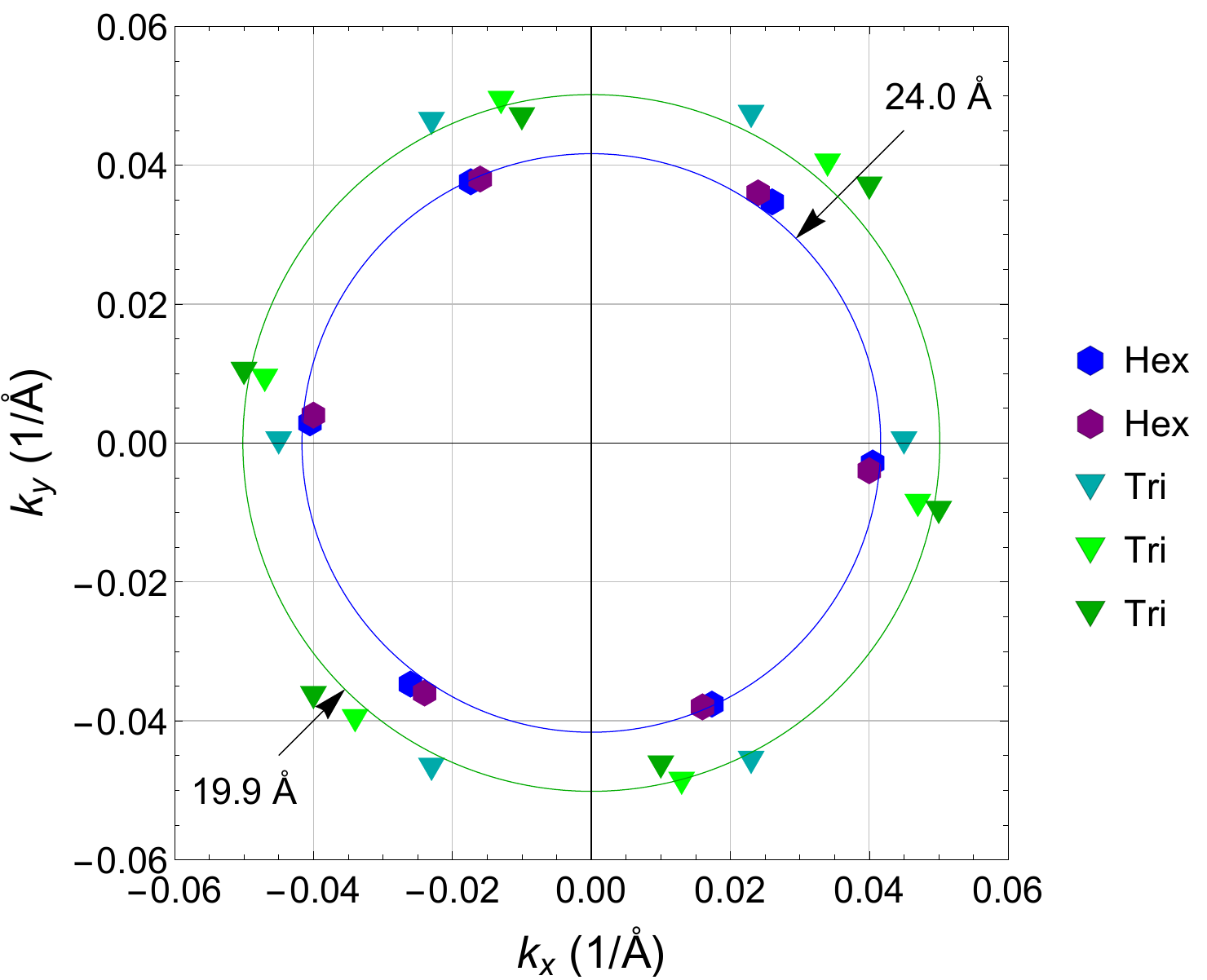}%
\caption{\CO Summary of the inner FFT spots for the \agih\ and \agit\ phase, respectively, clearly showing two different length scales of the respective superstructures. The large circles mark the average radii of the respective phase.}\label{fig:FFTsum2}%
\end{figure}
Figure \ref{fig:FFTsum2} shows an enlarged version of the inner FFT spots from Fig.~\ref{fig:FFTsum} representing the superstructure of the \agih\ and \agit\  phase. We found clearly different average length scales of $\overline{l}_\mathrm{\agih}=24.0$~\AA\ and $\overline{l}_\mathrm{\agit}=19.9$~\AA\, respectively. The clear difference points to different superstructure origins. The angular distribution is somewhat similar likely linked to the underlying Ag (111) surface. 

\section{Supplemental \protect\agih}

\subsection{\protect\agih\ superlattice structure}
\label{app:hexsuper}

\begin{figure}%
\includegraphics[width=.32\columnwidth]{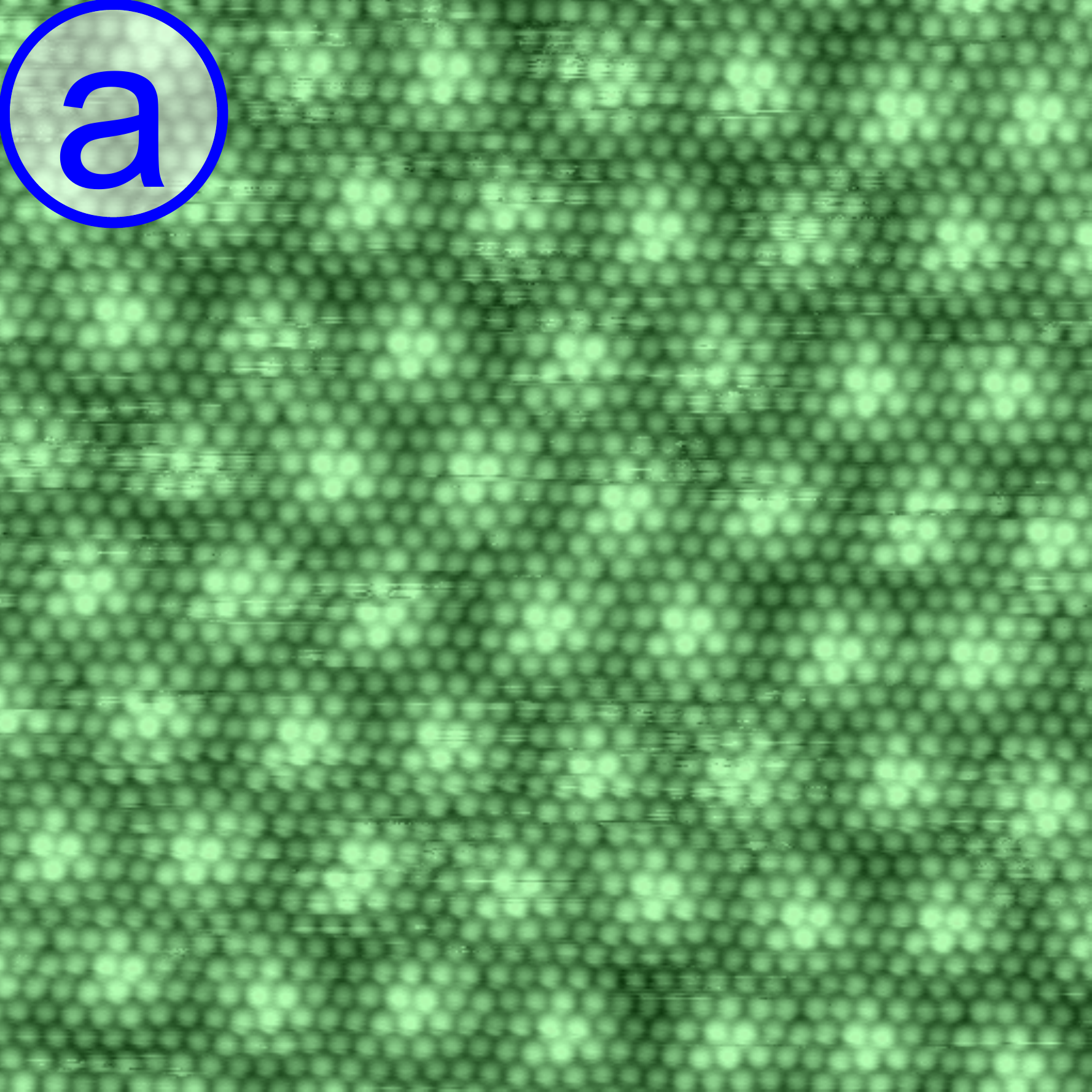}\hfill%
\includegraphics[width=.32\columnwidth]{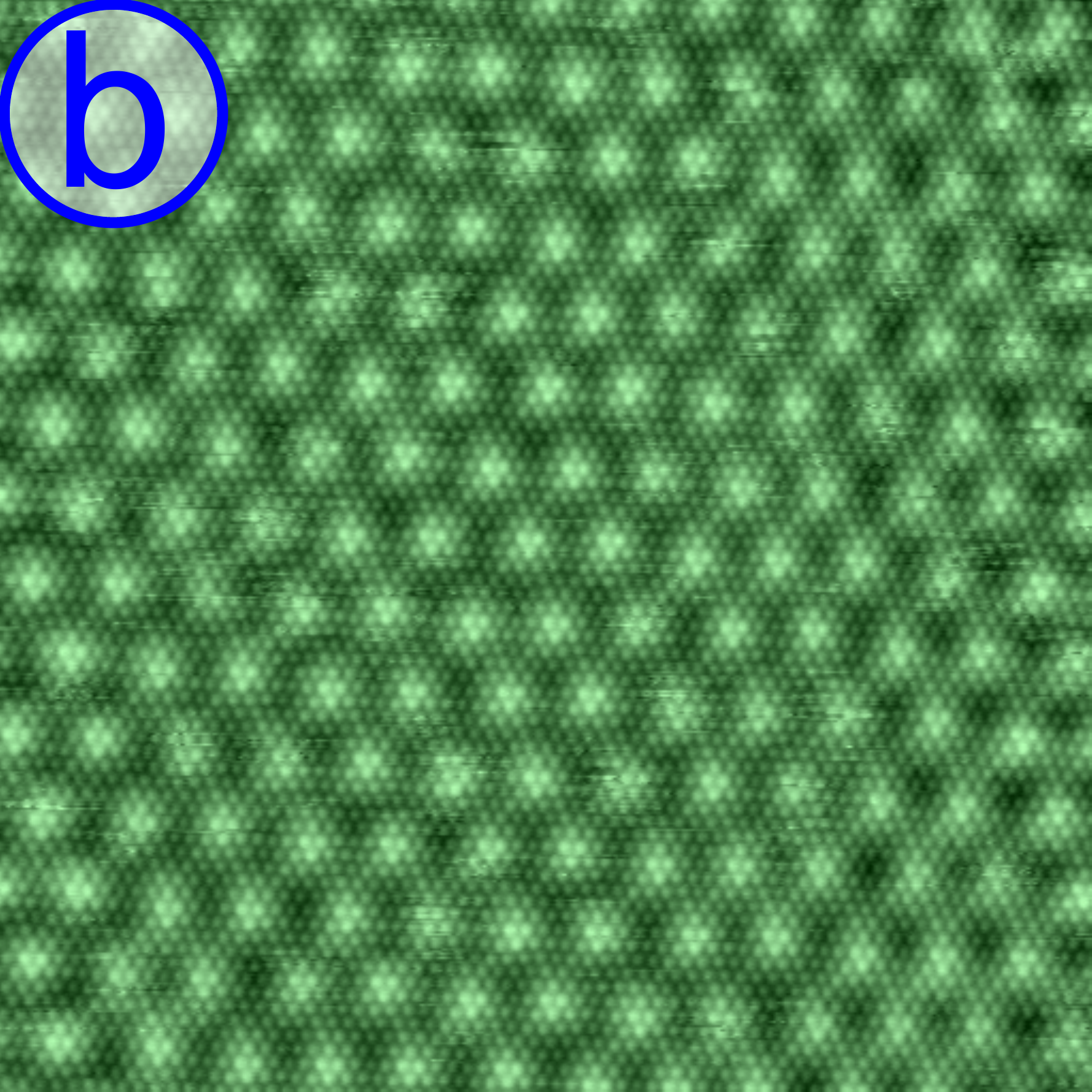}\hfill%
\includegraphics[width=.32\columnwidth]{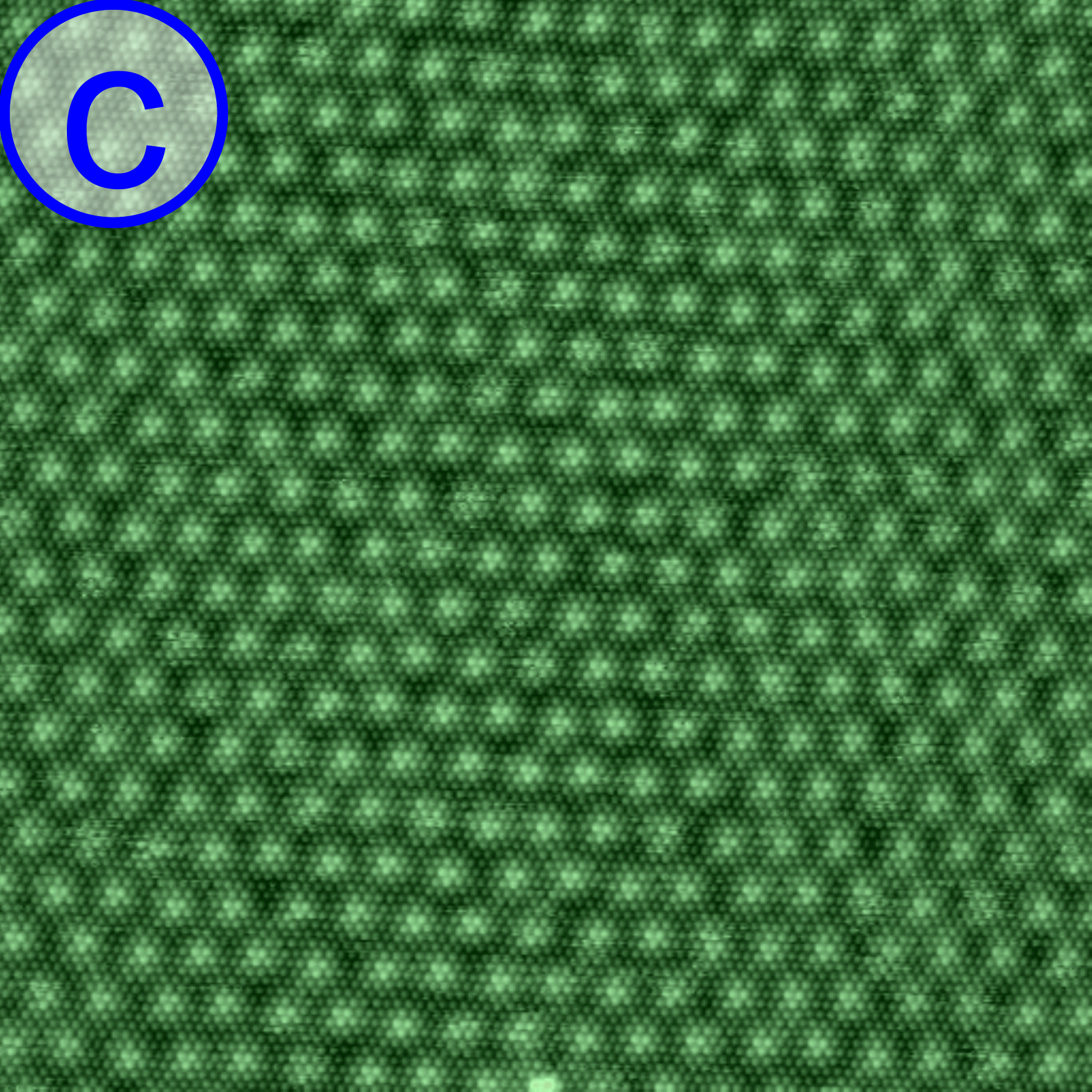}
\caption{\CO Source images for analyzing the \agih\ superstructure. Imaging parameters: (a) $20\times20$~nm$^2$, $\Vb=50$~mV, $I=0.1$~nA, (b) $35\times35$~nm$^2$, $\Vb=-500$~mV, $I=500$~pA, (c) $50\times50$~nm$^2$, $\Vb=-100$~mV, $I=100$~pA}\label{fig:AgIHexSource}%
\end{figure}

\begin{figure}%
\includegraphics[width=.4\columnwidth]{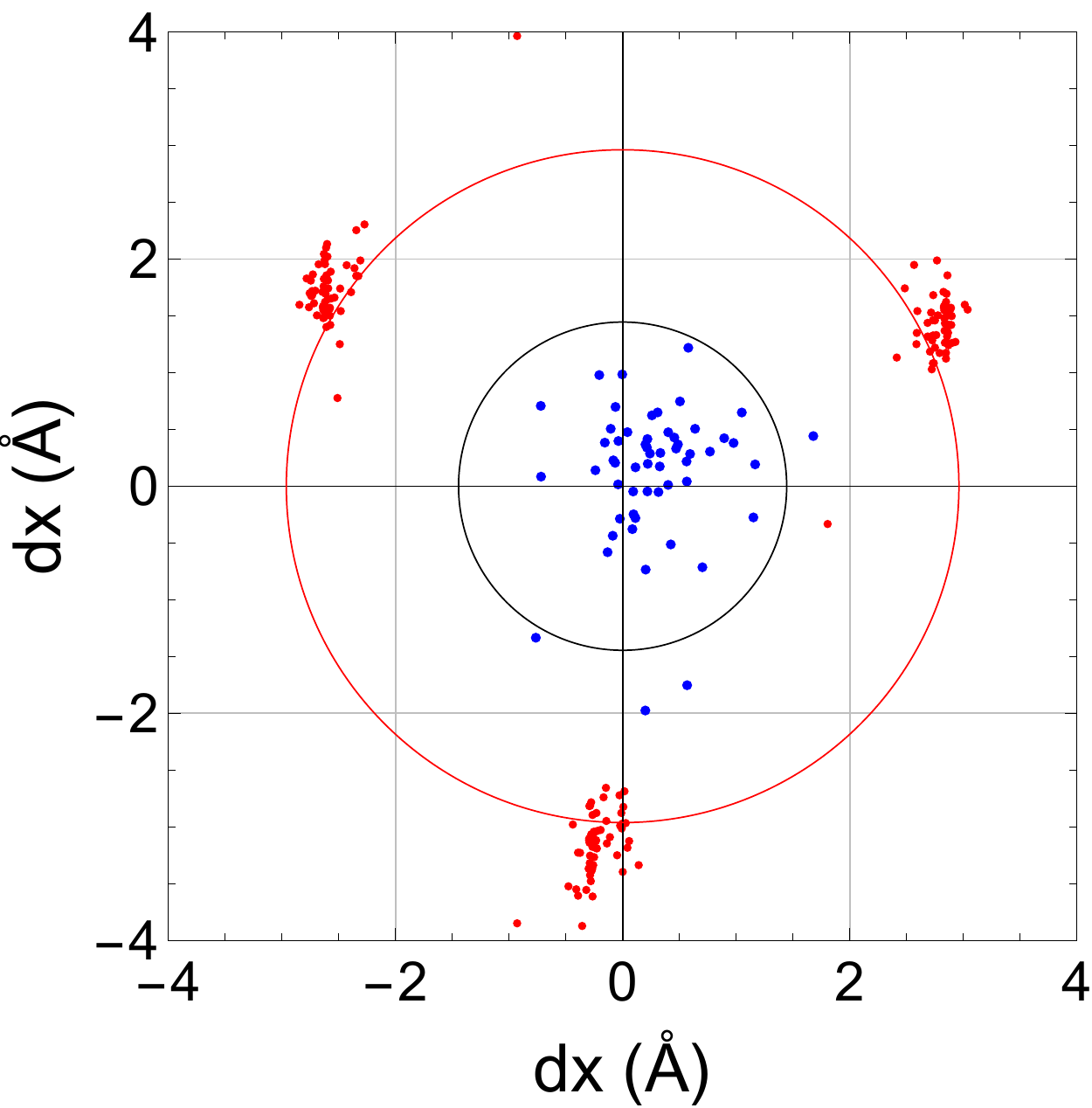}\hfill%
\includegraphics[width=.48\columnwidth]{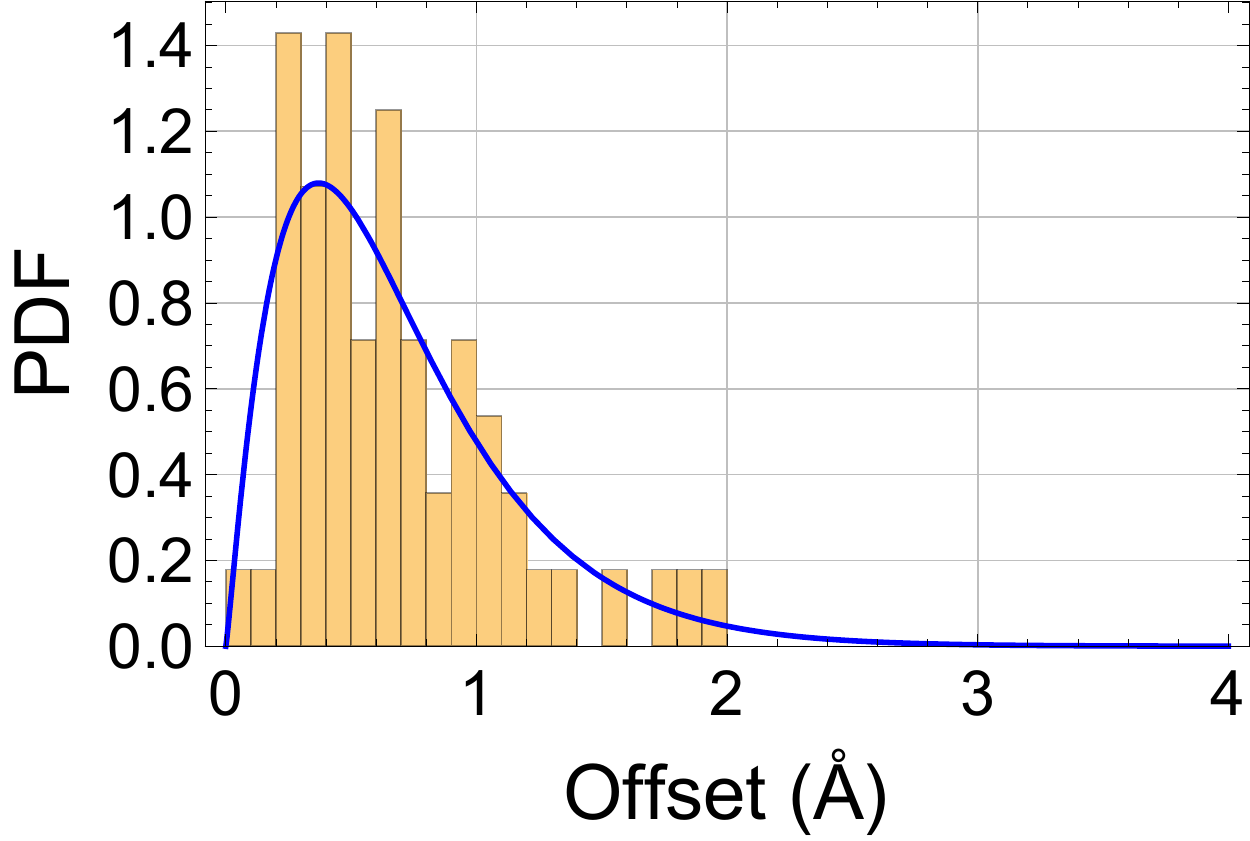}
\includegraphics[width=.4\columnwidth]{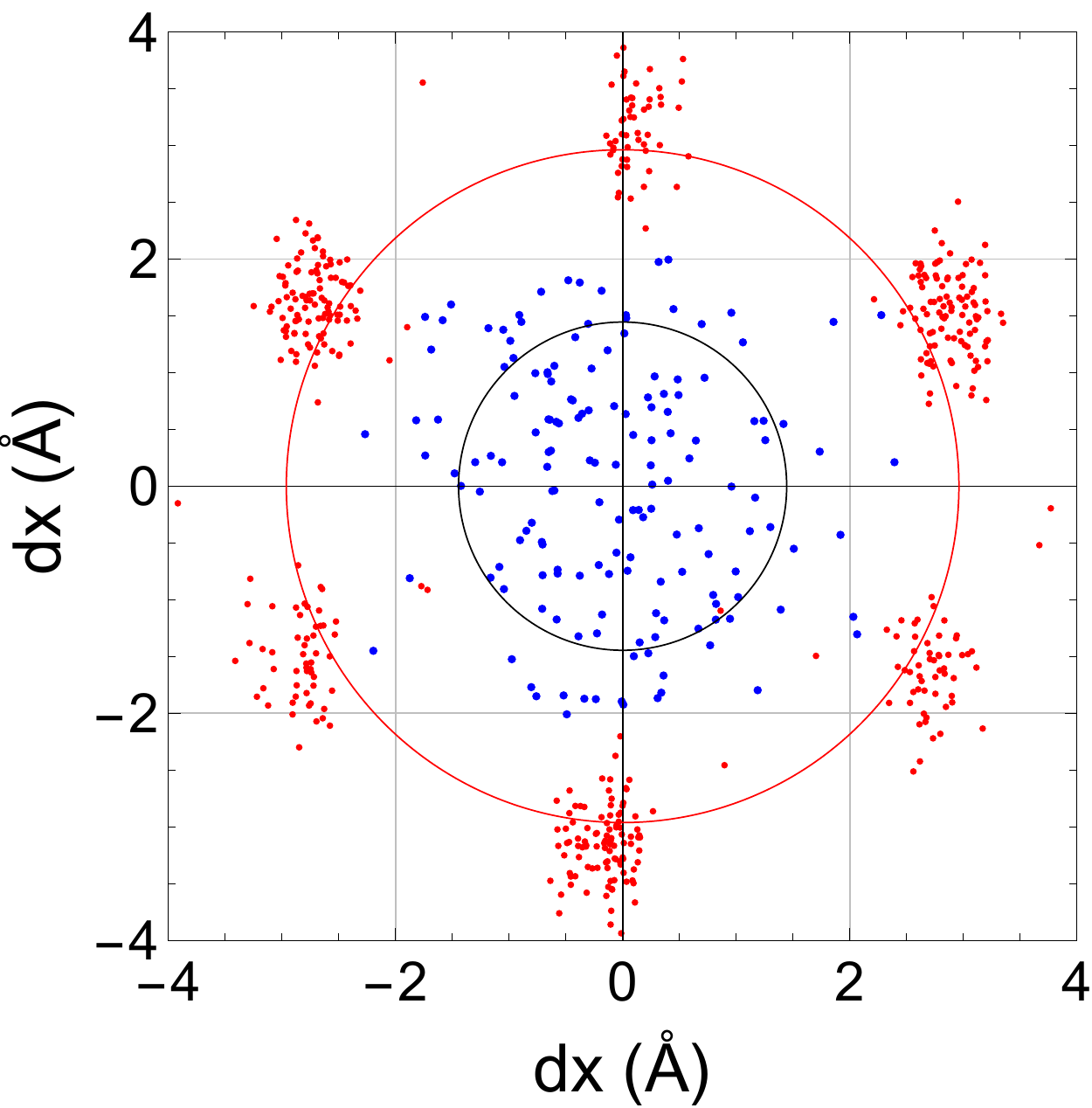}\hfill%
\includegraphics[width=.48\columnwidth]{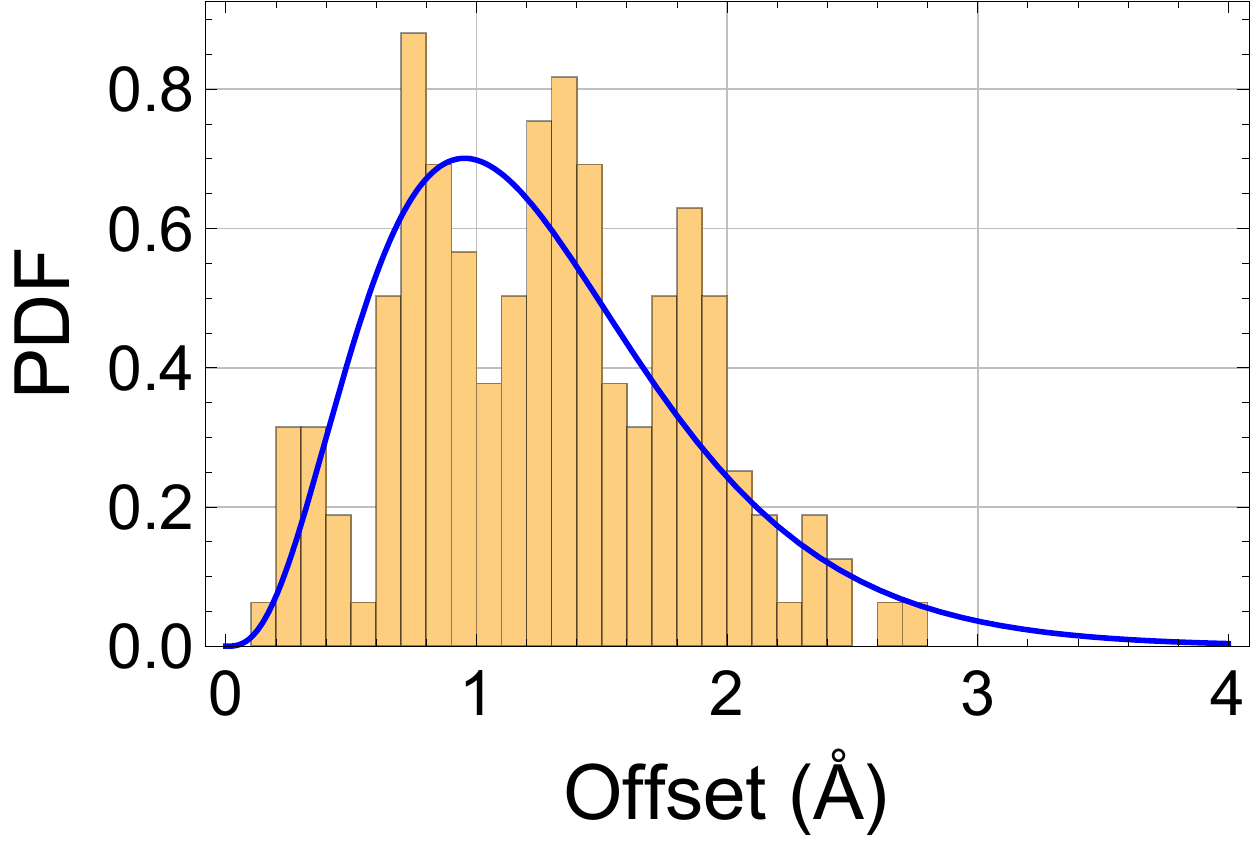}
\includegraphics[width=.4\columnwidth]{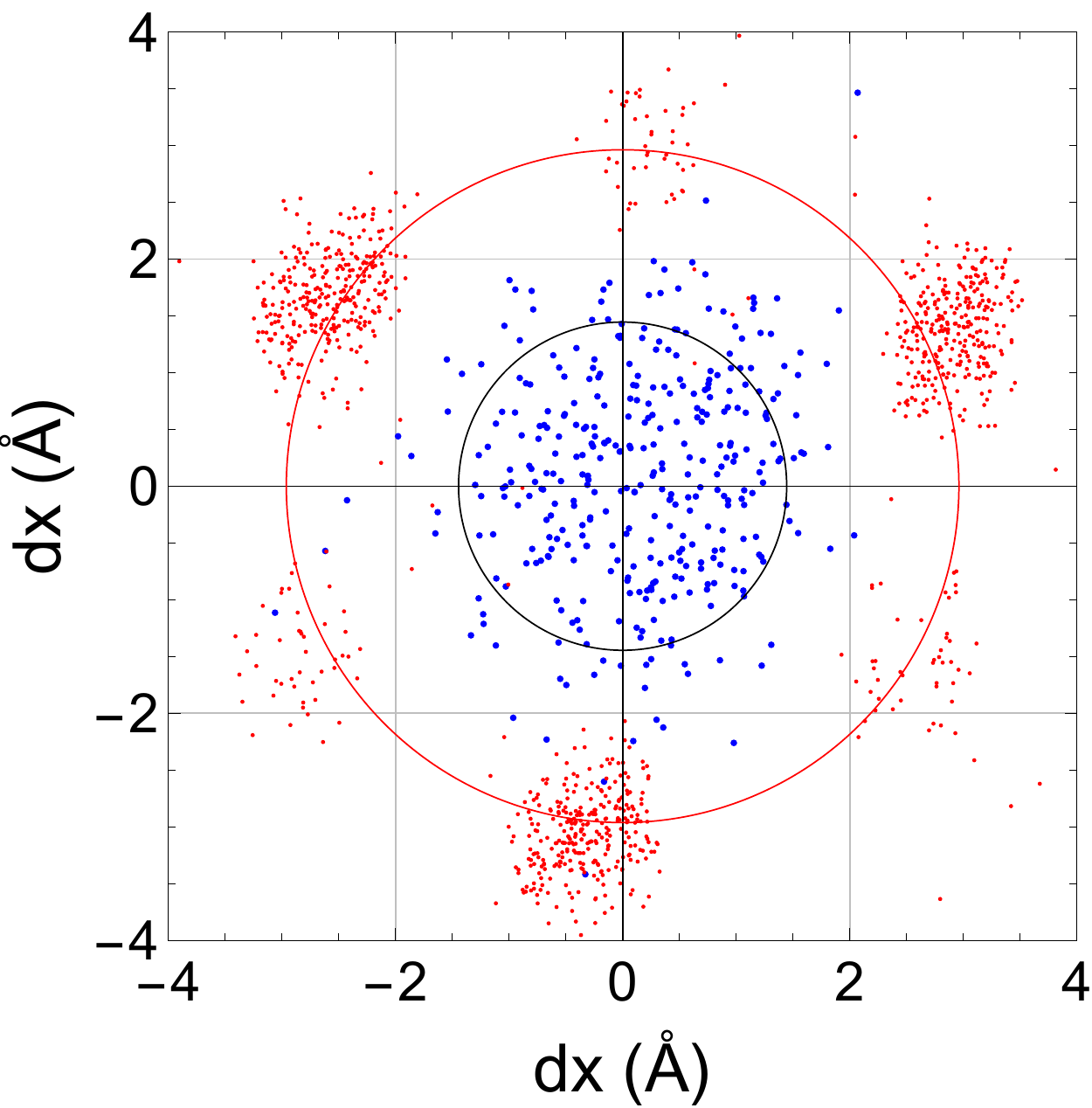}\hfill%
\includegraphics[width=.48\columnwidth]{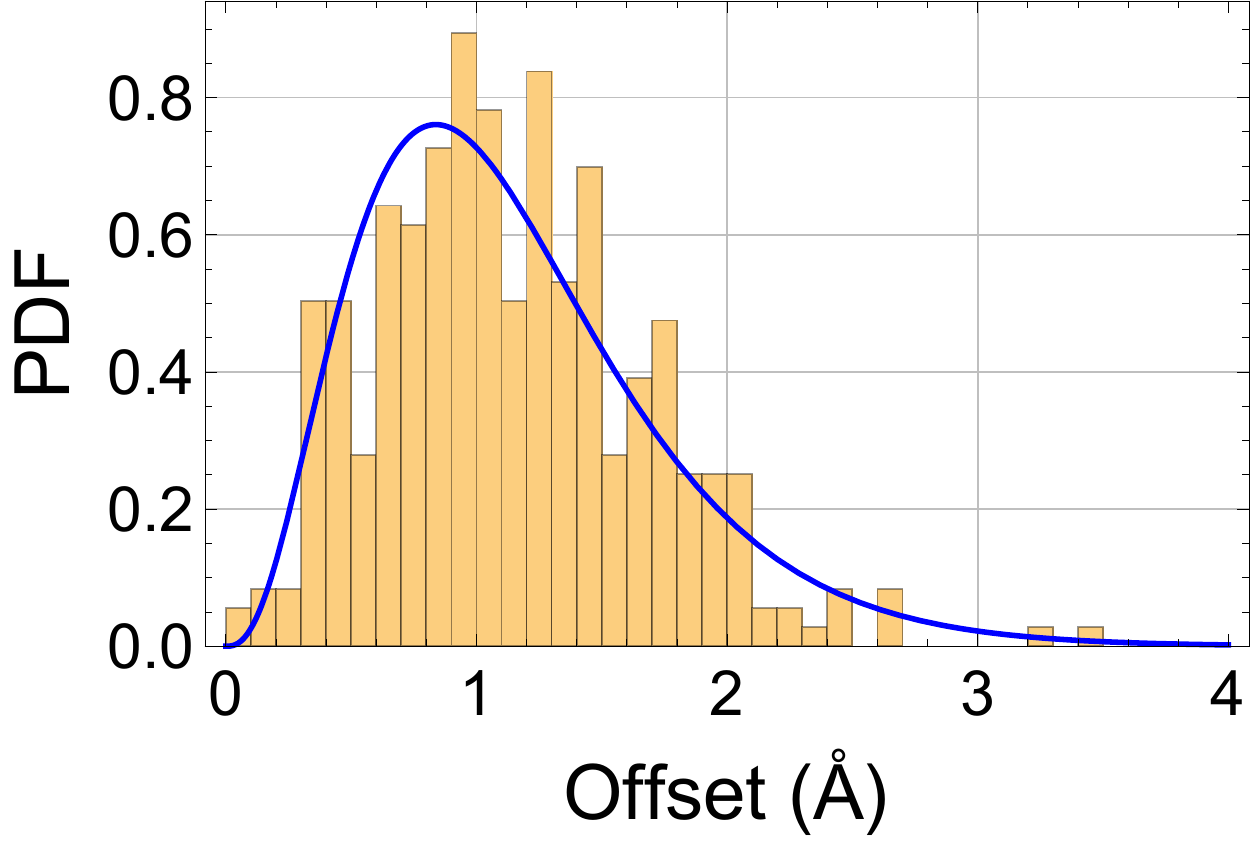}
\caption{\CO Distribution of the superlattice maxima (blue) in relation to the three nearest surface atoms (red). The histograms measure the radial distance to the geometric center of the three surrounding surface atoms and are fit to a gamma distribution.}\label{fig:HexDist1}%
\end{figure}

We determined the relative position of the superlattice maxima (SLM) with respect to the atomic lattice of the \agih\ phase. For each of the SLM within three STM images (see Fig.~\ref{fig:AgIHexSource}) we determined the location of the three nearest surface atoms. From the three locations we calculated the respective center $(x_\mathrm{c},y_\mathrm{c})$ and the distance to the SLM. The scatter plot in Fig.~\ref{fig:HexDist1} shows the surface atom locations in red and the SLM in blue. The locations are aligned by subtracting the respective $(x_\mathrm{c},y_\mathrm{c})$ from both the SLM position as well as the three atom positions. It is obvious that while the SLM location shows some scatter it lies mostly near the center of the three surrounding atoms. The histograms show the probability density function of the SLM distance from the center estimated by a gamma distribution
$F(x;\alpha,\beta)=\frac{\beta^\alpha x^{\alpha-1} e^{-\beta x}}{\Gamma(\alpha)}$.

\begin{figure}%
\includegraphics[width=.75\columnwidth]{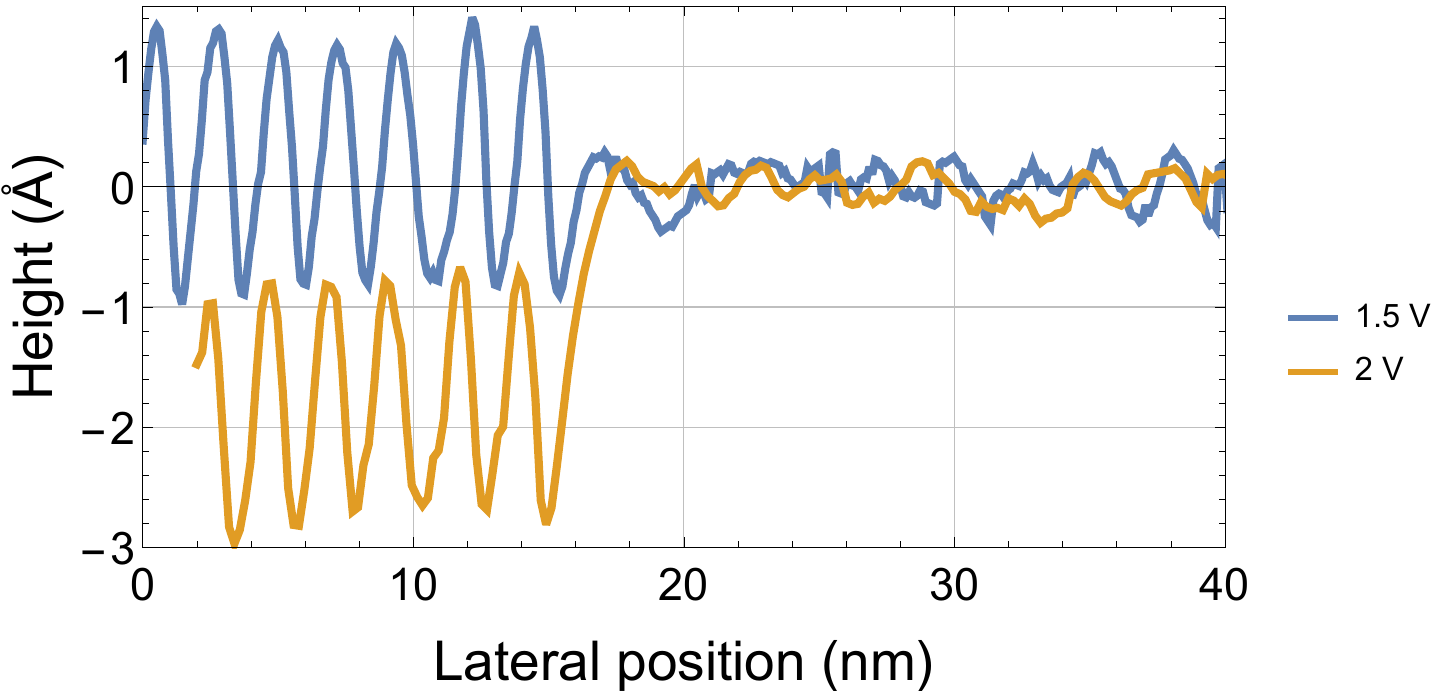}
\includegraphics[width=.75\columnwidth]{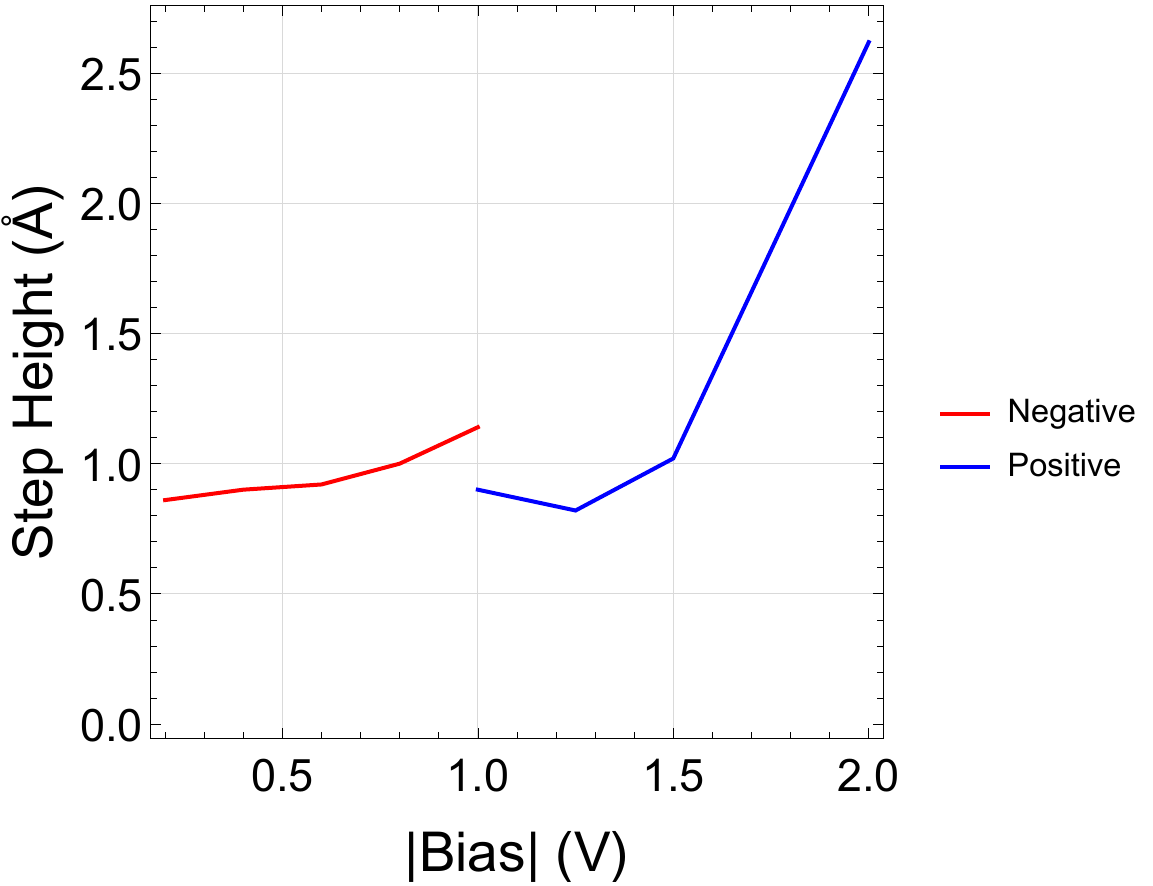}
\caption{\CO Variation of step height from \agit\ to \agih\ in Fig.~\ref{fig:AgIMixed} with bias voltage.}\label{fig:HvsV}%
\end{figure}
The apparent step height from the \agit\ to the \agih\ reconstruction varies from 0.7~\AA\ at low absolute bias voltage  to 2.5~\AA\ at bias voltages $\ge 2$~V (Fig.~\ref{fig:AgIMixed}). This indicates that the \agih\ phase is more insulating than the \agit\ phase.

\begin{figure}%
\includegraphics[width=.55\columnwidth]{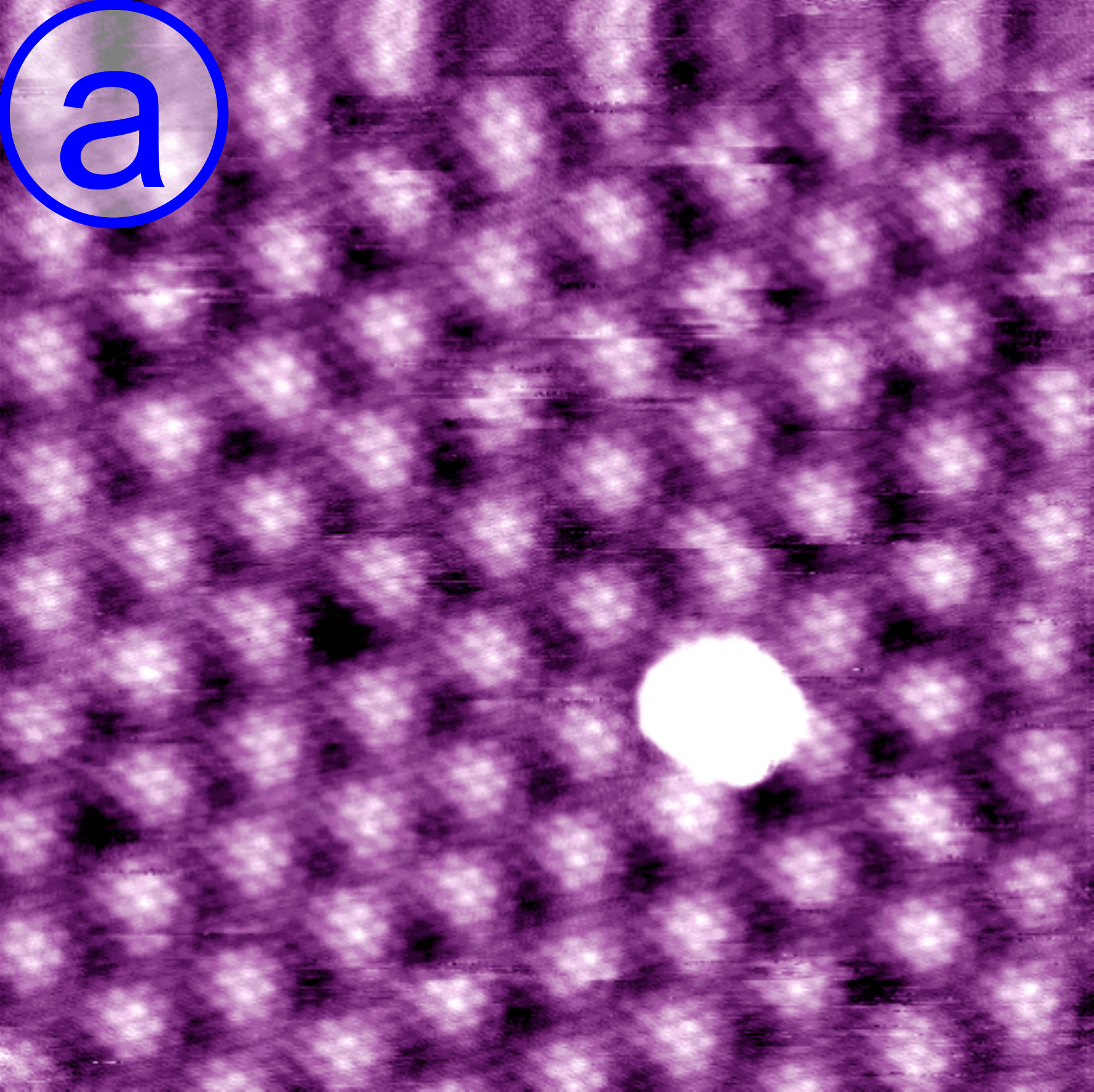}
\caption{\CO Imaging the \agih\ phase with a sub-optimal tip reproduces the STM image from a previous publication\cite{ANDRYUSHECHKIN201883}. (Imaging parameters:$25\times25$~nm$^2$, $\Vb=-650$~mV, $I=100$~pA)}\label{fig:hexbadtip}%
\end{figure}
Fig.~\ref{fig:hexbadtip} shows the \agih\ phase imaged with a suboptimal STM tip. The appearance resembles somewhat Fig.~23b) in a previous publication\cite{ANDRYUSHECHKIN201883}. Hence we assume both images to be of the \agih\ phase. 

\section{Tip induced modifications}
\label{app:tipmodi}

\begin{figure}%
\includegraphics[width=.485\columnwidth]{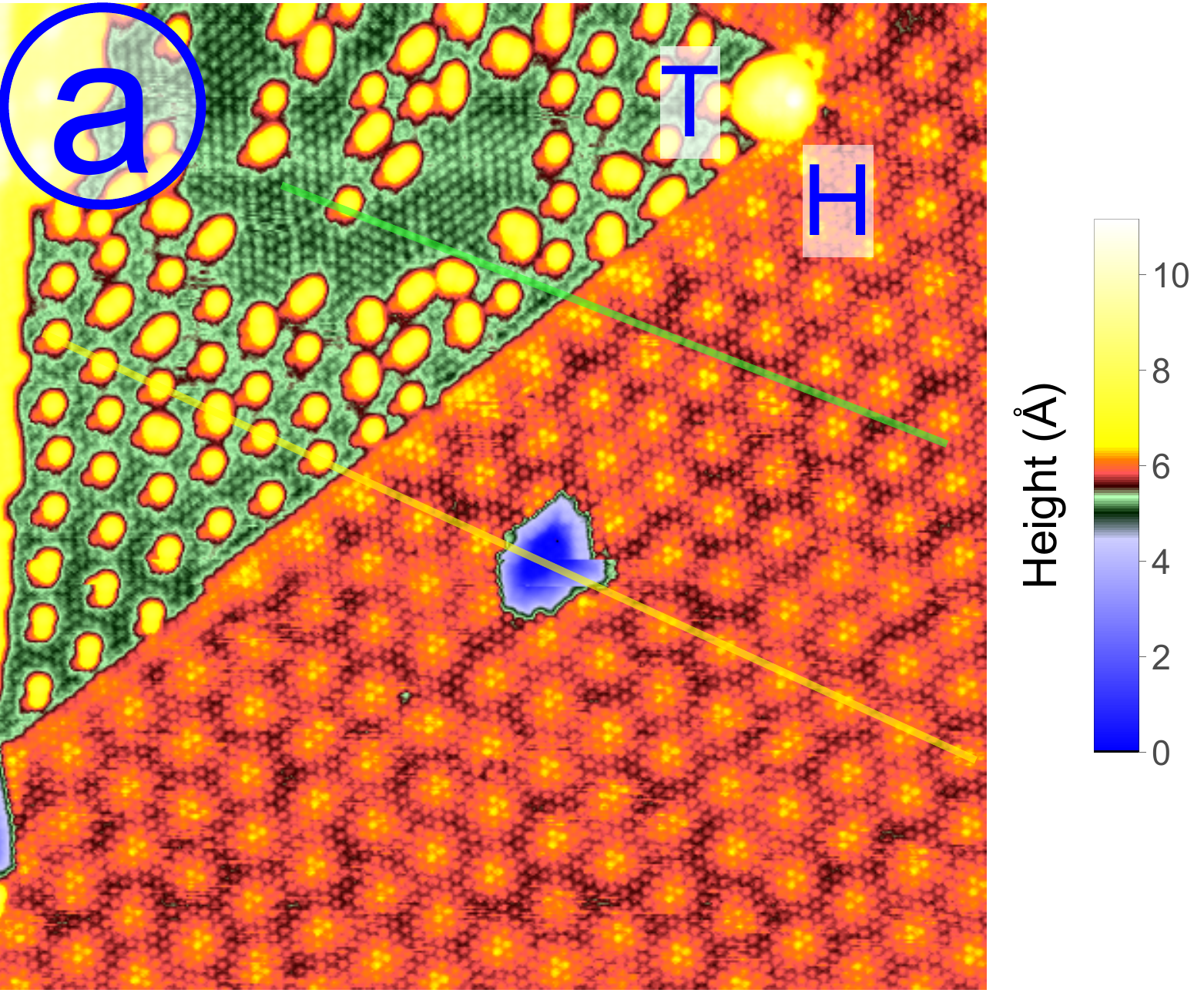}\hfill%
\includegraphics[width=.485\columnwidth]{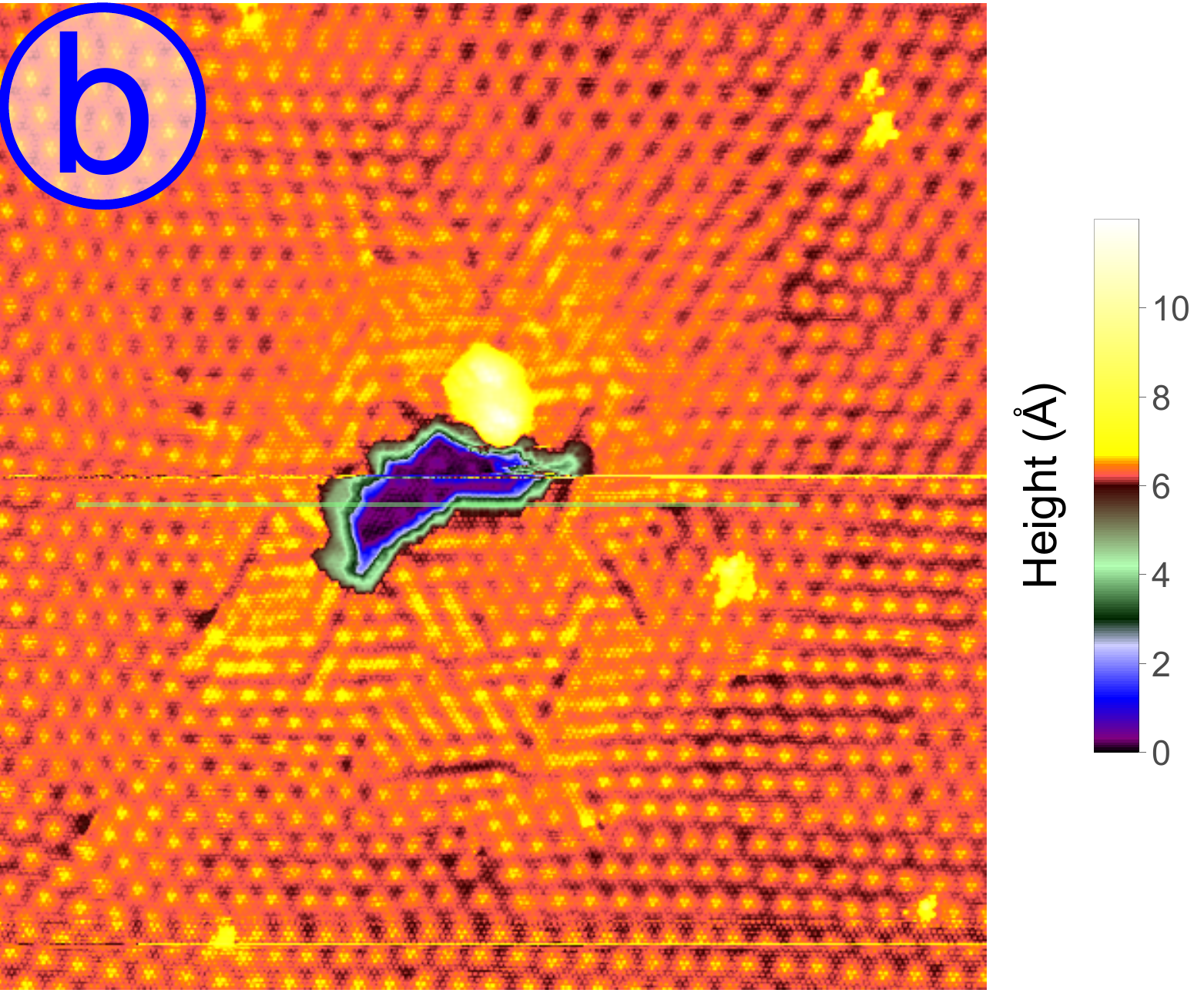}

\vspace{.1cm}

\parbox{.485\columnwidth}{\includegraphics[width=.485\columnwidth]{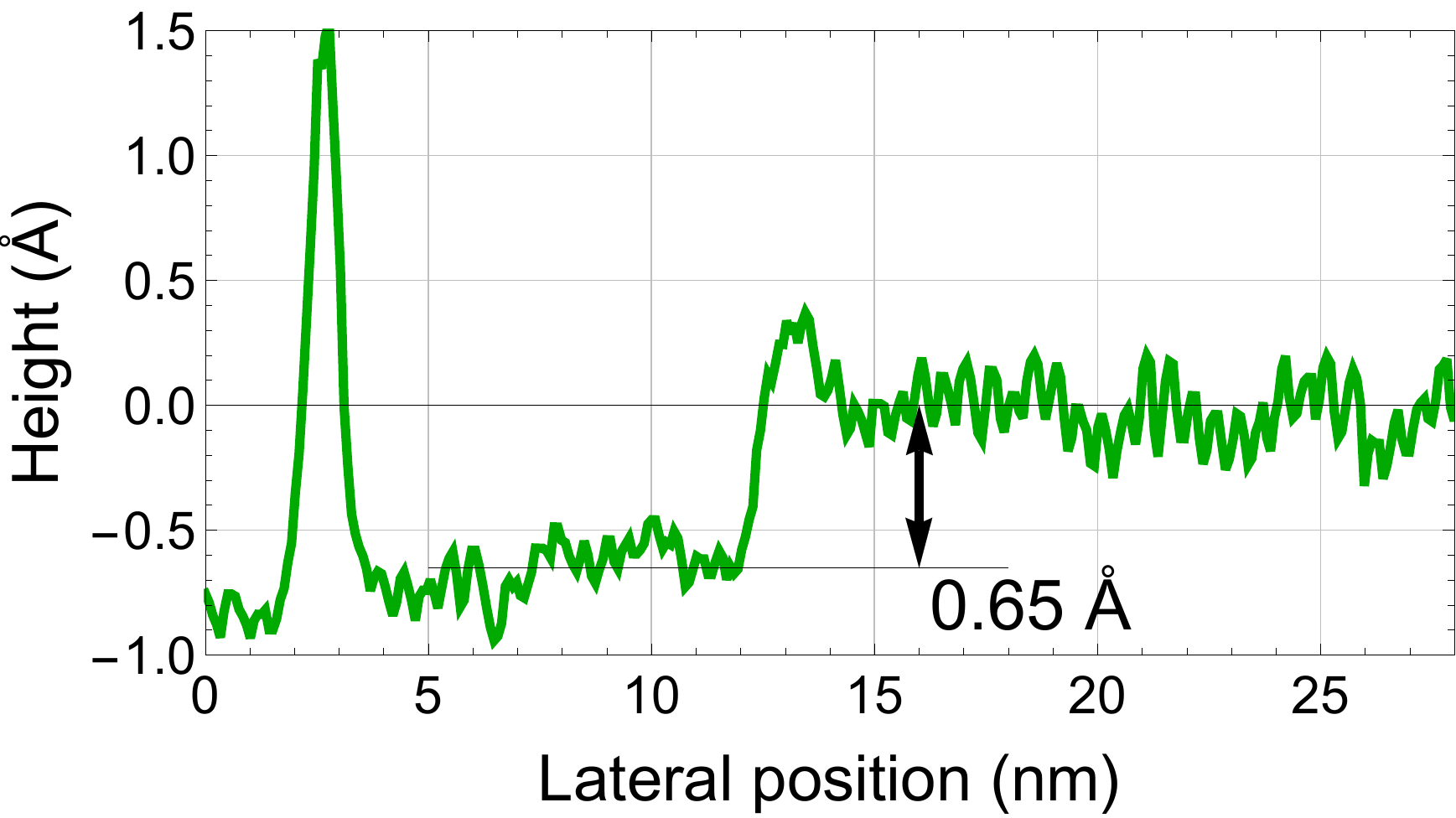}
\includegraphics[width=.485\columnwidth]{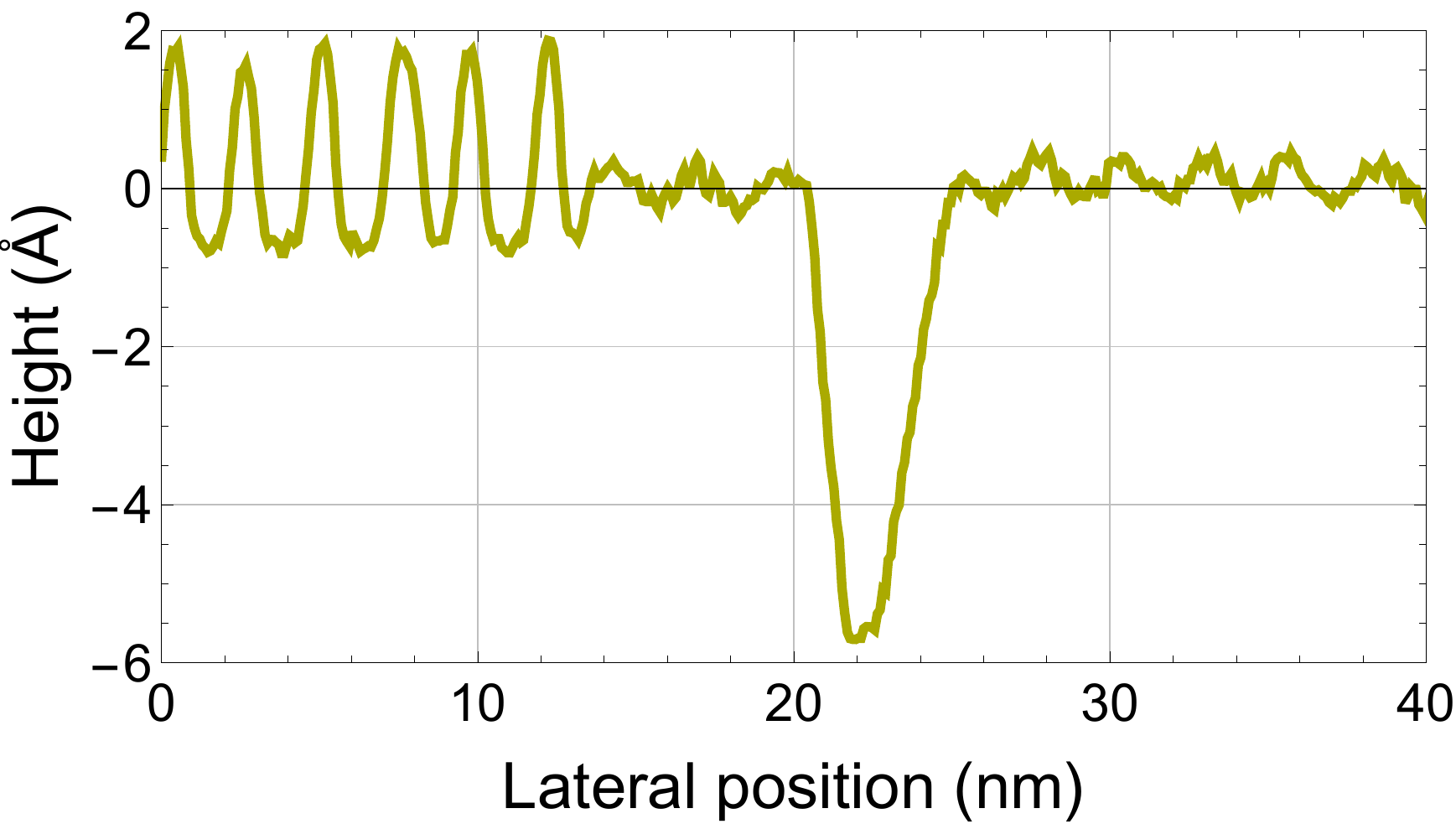}}\hfill%
\includegraphics[width=.485\columnwidth]{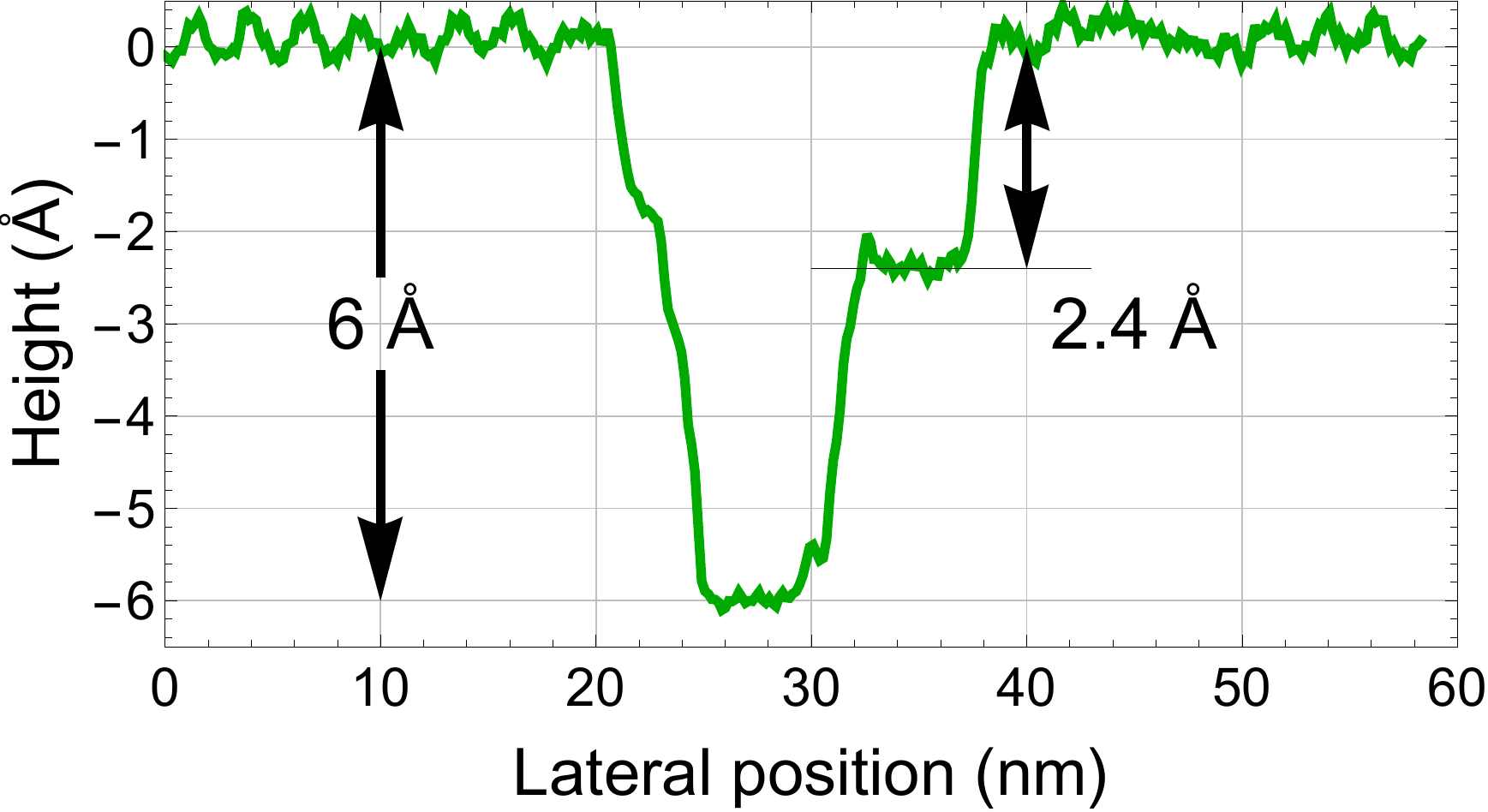}%
\caption{\CO Tip induced modifications. a) Both the \agit\ (T) and \agih\ (H) atomic structure can be modified by the tip. The green and yellow section show the apparent step height (green) and the depth of the hole (yellow). b) Deep hole produced by multiple passes on the \agih\ phase allows to measure a layer thickness of 6~\AA\ as shown in the section along the green line. (Imaging parameters: (a) $40\times40$~nm$^2$, $\Vb=-200$~mV, $I=100$~pA, (b) $80\times80$~nm$^2$, $\Vb=-750$~mV, $I=200$~pA)}\label{fig:AgImod2}%
\end{figure}

The \agih\ structure becomes unstable at bias voltages $\Vb\geq 3.0$~V
% (?) 
leading to pit formation when the STM tip repeatedly passes over an area. Two examples are shown in Fig.~\ref{fig:AgImod2} along with cross sections. Fig.~\ref{fig:AgImod2}a) shows a subset of the area shown in Fig.~\ref{fig:AgIMixed}b). The adatoms of the \agit\ phase have been disturbed after taking a spectroscopic map in the area with a voltage range of $\Vb=\left[1.0,-1.0\right]$~V. Topographic imaging at these voltages usually leaves the atomic structure unperturbed. The \agih\ phase was not disturbed by the STS measurement but subsequently modified by repeated imaging of a small area at $3$~V.
%(? might have been a pulse to start)
The cross sections show a height change of $0.65$~\AA\ from the \agit\ to the \agih\ phase (green). The step height measurement varies with voltage (App.~\ref{app:hexsuper}, Fig.~\ref{fig:HvsV}). This might indicate a reduced density of states of the \agih\ phase and makes it difficult to ascertain the layer thickness. The yellow cross section shows a corrugation of $0.6$~\AA\ for the \agit\ phase. The hole shows a depth of about $6$~\AA, but without a clear flat bottom. In a second area (Fig.~\ref{fig:AgImod2}b) we repeated the process to create a larger hole. Here the cross section clearly shows a flat bottom and a depth of $6.0$~\AA\ with a shelf at $2.4$~\AA. The later might be due to a double tip artifact. It is also noteworthy that in this case the superstructure showed distortions while for the smaller hole in Fig.~\ref{fig:AgImod2}a) the superstructure remained intact. Also, in both cases, the \agih\ phase is atomically resolved up to the edges of the holes as well as at the boundary with the \agit\ phase in Fig.~\ref{fig:AgImod2}a).

\section{Supplemental \protect\agit}
\label{app:trisuper}

\subsection{Selective FFT filtering}
\label{app:triselfilt}

\begin{figure}%
\includegraphics[width=.99\columnwidth]{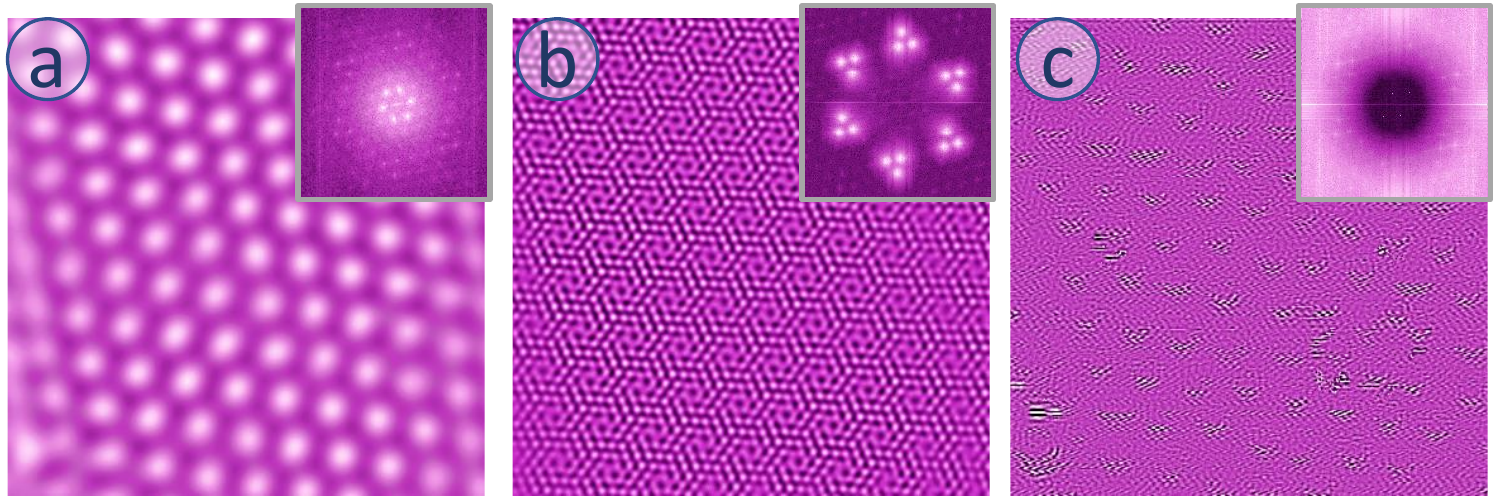}%
\caption{\CO Selective FFT filtering of the \agit\ phase. a) superstructure peaks. b) \agit\ peaks. c) 5\textsuperscript{th} order Butterworth high-pass filter.}\label{fig:triselfilt}%
\end{figure}
FFT filtering allows one to easily correlate peaks in the FFT spectrum to topographic features. Fig.~\ref{fig:triselfilt} shows an example for the \agit\ structure as filtered real space images together with its FFT after applying the respective filter. The inner spots in FFT (a) clearly belong to the superstructure, although the double occupancy at certain sites is lost. Filtering on the \agit\ FFT peaks (b) leaves the visible iodine monolayer showing the outer rosette structure. The center portion remains diffuse since it was covered up by the adatoms and thus cannot be reconstructed by inverse FFT. Finally, high-pass filtering with a cutoff length scale of $\sim 3.5$~\AA\ shows noisy features near the adatom position. This is the residual effect of variations of the adatom positions on the Ag (111) lattice. 

\subsection{\agit\ domains bound to step edge}
\label{app:tridom}

\begin{figure}%
\includegraphics[width=.32\columnwidth]{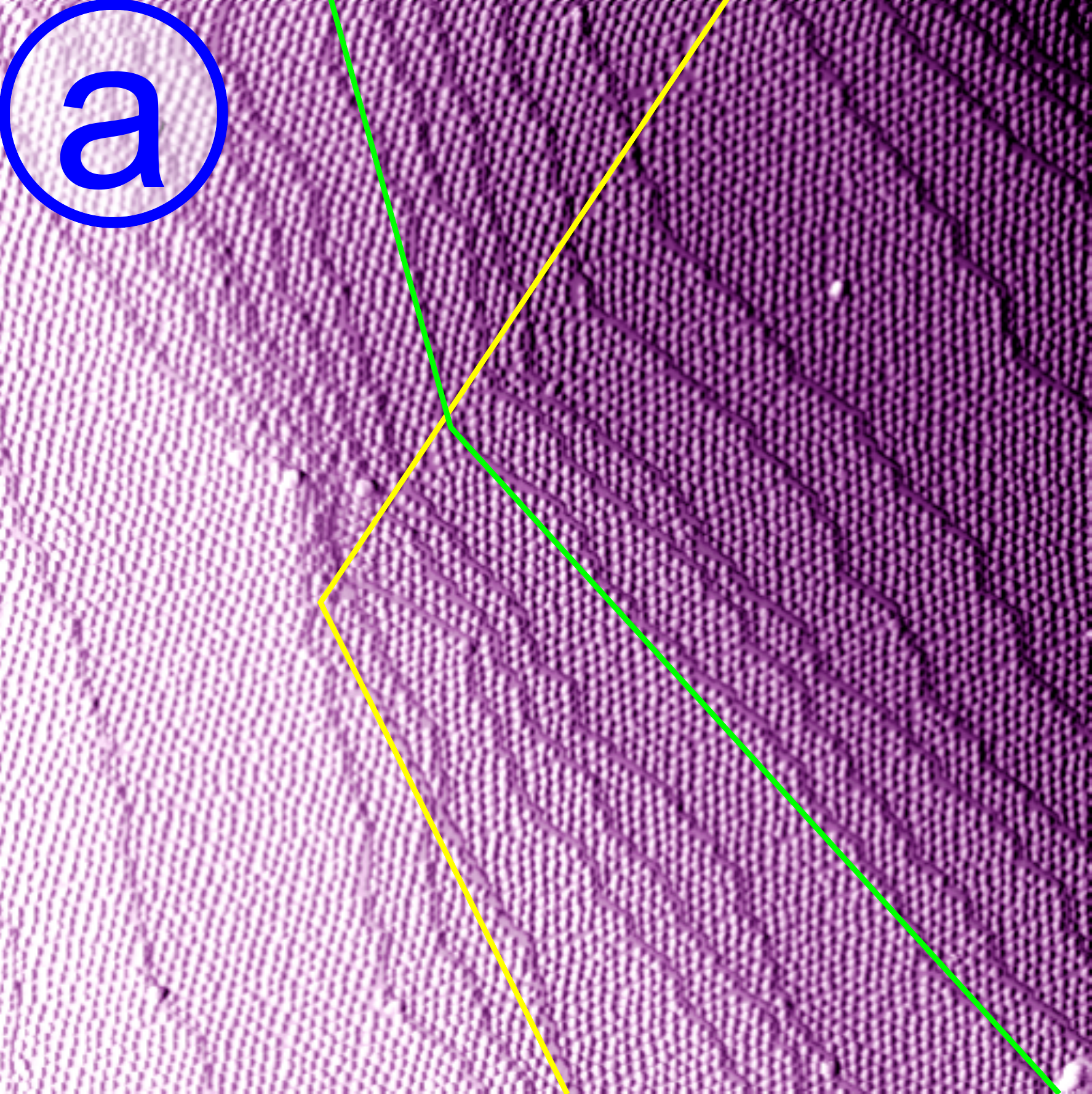}\hfill%
\includegraphics[width=.32\columnwidth]{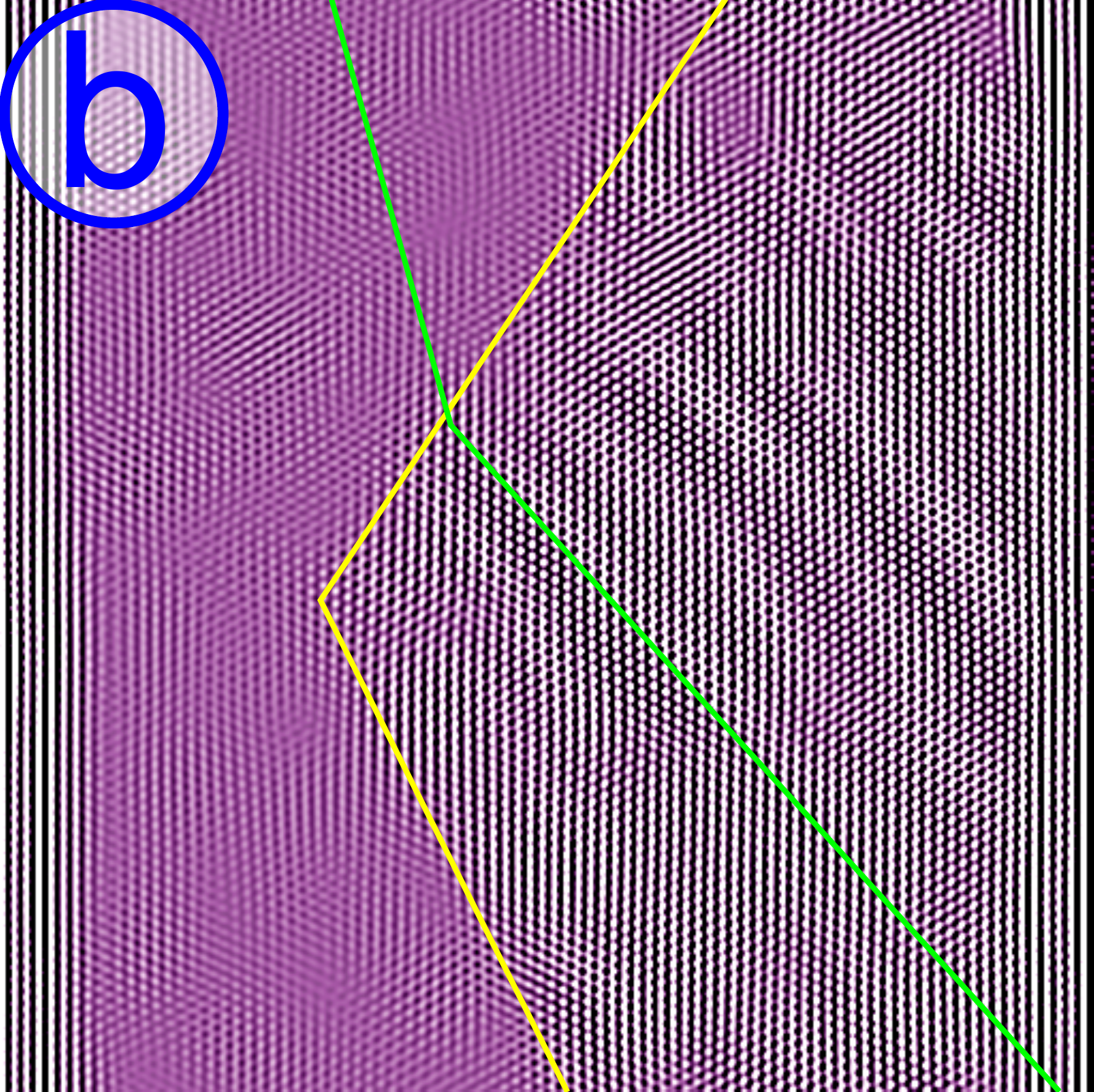}\hfill%
\includegraphics[width=.32\columnwidth]{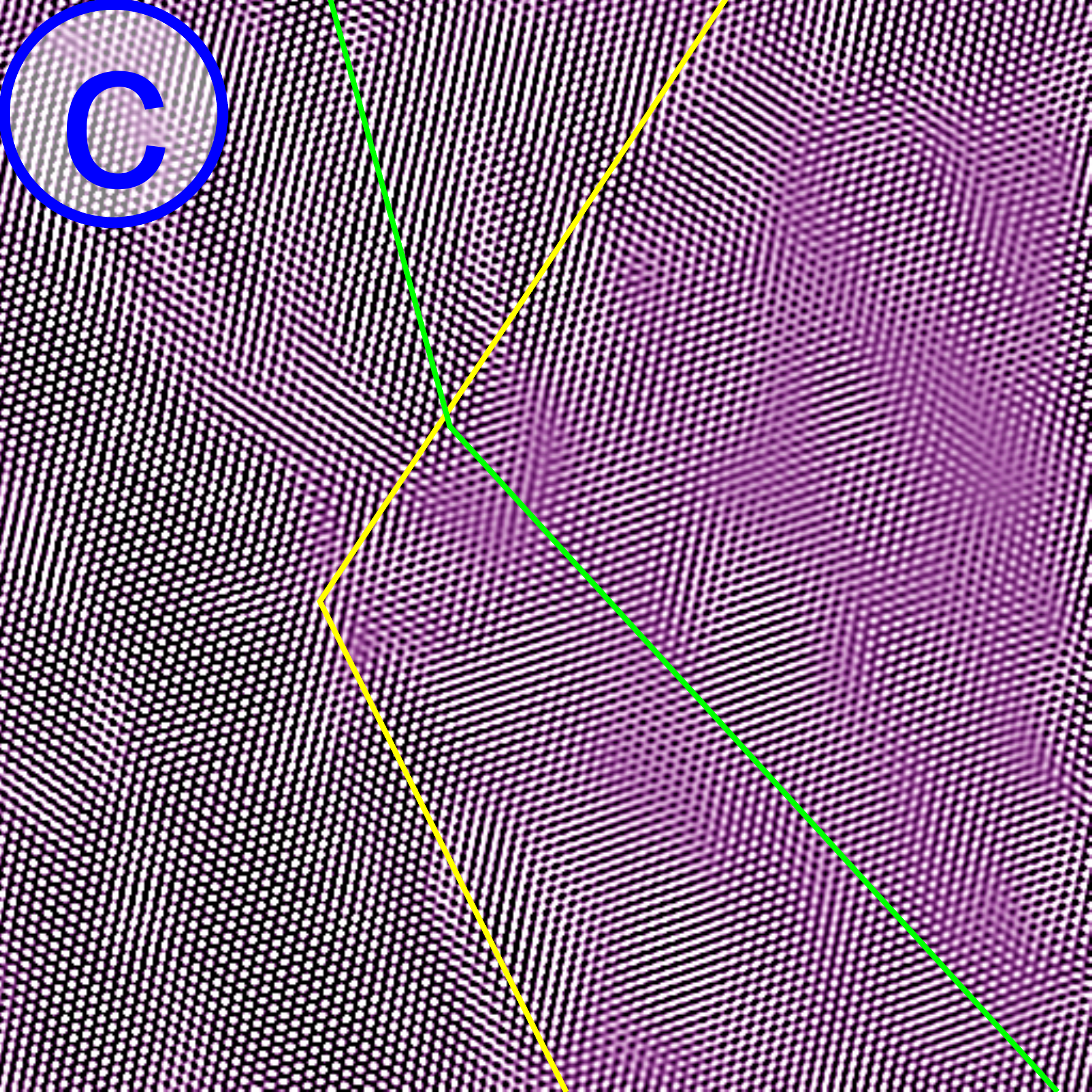}%
\caption{\CO a) Area showing the \agit\ phase with a slight change in prevalent Ag (111) step direction (green line) of $30^\circ$. b,c) Filtered images highlight two mirror domains of the \agit\ phase with estimated domain boundary shown in yellow. (Imaging parameters: $200\times200$~nm$^2$, $\Vb=750$~mV, $I=75$~pA)}\label{fig:apptds}%
\end{figure}

Figure \ref{fig:apptds}a) shows a stepped area with two \agit\ mirror domains. We used FFT filtering to highlight either domain, shown in Fig.~\ref{fig:apptds}b) and c). The estimated domain boundary is marked with a yellow line with an inscribed angle of $\alpha_\mathrm{domain}\approx 122^\circ$. The green line marks the estimated change in step direction of $\beta_\mathrm{step}\approx 27^\circ$. Comparing the domain boundary to the Ag (111) step direction in Fig.~\ref{fig:apptds}a) a linkage becomes apparent as the domain boundary (yellow) also demarcates the $\sim 30^\circ$ Ag (111) step domains.

\subsection{Semi-automatic Adatom Classification}
\label{app:adatoms}

\begin{table}%
\begin{tabular}{l|l|l|l|l|l|l}
data set&Total&single&double&ang1&ang2&ang3\\
\hline
% 90809B07.tfr
 1 & 165 & 62.4 \% & 37.6 \% & 
 \ot{25.8}{24.8} & \ot{45.2}{91.3} & \ot{29.0}{150.4}\\
% 90809b05.tfr
 2 & 1795 & 58.4 \% & 41.6 \% & 
 \ot{23.5}{30.2} & \ot{40.3}{87.0} & \ot{36.2}{152.6}\\
% 91009a08.tfr
 3 & 1849 & 70.4 \% & 29.6 \% & 
 \ot{35.1}{12.6} & \ot{38.1}{75.1} & \ot{26.7}{148.7}\\
% 91008A08.tfr
 4 & 2163 & 65.0 \% & 35.0 \% & 
 \ot{37.7}{13.7} & \ot{25.9}{97.8} & \ot{36.4}{156.0}\\
 91008A06.tfr
 5 & 517 & 66.2 \% & 33.8 \% & 
 \ot{34.3}{19.0} & \ot{35.4}{97.3} & \ot{30.3}{148.3}\\
\end{tabular}
\caption{Summary of adatom statistic over 5 data sets and two separate sample preparations. Given are the total number of adatom sites in the image as well as the fraction of single and double occupancy. For the double occupancies we give the the fraction for each of the three possible angles.}
\label{tab:adatoms}
\end{table}

\begin{figure}%
\includegraphics[width=.32\columnwidth]{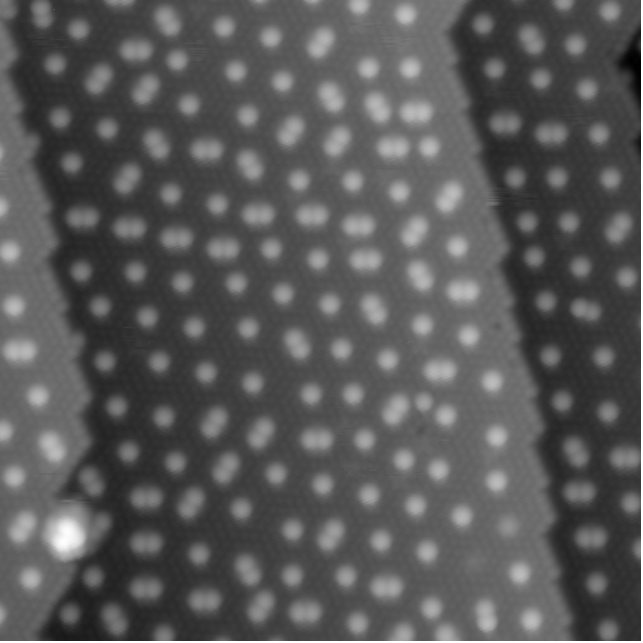}\hfill%
\includegraphics[width=.32\columnwidth]{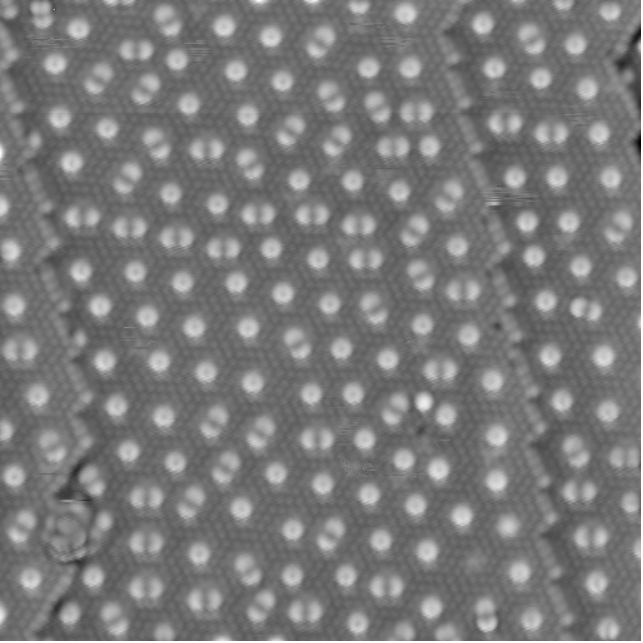}\hfill%
\includegraphics[width=.32\columnwidth]{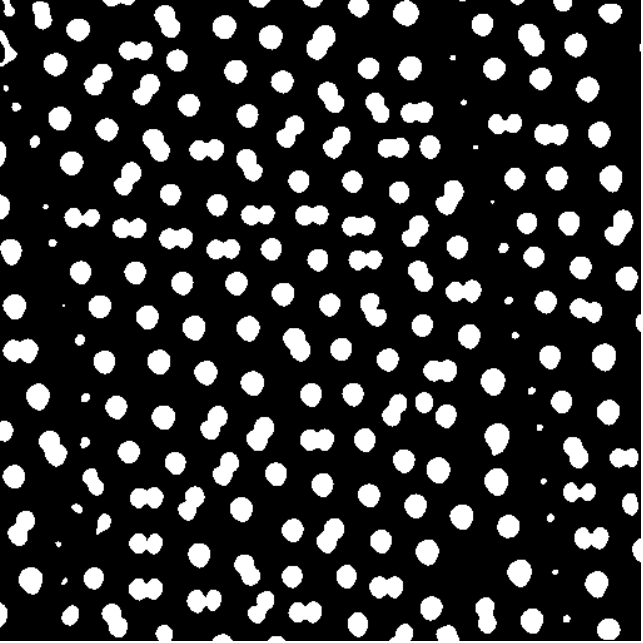}
\includegraphics[width=.32\columnwidth]{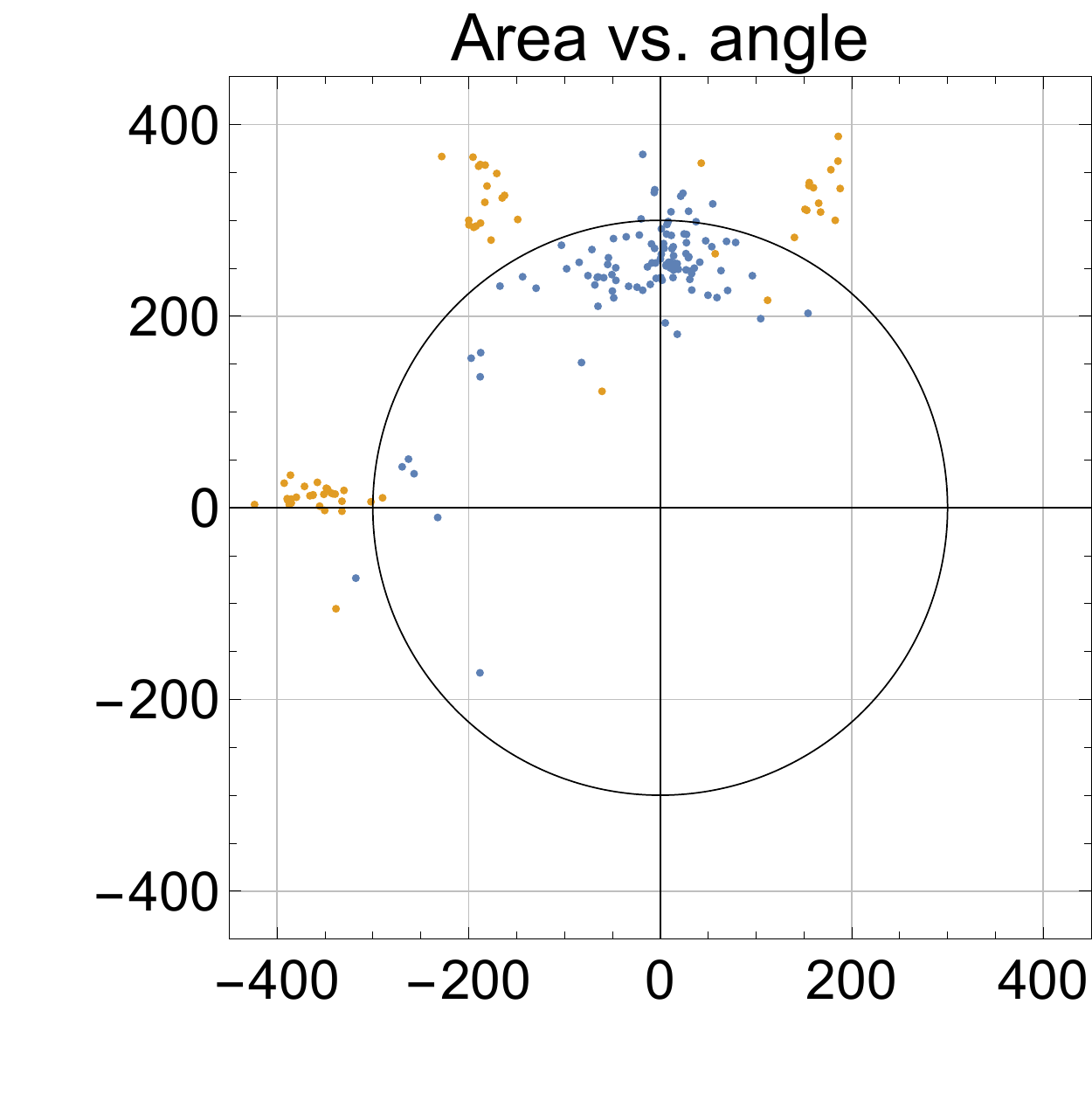}\hfill%
\includegraphics[width=.32\columnwidth]{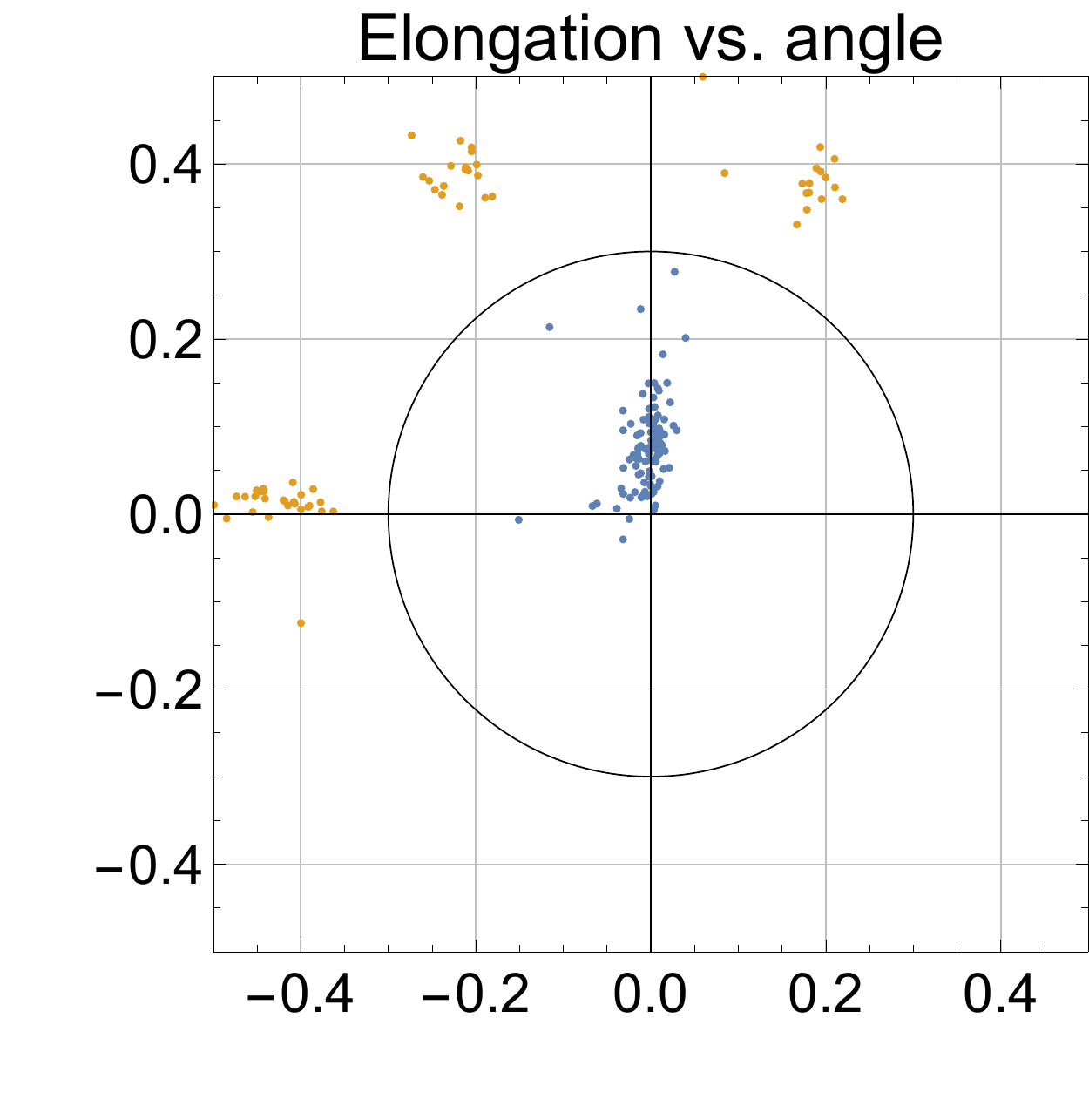}\hfill%
\includegraphics[width=.32\columnwidth]{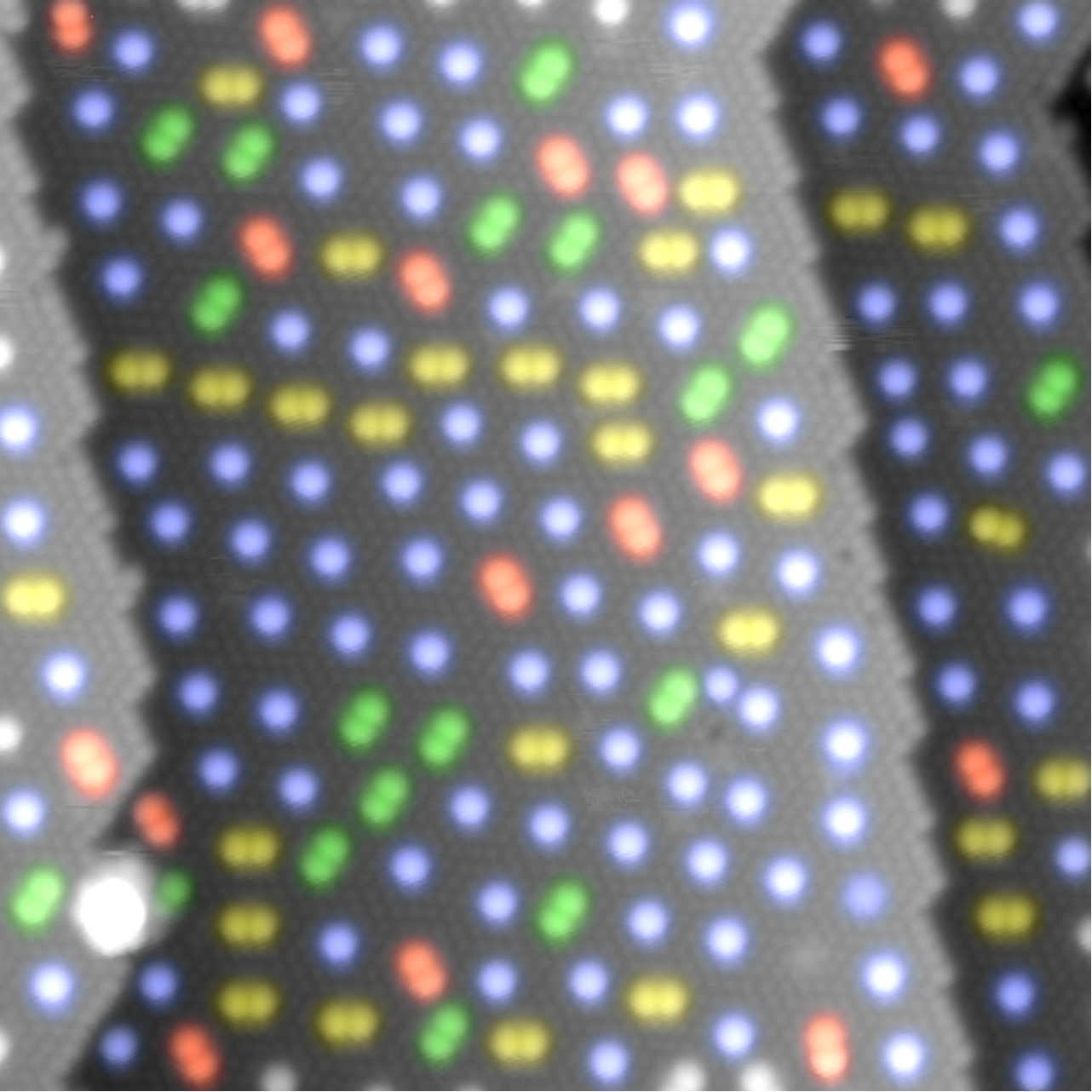}%
\caption{\CO Steps in semiautomatic adatom classification. left to right: STM image, background removal, binarization, and classification. The bottom row shows polar plots of the extracted area and elongation, respectively, vs. angle of the long axis. The bottom right color codes the adatoms according to classification. Blue: single adatom; green, yellow, red: dimers sorted by angle. (Imaging parameters: $30\times30$~nm$^2$, $\Vb=500$~mV, $I=200$~pA)}\label{fig:appada1}%
\end{figure}
We used Mathematica to identify and classify the adatoms found on the \agit\ phase. An example is shown in Fig.~\ref{fig:appada1}. The image was filter followed by binarization and extractions of cohesive areas. From areas above a pixel threshold of typically 5 we extracted the centroid position as well as other parameters such as area ($A$). The elongation ($E$) and angle of the long axis ($\theta$) where determined by a fit to an equal-area ellipse. Fig.~\ref{fig:appada1} shows polar plots of $A(\theta)$ and $E(\theta)$. The latter was used to distinguish between single or double occupancy and to classify $\theta$ for the adatom dimers. The angle of the dimers lies in three direction $\sim 30^\circ$ of the direction of the Ag (111) with a seemingly random distribution. The estimated inter-dimer distance was 7.7 \AA. All real space measurements followed this scheme. The filter function and binarization threshold had to be adapted for the respective data set. 

%(r7xr7)R
%Image set needs to meet criteria:
%-atomic resolution
%scan size: tradeoff statistics vs angle precision. 

\begin{figure}%
\includegraphics[width=.65\columnwidth]{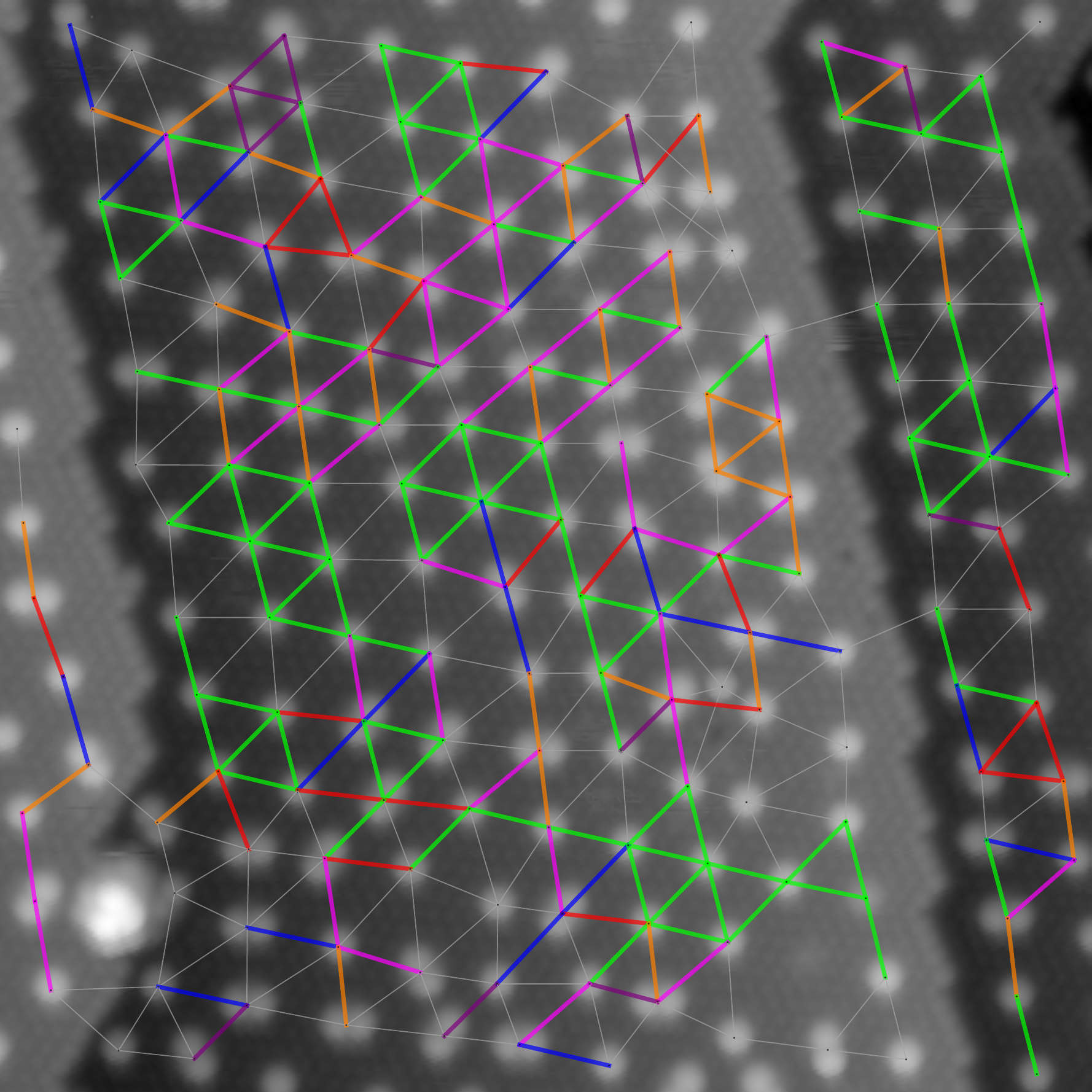}\hfill%
\includegraphics[width=.33\columnwidth]{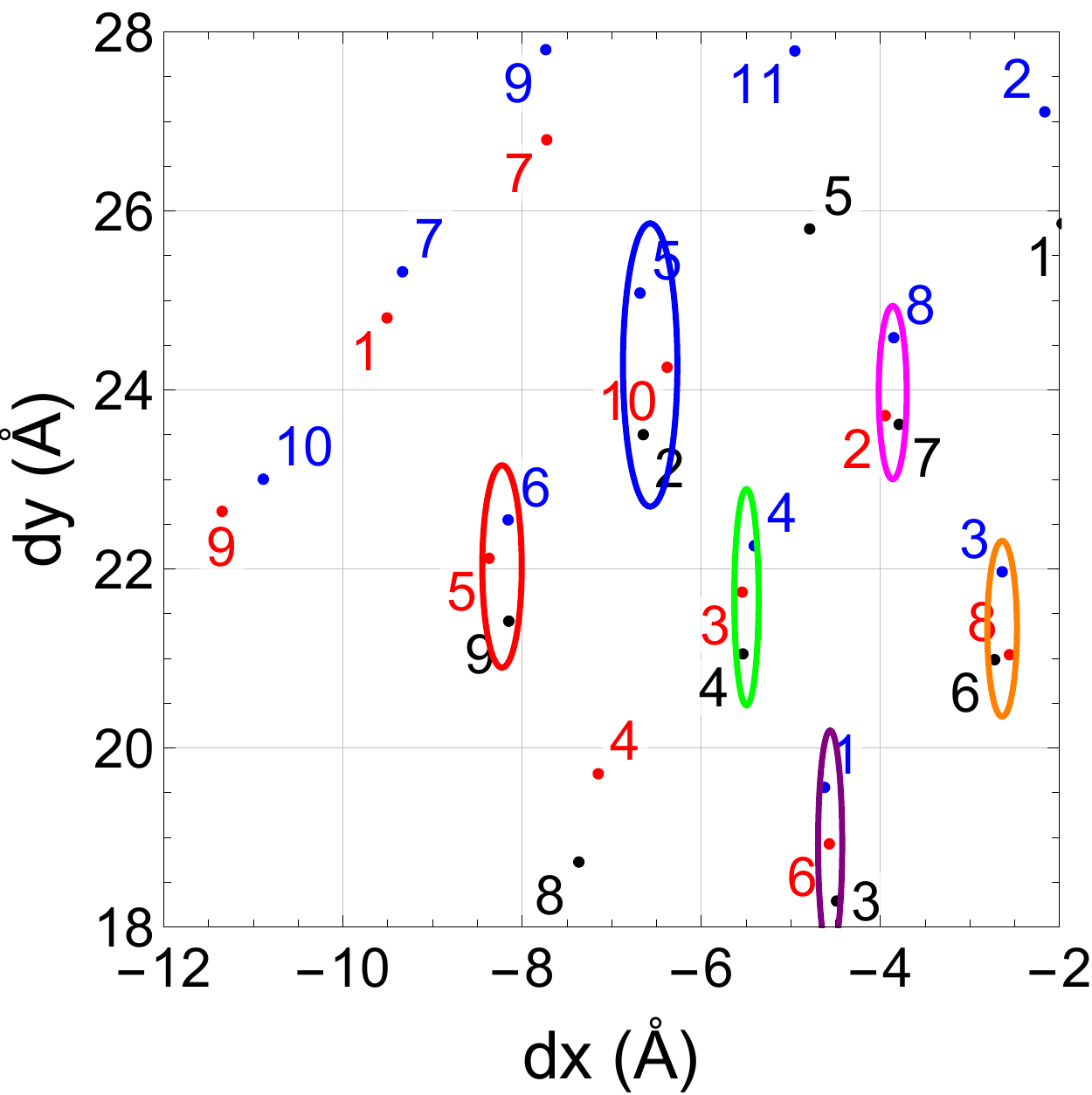}
\caption{\CO Nearest neighbor network of the \agit\ phase adatoms colorized by edge classification.}\label{fig:TriColAd}%
\end{figure}
For a deeper analysis we colored the edges of the adatom nearest neighbor network according to their length, similar to what we did for the \agih\ phase shown in Fig.~\ref{fig:AgIhl1}. The result is shown in Fig.~\ref{fig:TriColAd}. The three regions in the  NNN edge distribution plot were overlapped by rotating them by an angle of $\sim 60^\circ$ and $120^\circ$ degrees, respectively. However, since there are up to 15 different length scales to consider (the example shows coloring for just 6), this method is of limited use. 

\begin{figure}%
\includegraphics[width=.95\columnwidth]{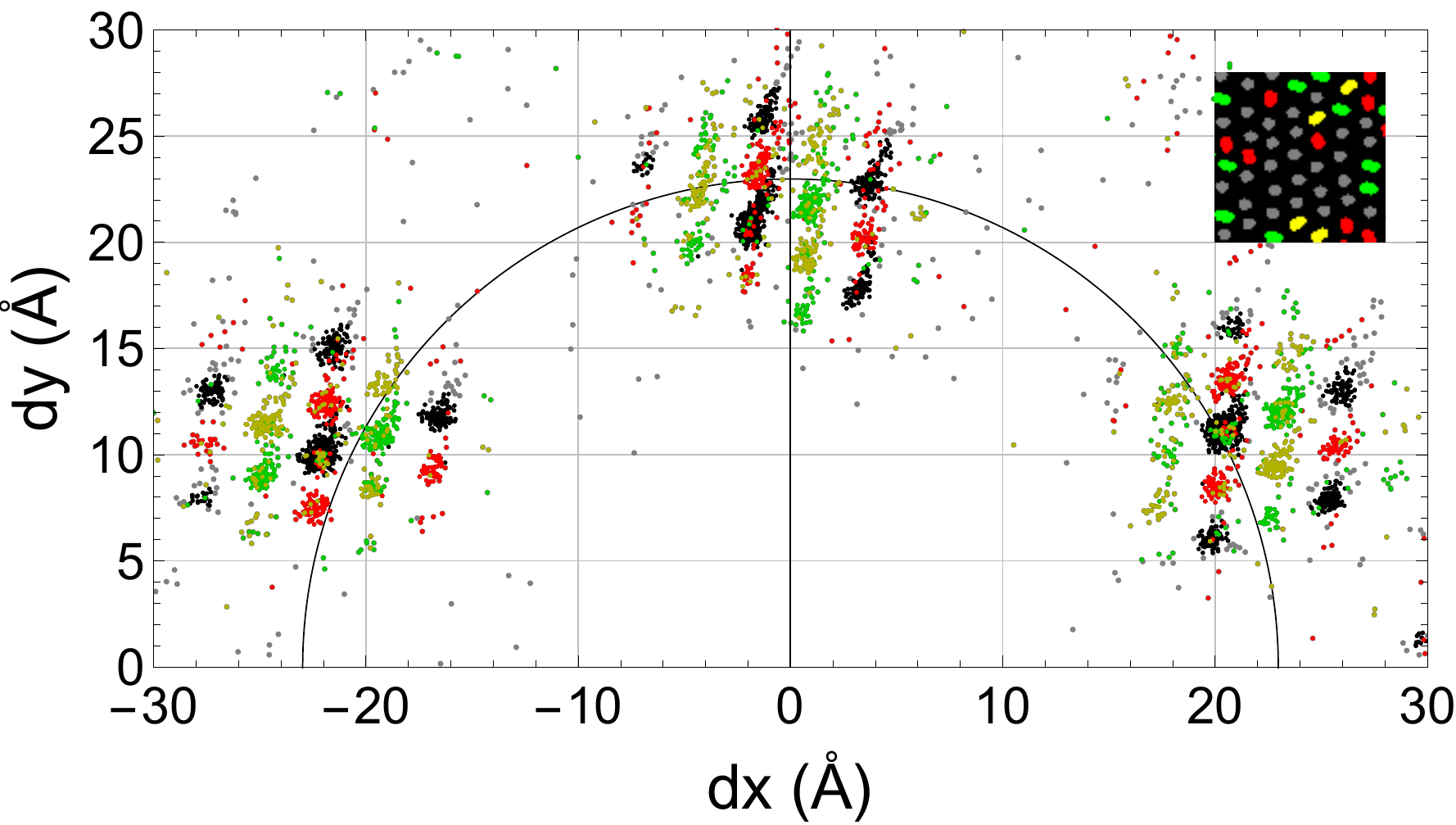}
\caption{\CO NNN distance vector plot of real-space measurements of the adatom distance for dual occupied sites (red, green, yellow) for the three possible dimer orientations.}\label{fig:Tridouble}%
\end{figure}
Some insight could be gained by finding the distances involving dimer-occupied sites. Fig.~\ref{fig:Tridouble} shows an example. The points in the NNN distance vector plot of the adatom lattice are colored if they involve a dimer site. The colors are chosen according to the major axis angle (see inset). This distinguishes the cluster points and hints at a coupling of the occupancy number and dimer orientation to the local strain build into the iodine monolayer. 

\begin{figure}%
\includegraphics[width=.95\columnwidth]{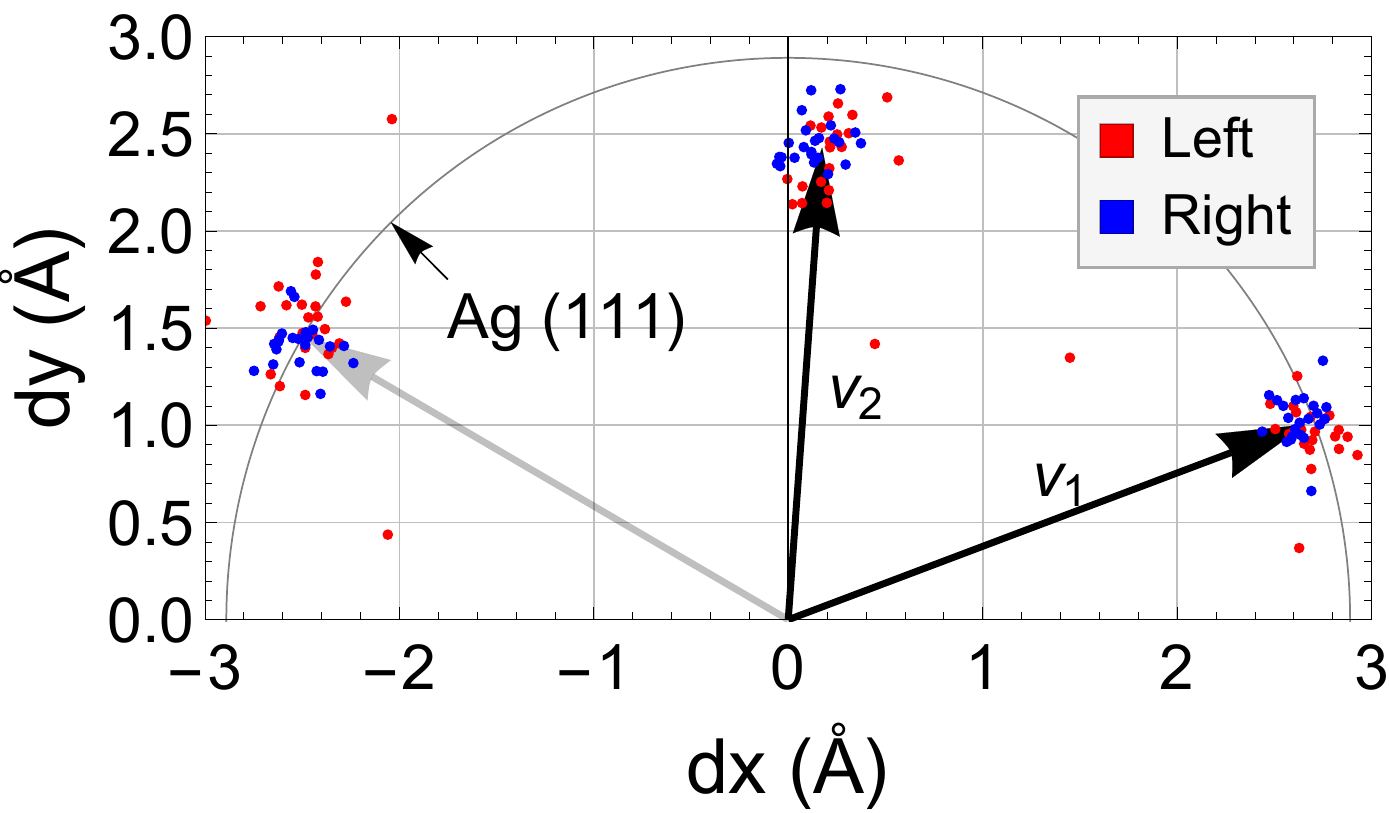}\hfill%
\includegraphics[width=.95\columnwidth]{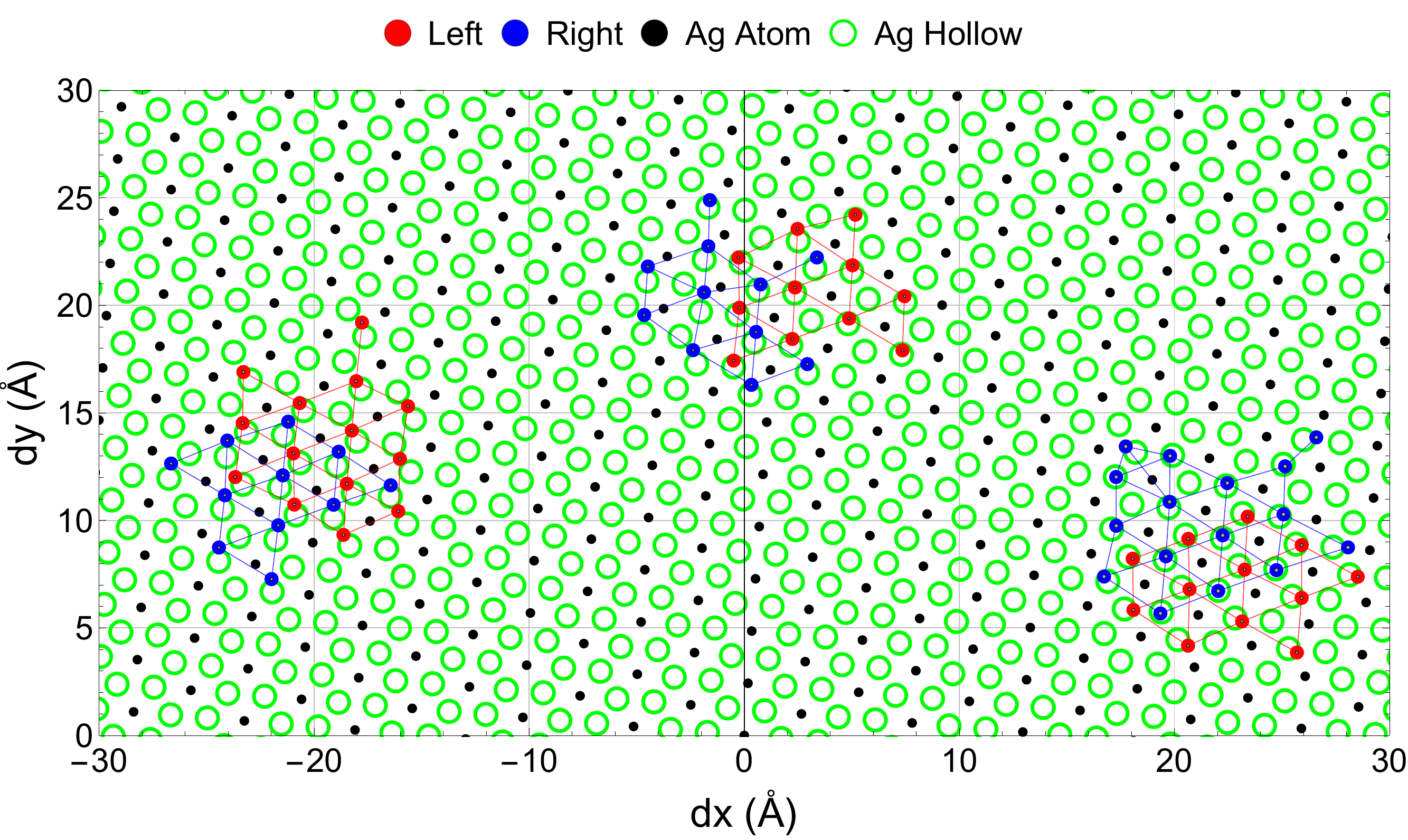}%
\caption{\CO Distance vector plot of the variations within the adatom distances from Fig.~\ref{fig:AgItlm1} used to reconstruct the Ag lattice. To accommodate both domains simultaneously the adatoms need to sit in the three fold corner positions.}\label{fig:Trimirat}%
\end{figure}
The mirror domains (Fig.~\ref{fig:Agtridom}) allow to locate the center of the rosettes on Ag (111) hollow sites. First, the distances within the NNN distance vector plots of the adatoms for both domains give three base vectors for a single Ag (111) surface (Fig.~\ref{fig:Trimirat}, top). Selecting two base vectors we can reconstruct the Ag (111) lattice including the x-y piezo asymmetry due to piezo creep. Overlaying the adatom NNN vector distribution shows hollow sites as a plausible location for the rosette centers.

\begin{figure}%
\includegraphics[width=.85\columnwidth]{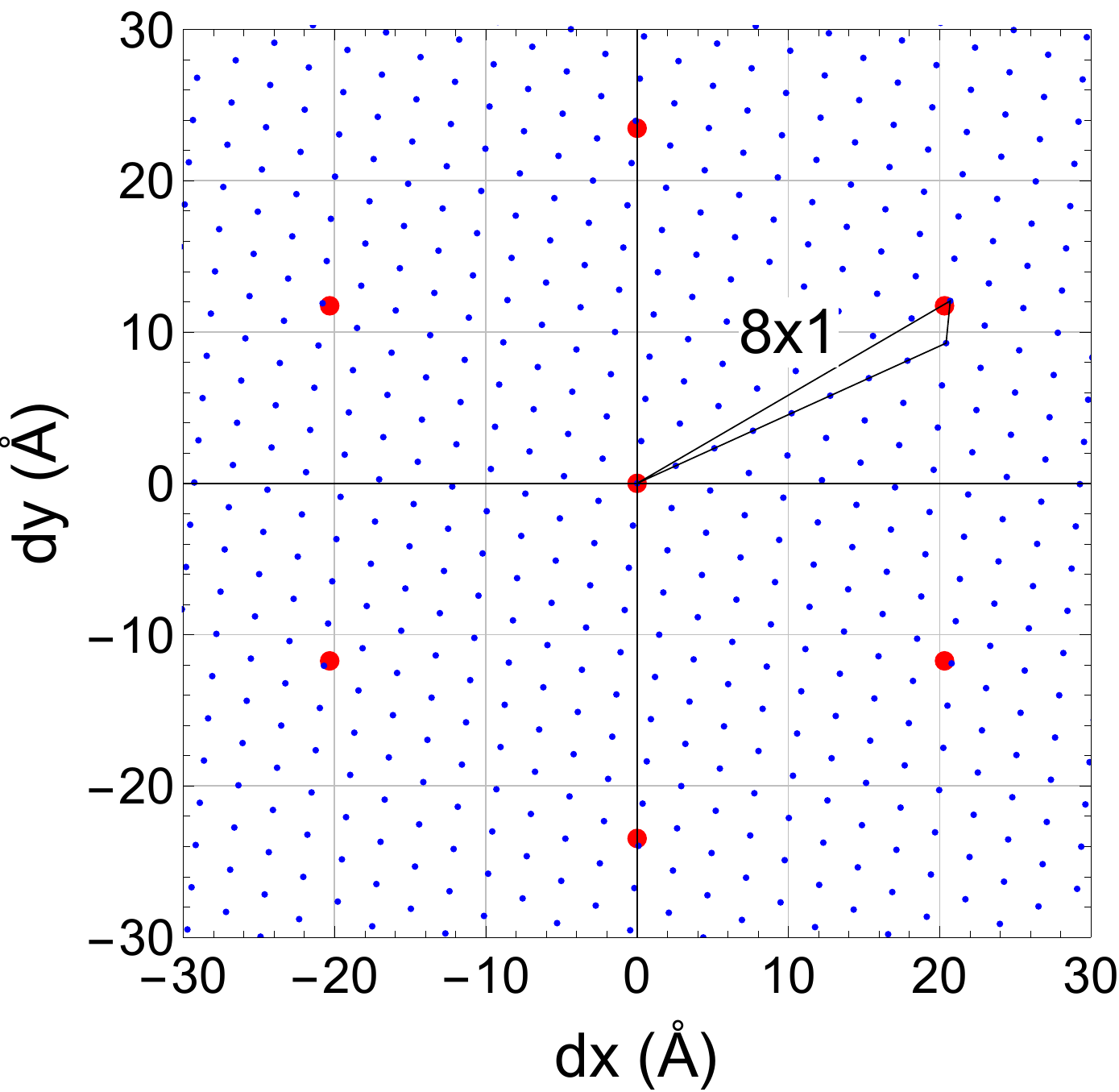}\hfill%
\includegraphics[width=.85\columnwidth]{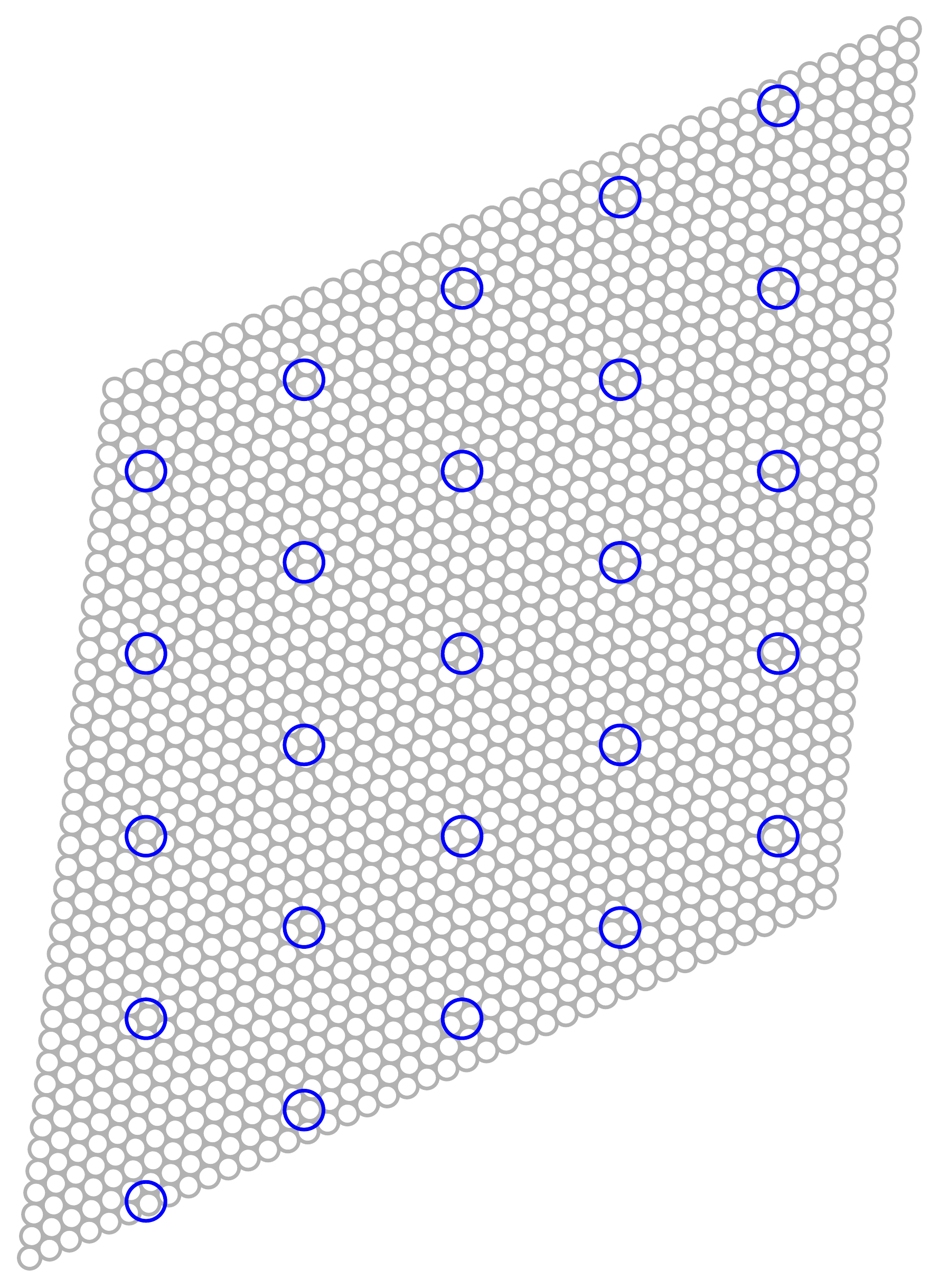}%
\caption{\CO Simple model of the Ag surface atoms with an almost $8\times1$ reconstruction of adatoms.}\label{fig:Trimodel}%
\end{figure}
Figure \ref{fig:Trimodel} shows a simple model for the rosette centers of the \agit\ phase. The blue points mark Ag (111) locations and the red points mark a $8\times1$ lattice --- the average lattice vector for the adatoms. The lower figure demonstrates that a simple $8\times1$ can not maintain hollow sites consistently. Therefore local distortions are expected.  

\end{document}